\documentclass[aps,prd,onecolumn,amsmath,11pt,superscriptaddress,floatfix,nofootinbib,preprintnumbers]{revtex4-2}

\usepackage{mathrsfs}
\usepackage{amsfonts}
\usepackage{amsmath,amssymb,bm}
\usepackage{array}
\usepackage{verbatim}
\usepackage{epsfig}
\usepackage{graphicx}
\usepackage{hyperref}
\hypersetup{colorlinks, linkcolor = [rgb]{0, 0, 0.5}, citecolor = [rgb]{0,0.0,0.5}, urlcolor = [rgb]{0,0.0,0.5}}
\usepackage[normalem]{ulem}
\usepackage{xcolor}
\usepackage{multirow}

\usepackage{graphicx}
\graphicspath{{./figs/}}
\usepackage{dcolumn}
\usepackage{bm}
\usepackage{slashed}
\usepackage{siunitx}
\usepackage{ulem,xpatch}
\usepackage{hyperref}
\usepackage[mathlines]{lineno}
\usepackage{comment}
\usepackage{soul}
\usepackage{xcolor}
\usepackage{xfrac}

\allowdisplaybreaks[4]

\begin{document}

\title{Semi-inclusive production of spin-3/2 hadrons in deep inelastic scattering}

\newcommand*{\SDU}{Key Laboratory of Particle Physics and Particle Irradiation (MOE), Institute of Frontier and Interdisciplinary Science, Shandong University, Qingdao, Shandong 266237, China}\affiliation{\SDU}
\newcommand*{\SCNT}{Southern Center for Nuclear-Science Theory (SCNT), Institute of Modern Physics, Chinese Academy of Sciences, Huizhou 516000, China}\affiliation{\SCNT}

\author{Jing~Zhao}\affiliation{\SDU}
\author{Zhe~Zhang}\affiliation{\SDU}
\author{Zuo-tang~Liang}\affiliation{\SDU}
\author{Tianbo~Liu}\email{liutb@sdu.edu.cn}\affiliation{\SDU}\affiliation{\SCNT}
\author{Ya-jin~Zhou}\email{zhouyj@sdu.edu.cn}\affiliation{\SDU}

\begin{abstract}

We investigate the production of spin-3/2 hadrons in semi-inclusive deep inelastic lepton-nucleon scatterings. The complete differential cross section is derived through the kinematic analysis and expressed in terms of 288 structure functions, corresponding to all polarization configurations and azimuthal modulations. For an unpolarized lepton beam, half of the 192 structure functions have nonzero leading order contributions in the parton model, among which 42 are from rank-3 tensor polarized fragmentation functions of the hadron. For a polarized lepton beam, one third of the 96 structure functions contribute at the leading order and 14 of them are from rank-3 tensor polarized fragmentation functions. In addition to the formalism, we perform a model estimation of the spin transfer to a $S_{hLLL}$ polarized hadron and sizable asymmetry is expected. Therefore, these newly defined observables for the production of a spin-3/2 hadron in a deep inelastic scattering process can be explored in future experiments to understand nucleon spin structures and spin-dependent fragmentation functions.  
 
\end{abstract}
\maketitle

\section{Introduction}
\label{s.intro}

Nucleons as building blocks of our visible world are bound states of the strong interaction, which is described by the underlying theory of quantum chromodynamics (QCD). However, the nonperturbative nature of QCD at low-energy scales makes it challenging to derive all properties of the nucleon from first principle, although much progress has been made in the past decade~\cite{Liu:1993cv,Detmold:2005gg,Braun:2007wv,Ji:2013dva,Ji:2014gla,Chambers:2017dov,Radyushkin:2017cyf,Ma:2014jla,Ma:2017pxb}. Understanding internal structures of the nucleon from quarks and gluons degrees of freedom has become an active frontier of nuclear and particle physics.

Deep inelastic scattering (DIS) is an established powerful tool to probe nucleon internal structures. Taking advantage of the asymptotic freedom~\cite{Gross:1973id,Politzer:1973fx}, a feature of QCD at high-energy scales, one can approximate the DIS cross section as the lepton-parton scattering cross section convoluted with parton distribution functions (PDFs), $f_{q/P}(x)$, which is interpreted as the probability density of a parton of flavor $q$ that carries a fraction $x$ of the nucleon momentum. This parton model is then formally developed into the QCD factorization~\cite{Collins:1989gx}, which nowadays serves as the framework in nearly all analyses of hadron-involved high-energy scattering processes. 

The inclusive DIS, as a single scale process, is not sensitive to the confined motion of partons in the nucleon. To access the transverse momentum distribution, one needs to tag the struck parton in the final state. Because of the color confinement, quarks and gluons cannot be isolated as free particles, and only hadrons or jets can be observed. As an analog to the PDF, the fragmentation function (FF), $D_{q\to h}(z)$, is introduced to describe the probability density of the identified hadron $h$ that carries a fraction $z$ of the parent parton $q$. Then a semi-inclusive DIS (SIDIS) process, where a hadron is detected in addition to the scattered lepton, can be factorized into the convolution of the PDFs of the nucleon, the FFs to the hadron, and the short-distance hard scattering of partons. When the transverse momentum $\bm{P}_{hT}$ of the observed hadron is much smaller than the hard scale $Q$ characterized by the virtual photon, it becomes a double scale process, one can apply the transverse momentum dependent (TMD) factorization~\cite{Collins:1981uw,Collins:1981uk,Aybat:2011zv}, and the cross section is expressed in terms of TMD PDFs, $f(x,\bm{k}_T^2)$ and TMD FFs, $D(z,\bm{p}_T^2)$, where $\bm{k}_T$ and $\bm{p}_T$ are the transverse momenta of the parton with respect to the hadron.

Taking the spin degrees of freedom of the struck parton and the nucleon into account, one can learn much richer information of nucleon structures. Historically, the polarized DIS experiment pioneered by the EMC Collaboration~\cite{EuropeanMuon:1987isl,EuropeanMuon:1989yki} led to the {\it proton spin crisis} and invoked great interest in the nucleon spin physics. After more than three decades, the nucleon spin structure is still not well understood. Although quark helicity distributions have been constrained to relatively good precision at least in the intermediate-$x$ region, or the so-called valence region, the contributions from orbital angular momenta, which are essentially the correlation between the spin of the nucleon and the transverse motion of the parton, are very little known. Hence, a three-dimensional imaging of the nucleon spin structure is desired to fully resolve the proton spin puzzle. The spin-dependent TMD PDFs, as a kind of 3dPDFs, encode the correlation of quark and gluon three-dimensional momentum distributions and the spin of the parton or the nucleon. They can be extracted from polarized SIDIS experiments and many measurements have been carried out by HERMES~\cite{HERMES:1999ryv,HERMES:2001hbj,HERMES:2002buj,HERMES:2004mhh,HERMES:2005mov,HERMES:2020ifk}, COMPASS~\cite{COMPASS:2005csq,COMPASS:2006mkl,COMPASS:2022jth,COMPASS:2012dmt,COMPASS:2014bze}, and JLab~\cite{JeffersonLabHallA:2011ayy,JeffersonLabHallA:2014yxb}. The precise determination of spin-dependent TMD PDFs is also one of the main goals of future electron-ion colliders~\cite{AbdulKhalek:2021gbh,Anderle:2021wcy,Accardi:2012qut}.

Apart from the nucleon spin, the polarization of some final-state hadrons can also be analyzed with certain techniques.  Through the self-analyzing weak decay, the polarization of the $\Lambda$ hyperon has been measured in $e^+e^-$ annihilation~\cite{ALEPH:1996oew,OPAL:1997oem,Belle:2018ttu}, $pp$ collision~\cite{Bellwied:2002rg,STAR:2002okv,STAR:2006nmo,ATLAS:2014ona}, and SIDIS~\cite{E665:1999fso,NOMAD:2000wdf,HERMES:1999buc,Belostotsky:1999kj,HERMES:2006lro,COMPASS:2009nhs,Sapozhnikov:2005sc,Alexakhin:2005dz,COMPASS:2021bws,McEneaney:2022bsf,Rith:2007zz,Ferrero:2007zz,HERMES:2014fmx,Karyan:2016rls} processes. These measurements not only allow us to learn the role of the spin in the hadronization process via spin-dependent FFs~\cite{DAlesio:2020wjq,Callos:2020qtu,Chen:2021hdn} but also provide a novel approach to study nucleon spin structures. For example, the spin transfer to the $\Lambda$ hyperon in polarized SIDIS is considered an observable sensitive to the strangeness components in the nucleon and has been experimentally measured~\cite{HERMES:1999buc,HERMES:2006lro,COMPASS:2009nhs,McEneaney:2022bsf,Karyan:2016rls} and phenomenologically studied~\cite{Ellis:1995fc,Lu:1995np,Chi:2013hka,Chi:2014xba,Du:2016irt,Du:2017nzy}. Similarly, one can also measure the polarization of a produced vector meson, such as $\rho$ and $K^*$. As a spin-1 particle, it has five rank-2 tensor polarized states in addition to the three vector polarized ones. These tensor polarizations can be analyzed from the spatial distribution of the decay particles, and have been measured in $e^+e^-$ annihilation experiments~\cite{DELPHI:1997ruo,OPAL:1997vmw,OPAL:1999hxs,OPAL:1997nwj}. Extensive phenomenological and theoretical studies have also been carried out~\cite{Bacchetta:2000jk,Mulders:2000sh,Bacchetta:2001rb,Gliske:2014wba,Boer:2016xqr,Ninomiya:2017ggn,Kumano:2020ijt,Cotogno:2017puy,Wei:2013csa,Wei:2014pma,Chen:2016moq,Chen:2020pty,Jiao:2022gzu,Kumano:2021xau,Kumano:2021fem}.

Recently, the BESIII Collaboration measured the polarization of the $\Omega$ through the chain of weak decays $\Omega^-\rightarrow K^-\Lambda$ and $\Lambda\rightarrow p\pi^-$~\cite{BESIII:2020lkm}. On the other hand, a complete set of TMD FFs to spin-3/2 hadrons has been derived in Ref.~\cite{Zhao:2022lbw} and applied in describing inclusive and semi-inclusive $\Omega$ productions in $e^+e^-$ annihilation. With the experimental techniques of analyzing the polarization of the $\Omega$ and the theoretical setup of its TMD FFs, we can study the polarization of $\Omega$ produced in the SIDIS process. This is not just an extension to spin-3/2 particles. The $\Omega$ baryon, which carries three valence strange quarks, is extremely sensitive to the strange sea of the nucleon, including its polarization. Although the yield of $\Omega$ production is expected to be much less than that of $\Lambda$ production, the contamination from target fragmentation will be highly suppressed because the generation of three strange quarks requires enough energy transferred from the virtual photon. Furthermore, tensor polarized states of the $\Omega$ provide additional observables to extract partonic structures of the nucleon. Therefore, the SIDIS production of $\Omega$ has unique advantages to study the nucleon spin structures, particularly the strange sea in the nucleon.

In this paper, we derive the differential cross section, taking into account all possible combinations of the polarization states of the lepton, the nucleon, and the produced spin-3/2 hadron. Applying the TMD formalism, we perform leading order calculations to express the structure functions as convolutions of TMD PDFs and TMD FFs. For an unpolarized lepton beam, half of the 192 structure functions are found nonvanishing and 42 of them are from rank-3 tensor polarized TMD FFs. For a polarized lepton beam, one third of the 96 structure functions are nonzero and 14 of them are for rank-3 tensor polarized hadron states. Besides, taking the spin transfer to a rank-3 tensor polarized state $S_{hLLL}$ as an example, we demonstrate with a model estimation that the newly defined structure functions related to the polarization of the produced $\Omega$ may generate sizable asymmetries to be measured in future experiments.

The paper is organized as follows. In Sec.~\ref{s.kinematic}, we derive the differential cross section of SIDIS with the production of a spin-3/2 hadron. In Sec.~\ref{s.partonmodel}, we calculate the parton model expressions of all structure functions at the leading order in terms of TMD PDFs and FFs. In Sec.~\ref{s.numericalresult}, we provide a model estimation of the spin transfer to produce an $S_{hLLL}$ tensor polarized $\Omega$. A summary and outlook is given in Sec.~\ref{s.summary}.

\section{The differential cross section}
\label{s.kinematic}

In this section, we derive the differential cross section of the electron-nucleon SIDIS process with the production of a spin-3/2 hadron.

\subsection{Kinematics and spin states}

We consider the SIDIS process as
\begin{align}
	e^{-}(l)+N(P)\rightarrow e^{-}(l^{\prime})+\Omega(P_{h})+X(P_{X}),
\end{align}
where the variables in parentheses indicate the four momenta of the corresponding particles. Although $\Omega$ is chosen to label the produced spin-3/2 hadron, the formalism and results derived in this section can be used for any spin-3/2 hadron. Instead of $l^\prime$, one usually uses the transferred four momentum $q=l-l^\prime$. Within one-photon-exchange approximation, $Q^2 = -q^2$ gives its virtuality and serves as the hard scale of the process. Some commonly used dimensionless variables to describe the SIDIS kinematics are defined as 
\begin{align}
    x=\frac{Q^{2}}{2P\cdot q},
    \quad
    y=\frac{P\cdot q}{P\cdot l},
    \quad
    z=\frac{P\cdot P_{h}}{P\cdot q},
    \quad
    \gamma=\frac{2Mx}{Q},
\end{align}
where $M$ is the mass of the nucleon target. Following the Trento conventions~\cite{Bacchetta:2004jz}, one often chooses the frame in which the virtual photon momentum $q$ and the nucleon momentum $P$ are along the longitudinal direction. It is convenient to define the transverse metric tensor,
\begin{align}
	g_{\perp}^{\mu\nu} 
	& =g^{\mu\nu}-\frac{q^{\mu}P^{\nu}+q^{\nu}P^{\mu}}{P\cdot q(1+\gamma^{2})}+\frac{\gamma^{2}}{1+\gamma^{2}}\left(\frac{q^{\mu}q^{\nu}}{Q^{2}}-\frac{P^{\mu}P^{\nu}}{M^2}\right),\label{e.gperp}
\end{align}
and the transverse antisymmetric tensor,
\begin{align}
	\epsilon_{\perp}^{\mu\nu} 
	& =\epsilon^{\mu\nu\rho\sigma}\frac{P_{\rho}q_{\sigma}}{P\cdot q\sqrt{1+\gamma^{2}}},\label{e.epsilonperp}
\end{align}
where the sign convention of the totally antisymmetric tensor is $\epsilon^{0123}=1$. 
Then, $l_{\perp}^\mu=g_{\perp}^{\mu \nu} l_\nu$ and $P_{h \perp}^\mu=g_{\perp}^{\mu \nu} P_{h \nu}$ give the transverse momenta of the electron and the final-state hadron in the photon-nucleon frame, and $\bm{P}_{h\perp}^2 = -P_{h \perp}^2$ provides a second and adjustable scale of the SIDIS process.
The azimuthal angle $\phi_h$ from the lepton plane to the hadron plane can be defined in an invariant form as
\begin{align}
    \cos\phi_h=-\frac{l_\mu P_{h\nu} g_\perp^{\mu\nu}}{\sqrt{l_\perp^2 P_{h\perp}^2}},
    \qquad
    \sin\phi_h=-\frac{l_\mu P_{h\nu} \epsilon_\perp^{\mu\nu}}{\sqrt{l_\perp^2 P_{h\perp}^2}}.\label{e.phi_h}
\end{align}

The polarization of the nucleon, as a spin-1/2 particle, is characterized by the spin vector $S^\mu$. In the photon-nucleon frame as chosen by the Trento conventions, it can be decomposed into longitudinal and transverse components as
\begin{align}
    S^\mu=S_L \frac{P^\mu-q^\mu M^2/(P\cdot q)}{M\sqrt{1+\gamma^2}}+S_T^\mu,
\end{align}
where
\begin{align}
    S_L=\frac{S\cdot q}{P\cdot q}\frac{M}{\sqrt{1+\gamma^2}},
    \qquad 
    S_T^\mu=g_\perp^{\mu\nu}S_\nu.
\end{align}
The azimuthal angle $\phi_T$ from the lepton plane to the direction of the nucleon transverse spin $S_T^\mu$ can be defined as 
\begin{align}
    \cos\phi_T=-\frac{l_\mu S_{\nu} g_\perp^{\mu\nu}}{\sqrt{l_\perp^2 S_T^2}},
	\qquad
    \sin\phi_T=-\frac{l_\mu S_{\nu} \epsilon_\perp^{\mu\nu}}{\sqrt{l_\perp^2 S_{T}^2}}.
\end{align}

The polarization of the spin-3/2 hadron is characterized by the spin vector and tensors, $S_h^\mu$, $T_h^{\mu\nu}$, and $R_h^{\mu\nu\rho}$, as defined in Ref.~\cite{Zhao:2022lbw}. 
They are orthogonal to the momentum of the hadron,
\begin{align}
    P_{h\mu}S_h^\mu=0,
    \quad 
    P_{h\mu}T_h^{\mu\nu}=0,
    \quad 
    P_{h\mu}R_h^{\mu\nu\rho}=0. 
\end{align} 
Similar to the decomposition of the nucleon spin vector, we decompose the longitudinal and transverse components of the hadron spin vector and tensors with respect to the hadron momentum $P_h$. For convenience, we introduce two lightlike basis vectors, $n$ and $\bar{n}$, satisfying $n\cdot n=0$, $\bar{n}\cdot\bar{n}=0$, and $n\cdot\bar{n}=1$, and express the hadron momentum as
\begin{align}
    P_h^\mu = (P_h\cdot \bar{n}) n^\mu + \frac{M_h^2}{2P_h\cdot \bar{n}} \bar{n}^\mu,
\end{align} 
where $M_h$ is the mass of the hadron. Then the spin tensors of the hadron are decomposed as~\cite{Zhao:2022lbw}
\begin{align}
	S^{\mu}_h &= 
	S_{hL}\left( \frac{M_h}{2 P_h \cdot \bar{n}}\bar{n}^\mu-\frac{P_h  \cdot \bar{n}}{M_h} n^\mu\right)
	+{S}_{hT}^\mu, \label{e.Sh}	\\ 
	T^{\mu \nu}_h&=
	\frac{1}{2}\Bigg\{  S_{hLL}\Bigg[\frac{1}{2}\left(\frac{M_h}{P_h \cdot \bar{n}}\right)^2 \bar{n}^\mu\bar{n}^\nu 
	+2\left(\frac{P_h \cdot \bar{n}}{M_h}\right)^2 n^\mu n^\nu
	-\bar{n}^{\{\mu}n^{\nu\}} + g_{T}^{\mu\nu} \Bigg] \nonumber \\ 
	&\quad +\frac{1}{2}\left(\frac{M_h}{P_h \cdot \bar{n}}\right)\bar{n}^{\{\mu}S_{hLT}^{\nu\}}-\left(\frac{P_h \cdot \bar{n}}{M_h}\right)n^{\{\mu}S_{hLT}^{\nu\}}+S_{hTT}^{\mu\nu} \Bigg\},
    \label{e.Th} \\ 
	R^{\mu \nu \rho}_h&=
	\frac{1}{4}\Bigg\{S_{hLLL}\Bigg[\frac{1}{2}\left(\frac{M_h}{P_h \cdot \bar{n}}\right)^3 \bar{n}^\mu\bar{n}^\nu\bar{n}^\rho
	-\frac{1}{2}\left(\frac{M_h}{P_h \cdot \bar{n}}\right)\left(\bar{n}^{\{\mu} \bar{n}^\nu n^{\rho\}}
	-\bar{n}^{\{\mu}g_{T}^{\nu\rho\}} \right)       \nonumber\\
	&\quad +\left(\frac{P_h \cdot \bar{n}}{M_h}\right)\left(\bar{n}^{\{\mu} n^\nu n^{\rho\}}-n^{\{\mu}g_{T}^{\nu\rho\}} \right)-4\left(\frac{P_h \cdot \bar{n}}{M_h}\right)^3 n^{\mu} n^\nu n^{\rho}\Bigg]     \nonumber\\
	&\quad +\frac{1}{2}\left(\frac{M_h}{P_h \cdot \bar{n}}\right)^2 \bar{n}^{\{\mu} \bar{n}^\nu S_{hLLT}^{\rho\}}+2\left(\frac{P_h \cdot \bar{n}}{M_h}\right)^2 n^{\{\mu} n^\nu S_{hLLT}^{\rho\}}-2\bar{n}^{\{\mu} n^\nu S_{hLLT}^{\rho\}}+\frac{1}{2}S_{hLLT}^{\{\mu}g_{T}^{\nu\rho\}}        \nonumber\\
	&\quad +\frac{1}{4}\left(\frac{M_h}{P_h \cdot \bar{n}}\right) \bar{n}^{\{\mu}  S_{hLTT}^{\nu\rho\}}-\frac{1}{2}\left(\frac{P_h \cdot \bar{n}}{M_h}\right) n^{\{\mu}  S_{hLTT}^{\nu\rho\}}+S_{hTTT}^{\mu\nu\rho}\Bigg\},
	\label{e.Rh}
\end{align}
where the transverse metric tensor with respect to $P_h$ is given by
\begin{align}
    g_T^{\mu\nu}=g^{\mu\nu}-\bar{n}^\mu n^\nu -n^\mu \bar{n}^\nu.
\end{align}
The transverse components of $S_{hT}^\mu$, $S_{hLT}^\mu$, $S_{hTT}^{\mu\nu}$, $S_{hLLT}^\mu$, $S_{hLTT}^{\mu\nu}$, and $S_{hTTT}^{\mu\nu\rho}$ can be expressed in matrix form as
\begin{align}
    S_{hT}^i &= (S_{hT}^x, S_{hT}^y),
    \quad
     S_{hLT}^i = (S_{hLT}^x, S_{hLT}^y),
    \quad
     S_{hLLT}^i = (S_{hLLT}^x, S_{hLLT}^y),
    \nonumber\\
     S_{hTT}^{ij} &=
    \begin{pmatrix}
        S_{hTT}^{xx} & S_{hTT}^{xy} \\
        S_{hTT}^{xy} & -S_{hTT}^{xx}
    \end{pmatrix},
    \quad
    S_{hLTT}^{ij} = 
    \begin{pmatrix}
        S_{hLTT}^{xx} & S_{hLTT}^{xy} \\
        S_{hLTT}^{xy} & -S_{hLTT}^{xx}
    \end{pmatrix},
    \nonumber\\
    S_{hTTT}^{ijk} &=
    \left[
    \begin{pmatrix}
        S_{hTTT}^{xxx} & S_{hTTT}^{yxx}\\
        S_{hTTT}^{yxx} & -S_{hTTT}^{xxx}
    \end{pmatrix},
    \begin{pmatrix}
        S_{hTTT}^{yxx} & -S_{hTTT}^{xxx}\\
        -S_{hTTT}^{xxx} & -S_{hTTT}^{yxx}
    \end{pmatrix}
    \right].
    \label{e.spintransverse}
\end{align}

\subsection{The cross section in terms of structure functions}\label{sub.SFs}
\begin{figure}[ht]
	\centering
	\includegraphics[width=0.3\textwidth]{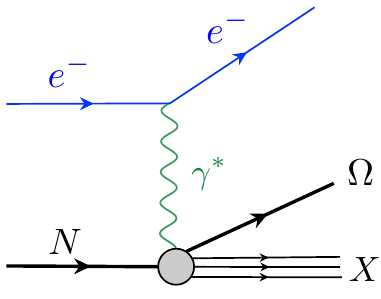}
	\caption{The process $e^-N\rightarrow e^-\Omega X$ with one-photon-exchange approximation.}
	\label{f.sidisdiagram}
\end{figure}

With the one-photon-exchange approximation as illustrated in Fig.~\ref{f.sidisdiagram}, one can express the differential cross section of the SIDIS process as
\begin{align}
	\frac{d\sigma}{dxdydz d\phi_h d\psi d P_{h\perp}^2}=\frac{\alpha^2 y}{8Q^4 z}L_{\mu\nu}W^{\mu\nu},\label{e.dsigma}
\end{align}
where $\alpha$ is the electromagnetic fine structure constant and $\psi$ is the azimuthal angle of the outgoing lepton around the lepton beam axis with respect to an arbitrary fixed direction. The leptonic tensor is 
\begin{align}
L_{\mu\nu}=2\left(l_{\mu}l_{\nu}^{\prime}+l_{\nu}l_{\mu}^{\prime}-g_{\mu\nu} l\cdot l^{\prime} + i\lambda_{e}\epsilon_{\mu\nu\rho\sigma}l^{\rho}l^{\prime\sigma}\right),
\end{align}
where $\lambda_e$ represents the helicity of the lepton beam. The hadronic tensor $W^{\mu\nu}$ is expressed as
\begin{align}
	W^{\mu\nu}\left( q ;P,S; P_h,S_h,T_h,R_h\right)=&\sum_{X}\delta^4\left(P+q-P_h-P_X\right)  \left\langle P,S | J^\mu(0) | P_X; P_h, S_h, T_h, R_h \right\rangle \nonumber \\
	&\times\left\langle P_X; P_h, S_h, T_h, R_h| J^\nu(0) |P,S\right\rangle,\label{e.hadronictensor}
\end{align}
where $J^\mu$ is the electromagnetic current operator and the symbol $\displaystyle{\sum_{X}}$ indicates the sum over the hadronic states $X$ with the momentum integration implicitly assumed. It satisfies the constraints required by the Hermiticity, parity invariance, and gauge invariance,
\begin{align}
	&W^{* \mu \nu}\left( q ;P,S; P_h,S_h,T_h,R_h\right)=W^{\nu \mu}\left( q ;P,S; P_h,S_h,T_h,R_h\right), \label{e.Hermiticity}\\ 
	&W^{\mu \nu}\left( q ;P,S; P_h,S_h,T_h,R_h\right)=W_{\mu \nu}\left(q ;P,-\bar{S}; P_h,-\bar{S}_h,\bar{T}_h,-\bar{R}_h\right),\label{e.parity}\\
	&q_\mu W^{\mu\nu}\left( q ;P,S; P_h,S_h,T_h,R_h\right)=W^{\mu\nu} \left( q ;P,S; P_h,S_h,T_h,R_h\right)q_\nu =0,\label{e.gaugeinvariace}
\end{align}
where $\bar{S}$, $\bar{S}_h$, $\bar{T}_h$, and $\bar{R}_h$ indicate a sign flip of all space components. 

With kinematic analysis, one usually expresses the hadronic tensor into terms as basis Lorentz tensors multiplied by scalar functions, known as the structure functions. 
To impose the gauge invariance, it is convenient to construct the basis Lorentz tensors using the so-called conserved vectors and metric tensor,
\begin{align}
	P_q^\mu&=P^\mu-\frac{P\cdot q}{q^2}q^\mu,\\
	P_{hq}^\mu&=P_h^\mu-\frac{P_h\cdot q}{q^2}q^\mu,\\
    g^{\mu\nu}_q &= g^{\mu\nu} - \frac{q^\mu q^\nu}{q^2},
\end{align}
which will vanish when contracted with the virtual photon momentum $q_\mu$. 
Then one can in principle exhaust all basis Lorentz tensors constructed from the conserved vectors, the metric tensor, and the spin tensors, but this approach becomes less efficient with the increasing number of the measured momenta. Instead of directly building all basis Lorentz tensors, one may tackle this problem by using a systematic procedure~\cite{Chen:2016moq,Jiao:2022gzu,Zhao:2022lbw}. We first group the nine basis tensors built from $P_q^\mu$, $P_{hq}^\mu$,  $g_q^{\mu\nu}$, and $\epsilon^{\mu \nu\rho\sigma}$ into four sets, 
\begin{align}
	h_{U}^{S\mu\nu} & =\left\{ g_q^{\mu\nu}, P_{q}^{\mu}P_{q}^{\nu}, P_{q}^{\{\mu}P_{hq}^{\nu\}},P_{hq}^{\mu}P_{hq}^{\nu}\right\} ,\label{e.hus}\\
	\tilde{h}_{U}^{S\mu\nu} & =\left\{ \epsilon^{\left\{ \mu qPP_{h}\right.}P_{q}^{\nu\}}, \epsilon^{\left\{ \mu qPP_{h}\right.}P_{hq}^{\nu\}}\right\}, \label{e.tildehus}\\
	h_{U}^{A\mu\nu} & =\left\{ P_{q}^{[\mu}P_{hq}^{\nu]}\right\}, \label{e.hua}\\
	\tilde{h}_{U}^{A\mu\nu} & =\left\{ \epsilon^{\mu\nu qP},\epsilon^{\mu\nu qP_{h}}\right\},\label{e.tildehua}
\end{align}
which are called the {\it basic Lorentz tensors}.
The superscripts $S$ and $A$ stand for the symmetric and antisymmetric tensors, respectively. The subscript $U$ indicates that both the target and the produced hadron are unpolarized. For conciseness, we adopt the shorthand notation $A^{\mu P}\equiv A^{\mu\nu}P_\nu$. Here the $h$ and $\tilde{h}$ represent the parity conserving and flipping terms, respectively. Restricted to the parity invariance~\eqref{e.parity}, unpolarized basis tensors are directly given by those in $h_{U}^{S\mu\nu}$ and $h_{U}^{A\mu\nu}$.

To construct the basis Lorentz tensors with polarized target and produced hadrons, the spin tensors $S^\mu$, $S_h^\mu$, $T_h^{\mu\nu}$, and $R_h^{\mu\nu\rho}$ are also involved. We claim that the spin-dependent basis Lorentz tensors in this case can be constructed from multiplying the basic Lorentz tensors given in Eqs.~\eqref{e.hus}--\eqref{e.tildehua} by spin-dependent scalars or pseudoscalars, because the four tensors $P^\mu$, $P_h^\mu$, $q^\mu$, and $\epsilon^{\mu \nu\rho\sigma}$ that construct the basic Lorentz tensors are sufficient to construct a four-dimensional spacetime, and the spin tensors can be expressed by these four tensors. To show this clearly we take $S^{\mu}$ and $T_h^{\mu\nu}$ as examples and express them in terms of $P^\mu$, $P_h^\mu$, $q^\mu$, and $\epsilon^{\mu \nu\rho\sigma}$, as shown in Appendix~\ref{a.Sdecompositon}. To conclude, the polarized basis Lorentz tensors can be constructed from the basic Lorentz tensors in $h_{U}^{S\mu\nu}$ and $h_{U}^{A\mu\nu}$ multiplied by spin-dependent scalars and those in $\tilde{h}_{U}^{S\mu\nu}$ and $\tilde{h}_{U}^{A\mu\nu}$ multiplied by spin-dependent pseudoscalars.

By applying the procedure above, one can obtain a complete set of polarized basis tensors. For an unpolarized nucleon, there are 13 basis tensors dependent on $S_h$, 
 \begin{align}
 	h_{UV_h}^{S\mu\nu} & =\epsilon^{PP_{h}qS_{h}}h_{U}^{S\mu\nu},\left\{ (P\cdot S_{h}),(q\cdot S_{h})\right\} \tilde{h}_{U}^{S\mu\nu}, \label{e.hvs}\\
 	h_{UV_h}^{A\mu\nu} & =\epsilon^{PP_{h}qS_{h}}h_{U}^{A\mu\nu},\left\{ (P\cdot S_{h}),(q\cdot S_{h})\right\} \tilde{h}_{U}^{A\mu\nu}; \label{e.hva}
 \end{align}
23 basis tensors dependent on $T_h$, 
\begin{align}
	h_{UT_{h}}^{S\mu\nu} & =\left\{ T_{h}^{PP},T_{h}^{Pq},T_{h}^{qq}\right\} h_{U}^{S\mu\nu}, \left\{ \epsilon^{T_{h}^{P}PP_{h}q},\epsilon^{T_{h}^{q}PP_{h}q}\right\} \tilde{h}_{U}^{S\mu\nu}, \\
	h_{UT_{h}}^{A\mu\nu} & =\left\{ T_{h}^{PP},T_{h}^{Pq},T_{h}^{qq}\right\} h_{U}^{A\mu\nu}, \left\{ \epsilon^{T_{h}^{P}PP_{h}q}, \epsilon^{T_{h}^{q}PP_{h}q}\right\} \tilde{h}_{U}^{A\mu\nu};
\end{align}
and 31 basis tensors dependent on $R_h$,
\begin{align}
	h_{UR_{h}}^{S\mu\nu} & =\left\{ \epsilon^{R_{h}^{PP}PP_{h}q},\:\epsilon^{R_{h}^{Pq}PP_{h}q},\epsilon^{R_{h}^{qq}PP_{h}q}\right\} h_{U}^{S\mu\nu},\left\{ R_{h}^{PPP},\:R_{h}^{PPq},\:R_{h}^{Pqq},R_{h}^{qqq}\right\} \tilde{h}_{U}^{S\mu\nu},\\
	h_{UR_{h}}^{A\mu\nu} & =\left\{ \epsilon^{R_{h}^{PP}PP_{h}q},\epsilon^{R_{h}^{Pq}PP_{h}q},\epsilon^{R_{h}^{qq}PP_{h}q}\right\} h_{U}^{A\mu\nu},\:\left\{ R_{h}^{PPP},R_{h}^{PPq},\:R_{h}^{Pqq},\:R_{h}^{qqq}\right\} \tilde{h}_{U}^{A\mu\nu},
\end{align}
where the first subscript ``$U$'' indicates the unpolarized state of the nucleon and the second subscripts ``$V_h$,'' ``$T_h$,'' and ``$R_h$'' represent the spin states of the produced hadron. 

For a polarized nucleon, there are 13 basis tensors if the produced hadron is unpolarized,
\begin{align}
	h_{VU}^{S\text{\ensuremath{\mu\nu}}} & =\epsilon^{PP_{h}qS}h_{U}^{S\mu\nu},\left\{ (P_{h}\cdot S),(q\cdot S)\right\} \tilde{h}_{U}^{S\mu\nu},\\
	h_{VU}^{A\mu\nu} & =\epsilon^{PP_{h}qS}h_{U}^{A\mu\nu},\left\{ (P_{h}\cdot S),(q\cdot S)\right\} \tilde{h}_{U}^{A\mu\nu},
\end{align}
which can be obtained by replacing the $S_h$ in Eqs.~\eqref{e.hvs} and \eqref{e.hva} by the nucleon spin vector $S$. Exhausting all possible spin-dependent scalars or pseudoscalars, one can obtain the basis tensors with both the nucleon and the produced hadron polarized. However, some of them are not linearly independent. The redundant ones can be eliminated by the identity, 
\begin{align}
g^{\alpha\beta}\epsilon^{\mu\nu\rho\sigma}=g^{\alpha\mu}\epsilon^{\beta\nu\rho\sigma}+g^{\alpha\nu}\epsilon^{\mu\beta\rho\sigma}+g^{\alpha\rho}\epsilon^{\mu\nu\beta\sigma}+g^{\alpha\sigma}\epsilon^{\mu\nu\rho\beta}.
	\label{e.gepsilon}
\end{align} 
Then we obtain 41 basis tensors dependent on $S$ and $S_h$,
\begin{align}
	h_{VV_h}^{S\mu\nu} & =\left\{\mathcal{S}_1 \right\} h_{U}^{S\mu\nu},
	\left\{ \mathcal{PS}_1 \right\} \tilde{h}_{U}^{S\mu\nu}, \\
	h_{VV_h}^{A\mu\nu} & =\left\{\mathcal{S}_1 \right\} h_{U}^{A\mu\nu},
	\left\{ \mathcal{PS}_1 \right\} \tilde{h}_{U}^{A\mu\nu}, 
\end{align}
where $\mathcal{S}_1$ stands for a set of spin-dependent scalars,
\begin{align}
    \mathcal{S}_1&: (S\cdot S_{h}),
    (P\cdot S_{h})(P_{h}\cdot S),
    (P_{h}\cdot S)(q\cdot S_{h}),
    (P\cdot S_{h})(q\cdot S), 
    (q\cdot S_{h})(q\cdot S),
\end{align}
and $\mathcal{PS}_1$ stands for a set of spin-dependent pseudoscalars,
\begin{align}
	\mathcal{PS}_1&: \epsilon^{PP_{h}SS_{h}}, 
    \epsilon^{PqSS_{h}}, 
    \epsilon^{P_{h}qSS_{h}}, 
    \epsilon^{PP_{h}qS_{h}}(S\cdot q).
\end{align}
Similarly, there are 67 basis tensors dependent on $S$ and $T_h$,
\begin{align}
	h_{VT_{h}}^{S\mu\nu} =&\left\{  \mathcal{S}_2\right\} h_{U}^{S\mu\nu},
	\left\{  \mathcal{PS}_2\right\} \tilde{h}_{U}^{S\mu\nu},\\
	h_{VT_{h}}^{A\mu\nu} =&\left\{  \mathcal{S}_2\right\} h_{U}^{A\mu\nu},
	 \left\{  \mathcal{PS}_2\right\} \tilde{h}_{U}^{A\mu\nu},
\end{align}
where the scalars in $\mathcal{S}_2$ and the pseudoscalars in $\mathcal{PS}_2$ are given by
\begin{align}
	\mathcal{S}_2& :\epsilon^{T_{h}^{P}SP_{h}q}, \epsilon^{T_{h}^{P}SPq}, \epsilon^{T_{h}^{P}SPP_{h}}, \epsilon^{T_{h}^{q}SP_{h}q}, \epsilon^{T_{h}^{q}SPq}, \epsilon^{T_{h}^{q}SPP_{h}},  \epsilon^{PP_{h}qS}T_{h}^{qq}, \\
	\mathcal{PS}_2&:  T_{h}^{SP},T_{h}^{Sq}, T_{h}^{PP}(S\cdot P_{h}), T_{h}^{PP}(S\cdot q), T_{h}^{Pq}(S\cdot P_{h}), T_{h}^{Pq}(S\cdot q), T_{h}^{qq}(S\cdot P_{h}), T_{h}^{qq}(S\cdot q);
\end{align}
and 95 basis tensors dependent on $S$ and $R_h$,
\begin{align}
	h_{V R_{h}}^{S\mu\nu} =&\left\{ \mathcal{S}_3\right\} h_{U}^{S\mu\nu}, \left\{ \mathcal{PS}_3\right\} \tilde{h}_{U}^{S\mu\nu}, \\
	h_{V R_{h}}^{A\mu\nu} =&\left\{ \mathcal{S}_3\right\} h_{U}^{A\mu\nu}, \left\{ \mathcal{PS}_3\right\} \tilde{h}_{U}^{A\mu\nu},
\end{align}
where the scalars in $\mathcal{S}_3$ and the pseudoscalars in $\mathcal{PS}_3$ are given by
\begin{align}
	\mathcal{S}_3 :&  R_{h}^{SPq}, R_{h}^{SPP}, R_{h}^{Sqq}, R_{h}^{PPP}\left(S\cdot P_{h}\right), R_{h}^{PPP}\left(S\cdot q\right), R_{h}^{PPq}\left(S\cdot P_{h}\right), 
 \nonumber\\
 &R_{h}^{PPq}\left(S\cdot q\right), R_{h}^{Pqq}\left(S\cdot P_{h}\right),
	R_{h}^{Pqq}\left(S\cdot q\right), R_{h}^{qqq}\left(S\cdot P_{h}\right), R_{h}^{qqq}\left(S\cdot q\right), \\
	\mathcal{PS}_3:& \epsilon^{R_{h}^{PP}SPq}, \epsilon^{R_{h}^{PP}SP_{h}P}, \epsilon^{R_{h}^{PP}SP_{h}q}, \epsilon^{R_{h}^{Pq}SP_{h}q}, \epsilon^{R_{h}^{Pq}SP_{h}P}, \epsilon^{R_{h}^{Pq}SPq}, 
 \nonumber\\
 &\epsilon^{R_{h}^{qq}SP_{h}q}, \epsilon^{R_{h}^{qq}SP_{h}P}, \epsilon^{R_{h}^{qq}SPq},
    \epsilon^{PP_{h}qS}R_{h}^{qqq}. 
\end{align}

By now, we have a complete set of 288 basis Lorentz tensors and the hadronic tensor can be expressed as
\begin{align}
	W^{\mu \nu}=&\sum_{i=1}^{4} V_{U, i}^S h_{U, i}^{S\mu\nu}
	+\sum_{i=1}^{8} V_{UV_h, i}^S  h_{UV_h, i}^{S\mu\nu} 
	+\sum_{i=1}^{16} V_{UT_h, i}^S h_{UT_h, i}^{S\mu\nu}+\sum_{i=1}^{20} V_{UR_h, i}^S h_{UR_h, i}^{S\mu\nu}\nonumber\\
    &+i\bigg(\sum_{i=1}^{1} V_{U, i}^A h_{U, i}^{A\mu\nu}
	+\sum_{i=1}^{5} V_{UV_h, i}^A  h_{UV_h, i}^{A\mu\nu} 
	+\sum_{i=1}^{7} V_{UT_h, i}^A h_{UT_h, i}^{A\mu\nu}+\sum_{i=1}^{11} V_{UR_h, i}^A h_{UR_h, i}^{A\mu\nu}\bigg)\nonumber\\
   &+\sum_{i=1}^{8} V_{VU, i}^S h_{VU, i}^{S\mu\nu}
	+\sum_{i=1}^{28} V_{VV_h, i}^S  h_{VV_h, i}^{S\mu\nu} 
	+\sum_{i=1}^{44} V_{VT_h, i}^S h_{VT_h, i}^{S\mu\nu}
	+\sum_{i=1}^{64} V_{VR_h, i}^S h_{VR_h, i}^{S\mu\nu}\nonumber\\
   &+i\bigg(\sum_{i=1}^{5} V_{VU, i}^A h_{VU, i}^{A\mu\nu}
	+\sum_{i=1}^{13} V_{VV_h, i}^A  h_{VV_h, i}^{A\mu\nu} 
	+\sum_{i=1}^{23} V_{VT_h, i}^A h_{VT_h, i}^{A\mu\nu}
	+\sum_{i=1}^{31} V_{VR_h, i}^A h_{VR_h, i}^{A\mu\nu}
    \bigg),
	 \label{e.totaltensor}
\end{align}
where the coefficients $V_{i}$ are referred to as structure functions, which are scalar functions of $q^2$, $P\cdot q$, $P_h\cdot q$, and $P\cdot P_h$. We note that the total number of independent structure functions is entirely determined by the number of basis tensors obtained from the kinematic analysis, although one may choose different linear combinations for convenience.

Contracting the hadronic tensor with the leptonic tensor, one can express the general expression of the differential cross section in Eq.~\eqref{e.dsigma} in terms of the structure functions $V_i$, whereas, it is practically convenient to specify a reference frame and to separate the contributions to various angular modulations. As a common convention, we decompose the spin components of the spin-3/2 hadron with respect to its momentum direction. Here we choose three orthogonal directions to project the spin components, i.e., the longitudinal direction along the $\Omega$ opposite momentum, the normal direction of the hadron plane, and the transverse direction in the production plane. We choose the temporal basis vector as
\begin{align}
	\hat{t}_{\Omega} & =  \left( 1,0,0,0\right),\label{e.tomega}
\end{align} 
and the spatial basis vector as
\begin{align}
	\hat{z}_{\Omega} & =  \left( 0,-\sin\theta_{h}\cos\phi_h,-\sin\theta_{h}\sin\phi_{h},\cos\theta_{h}\right).
\end{align}
In addition, the other two transverse spatial basis vectors are defined as
\begin{align}
	\hat{x}_{\Omega}&=\left( 0,\cos\theta_{h}\cos\phi_{h},\cos\theta_{h}\sin\phi_{h},\sin\theta_{h}\right), \\ 
	\hat{y}_{\Omega}&=\left( 0,-\sin\phi_{h},\cos\phi_{h},0\right).\label{e.yomega}
\end{align}
Here we use the $v^\mu=(v^0,v^x,v^y,v^z)$ representation to be distinguished from the light-cone component representation. The angle $\theta_{h}$ is spanned by the produced hadron momentum and the transverse component of the hadron momentum, which can be expressed in a Lorentz covariant form as 
\begin{align}
    \sin\theta_{h}  =\frac{\left| \bm{P}_{h\perp }\right|}{\left| \bm{P}_{h}\right|}=\frac{\sqrt{-P_{h\perp}^2}}
    {\sqrt{-P_{h\perp}^2+\left|\frac{P_h\cdot q}{Q}\right|^2}}.
\end{align}
The spin components in Eq.~\eqref{e.spintransverse} can be defined in Lorentz covariant forms as Eqs.~(78)--(92) in Ref.~\cite{Zhao:2022lbw}, with the same definitions for the quantities $S_{hL}$, $|S_{hT}|$, $\phi_{hT}$, $S_{hLL}$, $|S_{hLT}|$, $\phi_{hLT}$, $|S_{hTT}|$, $\phi_{hTT}$, $S_{hLLL}$, $|S_{hLLT}|$, $\phi_{hLLT}$, $|S_{hLTT}|$, $\phi_{hLTT}$, $|S_{hTTT}|$, and $\phi_{hTTT}$, while in our case the basis vectors are replaced with Eqs.~\eqref{e.tomega}--\eqref{e.yomega}. Since the basis vectors are constructed with the external momenta of the reaction, all these variables are Lorentz scalars, although the meanings of the variables are more clearly understood in this particular frame.

So far, we have obtained the expression of the differential cross section in terms of 288 structure functions in accordance with the angular distributions and the spin states of the hadron. For simplicity, we group the 288 terms in the cross section into 30 parts according to various combinations of the spin states as
\begin{align}
	\frac{d\sigma}{dxdydzd\phi_h d\psi dP_{h\perp}^2}&= \frac{\alpha^{2}}{xyQ^{2}}\frac{y^{2}}{2(1-\varepsilon)} \left(1+\frac{\gamma^{2}}{2x}\right) \Big[ \mathcal{F}_{U,U} + S_{hL}\mathcal{F}_{U,L} + \left|S_{hT}\right| \mathcal{F}_{U,T} \nonumber\\  
	&+ S_{hLL} \mathcal{F}_{U,LL} + \left|S_{hLT}\right| \mathcal{F}_{U,LT} + \left|S_{hTT}\right| \mathcal{F}_{U,TT} \nonumber\\
	&+ S_{hLLL} \mathcal{F}_{U,LLL} + \left|S_{hLLT}\right| \mathcal{F}_{U,LLT} + \left|S_{hLTT}\right| \mathcal{F}_{U,LTT} + \left|S_{hTTT}\right| \mathcal{F}_{U,TTT}\nonumber\\
   &+S_L \mathcal{F}_{L,U} +\left|S_{T}\right|\mathcal{F}_{T,U}\nonumber\\ 
	 &+S_L \big(S_{hL} \mathcal{F}_{L,L} +  \left|S_{hT}\right|\mathcal{F}_{L,T} \big)+ \left|S_{T}\right| \big(S_{hL}\mathcal{F}_{T,L} +  \left|S_{hT}\right| \mathcal{F}_{T,T}\big) \nonumber\\
	&+ S_L \big(S_{hLL} \mathcal{F}_{L,LL} +\left|S_{hLT}\right| \mathcal{F}_{L,LT} +  \left|S_{hTT}\right| \mathcal{F}_{L,TT}\big) \nonumber\\
	&+ \left|S_T\right| \big(S_{hLL} \mathcal{F}_{T,LL} +   \left|S_{hLT}\right| \mathcal{F}_{T,LT} +  \left|S_{hTT}\right| \mathcal{F}_{T,TT}\big) \nonumber\\
	&+S_L \big(S_{hLLL} \mathcal{F}_{L,LLL} +\left|S_{hLLT}\right| \mathcal{F}_{L,LLT} + \left|S_{hLTT}\right| \mathcal{F}_{L,LTT}+\left|S_{hTTT}\right| \mathcal{F}_{L,TTT}\big)\nonumber\\
	&+\left|S_T\right| \big(S_{hLLL} \mathcal{F}_{T,LLL} + \left|S_{hLLT}\right| \mathcal{F}_{T,LLT} + \left|S_{hLTT}\right| \mathcal{F}_{T,LTT} + \left|S_{hTTT}\right| \mathcal{F}_{T,TTT} \big)\Big],\label{e.totalcs}
\end{align}
where the two subscripts of $\mathcal{F}$ represent the spin states of the nucleon and the produced hadron, respectively. The ratio of the longitudinal and the transverse photon flux is expressed as
\begin{align}
	\varepsilon=\frac{1-y-\frac{1}{4}\gamma^2y^2}{1-y+\frac{1}{2}y^2+\frac{1}{4}\gamma^2y^2}.\label{e.ratio}
\end{align}

The explicit expression of each part can be written in terms of the corresponding structure functions.
Among the 288 structure functions in Eq.~\eqref{e.totalcs}, five are for unpolarized states,
\begin{align}
	\mathcal{F}_{U,U}=& F_{U,U}^{T}+ \varepsilon F_{U,U}^{L}+ \sqrt{2\varepsilon(1+\varepsilon)}\cos\phi_{h}F_{U,U}^{\cos\phi_{h}}+ \varepsilon\cos2\phi_{h}F_{U,U}^{\cos2\phi_{h}} \nonumber\\
    & +\lambda_{e} \sqrt{2\varepsilon(1-\varepsilon)}\sin\phi_{h}G_{U,U}^{\sin\phi_{h}};\label{e.FUU}
\end{align}
13 are for unpolarized target and vector polarized hadron states, 
\begin{align}
	\mathcal{F}_{U,L}=& \sqrt{2\varepsilon(1+\varepsilon)}\sin\phi_{h}F_{U,L}^{\sin\phi_{h}}+ \varepsilon\sin2\phi_{h}F_{U,L}^{\sin2\phi_{h}} \nonumber\\
    & +\lambda_{e}\left(\sqrt{1-\varepsilon^{2}}G_{ U,L}+ \sqrt{2\varepsilon(1-\varepsilon)}\cos\phi_{h}G_{ U,L}^{\cos\phi_{h}}\right),\\
	\mathcal{F}_{U,T}=&\sin\phi_{hT}\left( F_{U,T}^{T\sin\phi_{hT}}+ \varepsilon F_{U,T}^{L\sin\phi_{hT}}\right)\nonumber\\
	&+ \sqrt{2\varepsilon(1+\varepsilon)}\left(\sin\left(\phi_{h}-\phi_{hT}\right)F_{U,T}^{\sin\left(\phi_{h}-\phi_{hT}\right)}+\sin\left(\phi_{h}+\phi_{hT}\right)F_{U,T}^{\sin\left(\phi_{h}+\phi_{hT}\right)}\right)\nonumber\\
	&+ \varepsilon\left(\sin\left(2\phi_{h}-\phi_{hT}\right)F_{U,T}^{\sin\left(2\phi_{h}-\phi_{hT}\right)}+\sin\left(2\phi_{h}+\phi_{hT}\right)F_{U,T}^{\sin\left(2\phi_{h}+\phi_{hT}\right)}\right)\nonumber\\
	&+\lambda_{e}\left[ \sqrt{1-\varepsilon^{2}}\cos\phi_{hT}G_{U,T}^{\cos\phi_{hT}} \right.\nonumber\\
	&+\left. \sqrt{2\varepsilon(1-\varepsilon)}\left(\cos\left(\phi_{h}-\phi_{hT}\right)G_{U,T}^{\cos\left(\phi_{h}-\phi_{hT}\right)}+\cos\left(\phi_{h}+\phi_{hT}\right)G_{U,T}^{\cos\left(\phi_{h}+\phi_{hT}\right)}\right) \right];
\end{align}
23 are for unpolarized target and rank-2 tensor polarized hadron states,
\begin{align}
	\mathcal{F}_{U,LL}=& F_{U,LL}^{T}+ \varepsilon F_{U,LL}^{L}+ \sqrt{2\varepsilon(1+\varepsilon)}\cos\phi_{h}F_{U,LL}^{\cos\phi_{h}}+ \varepsilon\cos2\phi_{h}F_{U,LL}^{\cos2\phi_{h}} \nonumber\\
    & +\lambda_{e} \sqrt{2\varepsilon(1-\varepsilon)}\sin\phi_{h}G_{U, LL}^{\sin\phi_{h}},\\
	\mathcal{F}_{U,LT}=& \cos\phi_{hLT}\left(F_{U,LT}^{T\cos\phi_{hLT}}+ \varepsilon F_{U,LT}^{L\cos\phi_{hLT}}\right)\nonumber\\
	&+ \sqrt{2\varepsilon(1+\varepsilon)}\left(\cos\left(\phi_{h}-\phi_{hLT}\right)F_{U,LT}^{\cos\left(\phi_{h}-\phi_{hLT}\right)}+\cos\left(\phi_{h}+\phi_{hLT}\right)F_{U,LT}^{\cos\left(\phi_{h}+\phi_{hLT}\right)}\right)\nonumber\\
	&+ \varepsilon\left(\cos\left(2\phi_{h}-\phi_{hLT}\right)F_{U,LT}^{\cos\left(2\phi_{h}-\phi_{hLT}\right)}+\cos\left(2\phi_{h}+\phi_{hLT}\right)F_{U,LT}^{\cos\left(2\phi_{h}+\phi_{hLT}\right)}\right)\nonumber\\
	&+\lambda_{e} \left[\sqrt{1-\varepsilon^{2}}\sin\phi_{hLT}G_{U,LT}^{\sin\phi_{hLT}}\right.\nonumber\\
	&\left.+ \sqrt{2\varepsilon(1-\varepsilon)}\left(\sin\left(\phi_{h}-\phi_{hLT}\right)G_{U, LT}^{\sin\left(\phi_{h}-\phi_{hLT}\right)}+\sin\left(\phi_{h}+\phi_{hLT}\right)G_{U,LT}^{\sin\left(\phi_{h}+\phi_{hLT}\right)}\right)\right],\\
	 \mathcal{F}_{U,TT}=	&\cos2\phi_{hTT}\left( F_{U,TT}^{T\cos2\phi_{hTT}}+ \varepsilon F_{U,TT}^{L\cos2\phi_{hTT}}\right)\nonumber\\
	 &+ \sqrt{2\varepsilon(1+\varepsilon)}\left(\cos\left(\phi_{h}-2\phi_{hTT}\right)F_{U,TT}^{\cos\left(\phi_{h}-2\phi_{hTT}\right)}+\cos\left(\phi_{h}+2\phi_{hTT}\right)F_{U,TT}^{\cos\left(\phi_{h}+2\phi_{hTT}\right)}\right)\nonumber\\
	 &+ \varepsilon\left(\cos\left(2\phi_{h}-2\phi_{hTT}\right)F_{U,TT}^{\cos\left(2\phi_{h}-2\phi_{hTT}\right)}+\cos\left(2\phi_{h}+2\phi_{hTT}\right)F_{U,TT}^{\cos\left(2\phi_{h}+2\phi_{hTT}\right)}\right)\nonumber\\
	 &+\lambda_{e} \left[\sqrt{1-\varepsilon^{2}}\sin2\phi_{hTT}G_{U, TT}^{\sin2\phi_{hTT}}\right.\nonumber\\
	 &\left.+ \sqrt{2\varepsilon(1-\varepsilon)}\left(\sin\left(\phi_{h}-2\phi_{hTT}\right)G_{U, TT}^{\sin\left(\phi_{h}-2\phi_{hTT}\right)}+\sin\left(\phi_{h}+2\phi_{hTT}\right)G_{U, TT}^{\sin\left(\phi_{h}+2\phi_{hTT}\right)}\right)\right];
\end{align}
31 are for unpolarized target and rank-3 tensor polarized hadron states,
\begin{align}
\mathcal{F}_{U,LLL}=&\sqrt{2\varepsilon(1+\varepsilon)}\sin\phi_{h}F_{U,LLL}^{\sin\phi_{h}}+\varepsilon\sin2\phi_{h}F_{U,LLL}^{\sin2\phi_{h}}\nonumber\\
	&+\lambda_e \left(\sqrt{1-\varepsilon^{2}}G_{U,LLL}+\sqrt{2\varepsilon(1-\varepsilon)}\cos\phi_{h}G_{U,LLL}^{\cos\phi_{h}}\right),\\
	\mathcal{F}_{U,LLT}=&\sin\phi_{hLLT}\left(F_{U, LLT}^{T\sin\phi_{hLLT}}+\varepsilon F_{U, LLT}^{L\sin\phi_{hLLT}}\right)\nonumber\\
	&+\sqrt{2\varepsilon(1+\varepsilon)}\left(\sin\left(\phi_{h}-\phi_{hLLT}\right)F_{U, LLT}^{\sin\left(\phi_{h}-\phi_{hLLT}\right)}+\sin\left(\phi_{h}+\phi_{hLLT}\right)F_{U, LLT}^{\sin\left(\phi_{h}+\phi_{hLLT}\right)}\right)\nonumber\\
	&+\varepsilon\left(\sin\left(2\phi_{h}-\phi_{hLLT}\right)F_{U, LLT}^{\sin\left(2\phi_{h}-\phi_{hLLT}\right)}+\sin\left(2\phi_{h}+\phi_{hLLT}\right)F_{U, LLT}^{\sin\left(2\phi_{h}+\phi_{hLLT}\right)}\right)\nonumber\\
	&+\lambda_{e}\left[\sqrt{1-\varepsilon^{2}}\cos\phi_{hLLT}G_{ U,LLT}^{\cos\phi_{hLLT}}\right.\nonumber\\
	&\left.+\sqrt{2\varepsilon(1-\varepsilon)}\left(\cos\left(\phi_{h}-\phi_{hLLT}\right)G_{ U,LLT}^{\cos\left(\phi_{h}-\phi_{hLLT}\right)}+\cos\left(\phi_{h}+\phi_{hLLT}\right)G_{ U,LLT}^{\cos\left(\phi_{h}+\phi_{hLLT}\right)}\right)\right],\\
	\mathcal{F}_{U,LTT}=&\sin2\phi_{hLTT}\left(F_{U, LTT}^{T\sin2\phi_{hLTT}}+\varepsilon F_{U, LTT}^{L\sin2\phi_{hLTT}}\right)\nonumber\\
	&+\sqrt{2\varepsilon(1+\varepsilon)}\left(\sin\left(\phi_{h}-2\phi_{hLTT}\right)F_{U, LTT}^{\sin\left(\phi_{h}-2\phi_{hLTT}\right)}+\sin\left(\phi_{h}+2\phi_{hLTT}\right)F_{U, LTT}^{\sin\left(\phi_{h}+2\phi_{hLTT}\right)}\right)\nonumber\\
	&+\varepsilon\left(\sin\left(2\phi_{h}-2\phi_{hLTT}\right)F_{U, LTT}^{\sin\left(2\phi_{h}-2\phi_{hLTT}\right)}+\sin\left(2\phi_{h}+2\phi_{hLTT}\right)F_{U, LTT}^{\sin\left(2\phi_{h}+2\phi_{hLTT}\right)}\right)\nonumber\\
	&+\lambda_{e}\left[\sqrt{1-\varepsilon^{2}}\cos2\phi_{hLTT}G_{ U,LTT}^{\cos2\phi_{hLTT}}\right.\nonumber\\
	&\left.+\sqrt{2\varepsilon(1-\varepsilon)}\left(\cos\left(\phi_{h}-2\phi_{hLTT}\right)G_{ U,LTT}^{\cos\left(\phi_{h}-2\phi_{hLTT}\right)}+\cos\left(\phi_{h}+2\phi_{hLTT}\right)G_{ U,LTT}^{\cos\left(\phi_{h}+2\phi_{hLTT}\right)}\right)\right],\\
	\mathcal{F}_{U,TTT}=&\sin3\phi_{hTTT}\left(F_{ TTT}^{T\sin3\phi_{hTTT}}+\varepsilon F_{ U,TTT}^{L\sin3\phi_{hTTT}}\right)\nonumber\\
	&+\sqrt{2\varepsilon(1+\varepsilon)}\left(\sin\left(\phi_{h}-3\phi_{hTTT}\right)F_{ U,TTT}^{\sin\left(\phi_{h}-3\phi_{hTTT}\right)}+\sin\left(\phi_{h}+3\phi_{hTTT}\right)F_{ U,TTT}^{\sin\left(\phi_{h}+3\phi_{hTTT}\right)}\right)\nonumber\\
	&+\varepsilon\left(\sin\left(2\phi_{h}-3\phi_{hTTT}\right)F_{ U,TTT}^{\sin\left(2\phi_{h}-3\phi_{hTTT}\right)}+\sin\left(2\phi_{h}+3\phi_{hTTT}\right)F_{ U,TTT}^{\sin\left(2\phi_{h}+3\phi_{hTTT}\right)}\right)\nonumber\\
	&+\lambda_{e}\left[\sqrt{1-\varepsilon^{2}}\cos3\phi_{hTTT}G_{ U,TTT}^{\cos3\phi_{hTTT}}\right.\nonumber\\
	&\left.+\sqrt{2\varepsilon(1-\varepsilon)}\left(\cos\left(\phi_{h}-3\phi_{hTTT}\right)G_{ U,TTT}^{\cos\left(\phi_{h}-3\phi_{hTTT}\right)}+\cos\left(\phi_{h}+3\phi_{hTTT}\right)G_{ U,TTT}^{\cos\left(\phi_{h}+3\phi_{hTTT}\right)}\right)\right];\label{e.FUTTT}
\end{align}
13 are for vector polarized target states and unpolarized hadron,
\begin{align}
	\mathcal{F}_{L,U}=& \sqrt{2\varepsilon(1+\varepsilon)}\sin\phi_{h}F_{L,U}^{\sin\phi_{h}}+\varepsilon\sin2\phi_{h} F_{L,U}^{\sin2\phi_{h}}+\lambda_{e}\left(\sqrt{1-\varepsilon^{2}}G_{L,U}+ \sqrt{2\varepsilon(1-\varepsilon)}\cos\phi_{h}G_{L,U}^{\cos\phi_{h}}\right), \\
	\mathcal{F}_{T,U}=&\sin(\phi_{h}-\phi_{T})\left(F_{T,U}^{T \sin(\phi_{h}-\phi_{T})}+\varepsilon F_{T,U}^{L \sin(\phi_{h}-\phi_{T})}\right)+\varepsilon\sin(\phi_{h}+\phi_{T})F_{T,U}^{\sin(\phi_{h}+\phi_{T})}\nonumber\\
	&+\varepsilon\sin(3\phi_{h}+\phi_{T})F_{T,U}^{\sin(3\phi_{h}+\phi_{T})}+ \sqrt{2\varepsilon(1+\varepsilon)}\left(\sin\phi_{T}F_{T,U}^{\sin\phi_{T}}+ \sin(2\phi_{h}-\phi_{T})F_{T,U}^{\sin(2\phi_{h}-\phi_{T})}\right)\nonumber\\
	&+\lambda_{e}\left[\sqrt{1-\varepsilon^{2}}\cos(\phi_{h}-\phi_{T})G_{T,U}^{\cos(\phi_{h}-\phi_{T})}\right.\nonumber\\
	&\left.+ \sqrt{2\varepsilon(1-\varepsilon)}\left(\cos(2\phi_{h}-\phi_{T})G_{T,U}^{\cos(2\phi_{h}-\phi_{T})}+\cos\phi_{T}G_{T,U}^{\cos\phi_{T}}\right)\right];\label{e.ATU}
\end{align}
41 are for vector polarized target states and vector polarized hadron states,
\begin{align}
	\mathcal{F}_{L,L}&= F_{L,L}^{T}+ \varepsilon F_{L,L}^{L}+ \sqrt{2\varepsilon(1+\varepsilon)}\cos\phi_{h}F_{L,L}^{\cos\phi_{h}}+ \varepsilon\cos2\phi_{h} F_{L,L}^{\cos2\phi_{h}}+\lambda_{e} \sqrt{2\varepsilon(1-\varepsilon)}\sin\phi_{h}G_{L,L}^{\sin\phi_{h}},\\
    \mathcal{F}_{L,T}&=\cos\phi_{hT}\left(F_{L,T}^{T\cos\phi_{hT}}+\varepsilon F_{L,T}^{L\cos\phi_{hT}}\right)\nonumber\\
    &+\sqrt{2\varepsilon(1+\varepsilon)}\left(\cos\left(\phi_{h}-\phi_{hT}\right)F_{L,T}^{\cos\left(\phi_{h}-\phi_{hT}\right)}+\cos\left(\phi_{h}+\phi_{hT}\right)F_{L,T}^{\cos\left(\phi_{h}+\phi_{hT}\right)}\right)\nonumber\\
    &+\varepsilon\left(\cos\left(2\phi_{h}-\phi_{hT}\right)F_{L,T}^{\cos\left(2\phi_{h}-\phi_{hT}\right)}+\cos\left(2\phi_{h}+\phi_{hT}\right)F_{L,T}^{\cos\left(2\phi_{h}+\phi_{hT}\right)}\right)\nonumber\\
    &+\lambda_e \Big[\sqrt{1-\varepsilon^{2}}\sin\phi_{hT}G_{L,T}^{\sin\phi_{hT}}\nonumber\\
    &+\sqrt{2\varepsilon(1-\varepsilon)}\left(\sin\left(\phi_{h}-\phi_{hT}\right)G_{L,T}^{\sin\left(\phi_{h}-\phi_{hT}\right)}+\sin\left(\phi_{h}+\phi_{hT}\right)G_{L,T}^{\sin\left(\phi_{h}+\phi_{hT}\right)}\right)\Big],\\
	\mathcal{F}_{T,L}&=\cos\left(\phi_{h}-\phi_{T}\right)\left( F_{T,L}^{T\cos\left(\phi_{h}-\phi_{T}\right)}+ \varepsilon F_{T,L}^{L\cos\left(\phi_{h}-\phi_{T}\right)}\right)\nonumber\\
	&+ \sqrt{2\varepsilon(1+\varepsilon)}\left(\cos\phi_{T}F_{T,L}^{\cos\phi_{T}}+\cos\left(2\phi_{h}-\phi_{T}\right)F_{T,L}^{\cos\left(2\phi_{h}-\phi_{T}\right)}\right)\nonumber\\
	&+ \varepsilon\left(\cos\left(\phi_{h}+\phi_{T}\right)F_{T,L}^{\cos\left(\phi_{h}+\phi_{T}\right)}+\cos\left(3\phi_{h}-\phi_{T}\right)F_{T,L}^{\cos\left(3\phi_{h}-\phi_{T}\right)}\right)\nonumber\\
	&+\lambda_{e}\left[\sqrt{1-\varepsilon^{2}}\sin\left(\phi_{h}-\phi_{T}\right)G_{T,L}^{\sin\left(\phi_{h}-\phi_{T}\right)}\right.\nonumber\\
	&+ \left.\sqrt{2\varepsilon(1-\varepsilon)}\left(\sin\phi_{T}G_{T,L}^{\sin\phi_{T}}+\sin\left(2\phi_{h}-\phi_{T}\right)G_{T,L}^{\sin\left(2\phi_{h}-\phi_{T}\right)}\right)\right],\\
	\mathcal{F}_{T,T}&=\cos\left(\phi_{h}-\phi_{hT}-\phi_{T}\right)\left( F_{T,T}^{T\cos\left(\phi_{h}-\phi_{hT}-\phi_{T}\right)}+ \varepsilon F_{T,T}^{L\cos\left(\phi_{h}-\phi_{hT}-\phi_{T}\right)}\right)\nonumber\\
	&+\cos\left(\phi_{h}+\phi_{hT}-\phi_{T}\right)\left( F_{T,T}^{T\cos\left(\phi_{h}+\phi_{hT}-\phi_{T}\right)}+ \varepsilon F_{T,T}^{L\cos\left(\phi_{h}+\phi_{hT}-\phi_{T}\right)}\right)\nonumber\\
	&+ \sqrt{2\varepsilon(1+\varepsilon)}\left(\cos\left(\phi_{hT}-\phi_{T}\right)F_{T,T}^{\cos\left(\phi_{hT}-\phi_{T}\right)}+\cos\left(\phi_{hT}+\phi_{T}\right)F_{T,T}^{\cos\left(\phi_{hT}+\phi_{T}\right)}\right.\nonumber\\
	&\left.+\cos\left(2\phi_{h}-\phi_{hT}-\phi_{T}\right)F_{T,T}^{\cos\left(2\phi_{h}-\phi_{hT}-\phi_{T}\right)}+\cos\left(2\phi_{h}+\phi_{hT}-\phi_{T}\right)F_{T,T}^{\cos\left(2\phi_{h}+\phi_{hT}-\phi_{T}\right)}\right)\nonumber\\
	&+ \varepsilon\left(\cos\left(\phi_{h}-\phi_{hT}+\phi_{T}\right)F_{T,T}^{\cos\left(\phi_{h}-\phi_{hT}+\phi_{T}\right)}+\cos\left(\phi_{h}+\phi_{hT}+\phi_{T}\right)F_{T,T}^{\cos\left(\phi_{h}+\phi_{hT}+\phi_{T}\right)}\right.\nonumber\\
	&\left.+\cos\left(3\phi_{h}-\phi_{hT}-\phi_{T}\right)F_{T,T}^{\cos\left(3\phi_{h}-\phi_{hT}-\phi_{T}\right)}+\cos\left(3\phi_{h}+\phi_{hT}-\phi_{T}\right)F_{T,T}^{\cos\left(3\phi_{h}+\phi_{hT}-\phi_{T}\right)}\right)\nonumber\\
	&+\lambda_{e}\left[ \sqrt{2\varepsilon(1-\varepsilon)}\left(\sin\left(\phi_{hT}-\phi_{T}\right)G_{T,T}^{\sin\left(\phi_{hT}-\phi_{T}\right)}+\sin\left(\phi_{hT}+\phi_{T}\right)G_{T,T}^{\sin\left(\phi_{hT}+\phi_{T}\right)}\right.\right.\nonumber\\
	&\left.+\sin\left(2\phi_{h}-\phi_{hT}-\phi_{T}\right)G_{T,T}^{\sin\left(2\phi_{h}-\phi_{hT}-\phi_{T}\right)}+\sin\left(2\phi_{h}+\phi_{hT}-\phi_{T}\right)G_{T,T}^{\sin\left(2\phi_{h}+\phi_{hT}-\phi_{T}\right)}\right)\nonumber\\
	&\left.+\sqrt{1-\varepsilon^{2}}\left(\sin\left(\phi_{h}-\phi_{hT}-\phi_{T}\right)G_{T,T}^{\sin\left(\phi_{h}-\phi_{hT}-\phi_{T}\right)}+\sin\left(\phi_{h}+\phi_{hT}-\phi_{T}\right)G_{T,T}^{\sin\left(\phi_{h}+\phi_{hT}-\phi_{T}\right)}\right)\right];
\end{align}
 67 are for vector polarized target states and rank-2 tensor polarized hadron states,
\begin{align}
	\mathcal{F}_{L,LL}=& \sqrt{2\varepsilon(1+\varepsilon)}\sin\phi_{h}F_{L,LL}^{\sin\phi_{h}}+ \varepsilon\sin2\phi_{h}F_{L,LL}^{\sin2\phi_{h}}+\lambda_{e}\left(\sqrt{1-\varepsilon^{2}}G_{L,LL}+ \sqrt{2\varepsilon(1-\varepsilon)}\cos\phi_{h}G_{L,LL}^{\cos\phi_{h}}\right),\label{e.FLLL}\\
	\mathcal{F}_{L,LT}=&\sin\phi_{hLT}\left( F_{L,LT}^{T\sin\phi_{hLT}}+ \varepsilon F_{L,LT}^{L\sin\phi_{hLT}}\right)\nonumber\\
	&+ \sqrt{2\varepsilon(1+\varepsilon)}\left(\sin\left(\phi_{h}+\phi_{hLT}\right)F_{L,LT}^{\sin\left(\phi_{h}+\phi_{hLT}\right)}+\sin\left(\phi_{h}-\phi_{hLT}\right)F_{L,LT}^{\sin\left(\phi_{h}-\phi_{hLT}\right)}\right)\nonumber\\
	&+ \varepsilon\left(\sin\left(2\phi_{h}+\phi_{hLT}\right)F_{L,LT}^{\sin\left(2\phi_{h}+\phi_{hLT}\right)}+\sin\left(2\phi_{h}-\phi_{hLT}\right)F_{L,LT}^{\sin\left(2\phi_{h}-\phi_{hLT}\right)}\right)\nonumber\\
	&+\lambda_{e}\Big[\sqrt{1-\varepsilon^{2}}\cos\phi_{hLT}G_{L,LT}^{\cos\phi_{hLT}}\nonumber\\
	&+\sqrt{2\varepsilon(1-\varepsilon)}\left(\cos\left(\phi_{h}-\phi_{hLT}\right)G_{L,LT}^{\cos\left(\phi_{h}-\phi_{hLT}\right)}+\cos\left(\phi_{h}+\phi_{hLT}\right)G_{L,LT}^{\cos\left(\phi_{h}+\phi_{hLT}\right)}\right)\Big],\\
	 \mathcal{F}_{L,TT}=&\sin2\phi_{hTT}\left(F_{L,TT}^{T\sin2\phi_{hTT}}+ \varepsilon F_{L,TT}^{L\sin2\phi_{hTT}}\right)\nonumber\\
	 &+ \sqrt{2\varepsilon(1+\varepsilon)}\left(\sin\left(\phi_{h}+2\phi_{hTT}\right)F_{L,TT}^{\sin\left(\phi_{h}+2\phi_{hTT}\right)}+\sin\left(\phi_{h}-2\phi_{hTT}\right)F_{L,TT}^{\sin\left(\phi_{h}-2\phi_{hTT}\right)}\right)\nonumber\\
	 &+ \varepsilon\left(\sin\left(2\phi_{h}+2\phi_{hTT}\right)F_{L,TT}^{\sin\left(2\phi_{h}+2\phi_{hTT}\right)}+\sin\left(2\phi_{h}-2\phi_{hTT}\right)F_{L,TT}^{\sin\left(2\phi_{h}-2\phi_{hTT}\right)}\right)\nonumber\\
	 &+\lambda_{e}\left[\sqrt{1-\varepsilon^{2}}\cos2\phi_{hTT}G_{L,TT}^{\cos2\phi_{hTT}}\right.\nonumber\\
	 &\left.+ \sqrt{2\varepsilon(1-\varepsilon)}\left(\cos\left(\phi_{h}-2\phi_{hTT}\right)G_{L,TT}^{\cos\left(\phi_{h}-2\phi_{hTT}\right)}+\cos\left(\phi_{h}+2\phi_{hTT}\right)G_{L,TT}^{\cos\left(\phi_{h}+2\phi_{hTT}\right)}\right)\right],\\
	\mathcal{F}_{T,LL}=&\sin\left(\phi_{h}-\phi_{T}\right)\left( F_{T,LL}^{T\sin\left(\phi_{h}-\phi_{T}\right)}+ \varepsilon F_{T,LL}^{L\sin\left(\phi_{h}-\phi_{T}\right)}\right)\nonumber\\
	&+ \sqrt{2\varepsilon(1+\varepsilon)}\left(\sin\phi_{T}F_{T,LL}^{\sin\phi_{T}}+\sin\left(2\phi_{h}-\phi_{T}\right)F_{T,LL}^{\sin\left(2\phi_{h}-\phi_{T}\right)}\right)\nonumber\\
	&+ \varepsilon\left(\sin\left(\phi_{h}+\phi_{T}\right)F_{T,LL}^{\sin\left(\phi_{h}+\phi_{T}\right)}+\sin\left(3\phi_{h}-\phi_{T}\right)F_{T,LL}^{\sin\left(3\phi_{h}-\phi_{T}\right)}\right)\nonumber\\
	&+\lambda_{e}\left[\sqrt{1-\varepsilon^{2}}\cos\left(\phi_{h}-\phi_{T}\right)G_{T,LL}^{\cos\left(\phi_{h}-\phi_{T}\right)}\right.\nonumber\\
	&\:\left.+ \sqrt{2\varepsilon(1-\varepsilon)}\left(\cos\phi_{T}G_{T,LL}^{\cos\phi_{T}}+\cos\left(2\phi_{h}-\phi_{T}\right)G_{T,LL}^{\cos\left(2\phi_{h}-\phi_{T}\right)}\right)\right],\\
	\mathcal{F}_{T,LT}&=
	\sin\left(\phi_{h}-\phi_{hLT}-\phi_{T}\right)\left( F_{T,LT}^{T \sin \left(\phi_{h}-\phi_{hLT}-\phi_{T}\right)}+ \varepsilon F_{T,LT}^{L\sin\left(\phi_{h}-\phi_{hLT}-\phi_{T}\right)}\right)\nonumber\\
	& +\sin\left(\phi_{h}+\phi_{hLT}-\phi_{T}\right)\left(F_{T,LT}^{T \sin \left(\phi_{h}+\phi_{hLT}-\phi_{T}\right)}+ \varepsilon F_{T,LT}^{L\sin\left(\phi_{h}+\phi_{hLT}-\phi_{T}\right)}\right)\nonumber\\
	&+ \sqrt{2\varepsilon(1+\varepsilon)}\left(\sin\left(\phi_{hLT}-\phi_{T}\right)F_{T,LT}^{\sin\left(\phi_{hLT}-\phi_{T}\right)}+\sin\left(\phi_{hLT}+\phi_{T}\right)F_{T,LT}^{\sin\left(\phi_{hLT}+\phi_{T}\right)}\right.\nonumber\\
	&\left.+\sin\left(2\phi_{h}+\phi_{hLT}-\phi_{T}\right)F_{T,LT}^{\sin\left(2\phi_{h}+\phi_{hLT}-\phi_{T}\right)}+\sin\left(2\phi_{h}-\phi_{hLT}-\phi_{T}\right)F_{T,LT}^{\sin\left(2\phi_{h}-\phi_{hLT}-\phi_{T}\right)}\right)\nonumber\\
	&+ \varepsilon\left(\sin\left(\phi_{h}+\phi_{hLT}+\phi_{T}\right)F_{T,LT}^{\sin\left(\phi_{h}+\phi_{hLT}+\phi_{T}\right)}+\sin\left(\phi_{h}-\phi_{hLT}+\phi_{T}\right)F_{T,LT}^{\sin\left(\phi_{h}-\phi_{hLT}+\phi_{T}\right)}\right.\nonumber\\
	&\left.+\sin\left(3\phi_{h}-\phi_{hLT}-\phi_{T}\right)F_{T,LT}^{\sin\left(3\phi_{h}-\phi_{hLT}-\phi_{T}\right)}+\sin\left(3\phi_{h}+\phi_{hLT}-\phi_{T}\right)F_{T,LT}^{\sin\left(3\phi_{h}+\phi_{hLT}-\phi_{T}\right)}\right)\nonumber\\
	&+\lambda_{e}\left[ \sqrt{2\varepsilon(1-\varepsilon)}\left(\cos\left(\phi_{hLT}-\phi_{T}\right)G_{T,LT}^{\cos\left(\phi_{hLT}-\phi_{T}\right)}+\cos\left(\phi_{hLT}+\phi_{T}\right)G_{T,LT}^{\cos\left(\phi_{hLT}+\phi_{T}\right)}\right.\right.\nonumber\\
	&\left.+\cos\left(2\phi_{h}-\phi_{hLT}-\phi_{T}\right)G_{T,LT}^{\cos\left(2\phi_{h}-\phi_{hLT}-\phi_{T}\right)}+\cos\left(2\phi_{h}+\phi_{hLT}-\phi_{T}\right)G_{T,LT}^{\cos\left(2\phi_{h}+\phi_{hLT}-\phi_{T}\right)}\right)\nonumber\\
	&\left.+\sqrt{1-\varepsilon^{2}}\left(\cos\left(\phi_{h}-\phi_{hLT}-\phi_{T}\right)G_{T,LT}^{\cos\left(\phi_{h}-\phi_{hLT}-\phi_{T}\right)}+\cos\left(\phi_{h}+\phi_{hLT}-\phi_{T}\right)G_{T,LT}^{\cos\left(\phi_{h}+\phi_{hLT}-\phi_{T}\right)}\right)\right],\\
	\mathcal{F}_{T,TT}&=\sin\left(\phi_{h}-2\phi_{hTT}-\phi_{T}\right)\left(F_{T,TT}^{T\sin\left(\phi_{h}-2\phi_{hTT}-\phi_{T}\right)}+ \varepsilon F_{T,TT}^{L\sin\left(\phi_{h}-2\phi_{hTT}-\phi_{T}\right)}\right)\nonumber\\
	&+\sin\left(\phi_{h}+2\phi_{hTT}-\phi_{T}\right)\left(F_{T,TT}^{T\sin\left(\phi_{h}+2\phi_{hTT}-\phi_{T}\right)}+ \varepsilon F_{T,TT}^{L\sin\left(\phi_{h}+2\phi_{hTT}-\phi_{T}\right)}\right)\nonumber\\
	&+ \sqrt{2\varepsilon(1+\varepsilon)}\left(\sin\left(2\phi_{hTT}-\phi_{T}\right)F_{T,TT}^{\sin\left(2\phi_{hTT}-\phi_{T}\right)}+\sin\left(2\phi_{hTT}+\phi_{T}\right)F_{T,TT}^{\sin\left(2\phi_{hTT}+\phi_{T}\right)}\right.\nonumber\\
	&\left.+\sin\left(2\phi_{h}+2\phi_{hTT}-\phi_{T}\right)F_{T,TT}^{\sin\left(2\phi_{h}+2\phi_{hTT}-\phi_{T}\right)}+\sin\left(2\phi_{h}-2\phi_{hTT}-\phi_{T}\right)F_{T,TT}^{\sin\left(2\phi_{h}-2\phi_{hTT}-\phi_{T}\right)}\right)\nonumber\\
	&+ \varepsilon\left(\sin\left(\phi_{h}+2\phi_{hTT}+\phi_{T}\right)F_{T,TT}^{\sin\left(\phi_{h}+2\phi_{hTT}+\phi_{T}\right)}+\sin\left(\phi_{h}-2\phi_{hTT}+\phi_{T}\right)F_{T,TT}^{\sin\left(\phi_{h}-2\phi_{hTT}+\phi_{T}\right)}\right.\nonumber\\
	&\left.+\sin\left(3\phi_{h}-2\phi_{hTT}-\phi_{T}\right)F_{T,TT}^{\sin\left(3\phi_{h}-2\phi_{hTT}-\phi_{T}\right)}+\sin\left(3\phi_{h}+2\phi_{hTT}-\phi_{T}\right)F_{T,TT}^{\sin\left(3\phi_{h}+2\phi_{hTT}-\phi_{T}\right)}\right)\nonumber\\
	&+\lambda_{e}\left[ \sqrt{2\varepsilon(1-\varepsilon)}\left(\cos\left(2\phi_{hTT}-\phi_{T}\right)G_{T,TT}^{\cos\left(2\phi_{hTT}-\phi_{T}\right)}+\cos\left(2\phi_{hTT}+\phi_{T}\right)G_{T,TT}^{\cos\left(2\phi_{hTT}+\phi_{T}\right)}\right.\right.\nonumber\\
	&\left.+\cos\left(2\phi_{h}-2\phi_{hTT}-\phi_{T}\right)G_{T,TT}^{\cos\left(2\phi_{h}-2\phi_{hTT}-\phi_{T}\right)}+\cos\left(2\phi_{h}+2\phi_{hTT}-\phi_{T}\right)G_{T,TT}^{\cos\left(2\phi_{h}+2\phi_{hTT}-\phi_{T}\right)}\right)\nonumber\\
	&\left.+\sqrt{1-\varepsilon^{2}}\left(\cos\left(\phi_{h}-2\phi_{hTT}-\phi_{T}\right)G_{T,TT}^{\cos\left(\phi_{h}-2\phi_{hTT}-\phi_{T}\right)}+\cos\left(\phi_{h}+2\phi_{hTT}-\phi_{T}\right)G_{T,TT}^{\cos\left(\phi_{h}+2\phi_{hTT}-\phi_{T}\right)}\right)\right];
\end{align}
and 95 are for vector polarized target states and rank-3 tensor polarized hadron states,
\begin{align}
	\mathcal{F}_{L,LLL}=&F_{L,LLL}^{T}+\varepsilon F_{L,LLL}^{L}+\sqrt{2\varepsilon(1+\varepsilon)}\cos\phi_{h}F_{L,LLL}^{\cos\phi_{h}}+\varepsilon\cos2\phi_{h}F_{L,LLL}^{\cos2\phi_{h}}+\lambda_{e} 
    \sqrt{2\varepsilon(1-\varepsilon)}\sin\phi_{h}G_{L,LLL}^{\sin\phi_{h}},\\
   \mathcal{F}_{L,LLT}=&\cos\phi_{hLLT}\left(F_{L,LLT}^{T\cos\phi_{hLLT}}+\varepsilon F_{L,LLT}^{L\cos\phi_{hLLT}}\right)\nonumber\\
    &+\sqrt{2\varepsilon(1+\varepsilon)}\left(\cos\left(\phi_{h}-\phi_{hLLT}\right)F_{L,LLT}^{\cos\left(\phi_{h}-\phi_{hLLT}\right)}+\cos\left(\phi_{h}+\phi_{hLLT}\right)F_{L,LLT}^{\cos\left(\phi_{h}+\phi_{hLLT}\right)}\right)\nonumber\\
    &+\varepsilon\left(\cos\left(2\phi_{h}-\phi_{hLLT}\right)F_{L,LLT}^{\cos\left(2\phi_{h}-\phi_{hLLT}\right)}+\cos\left(2\phi_{h}+\phi_{hLLT}\right)F_{L,LLT}^{\cos\left(2\phi_{h}+\phi_{hLLT}\right)}\right)\nonumber\\
    &+\lambda_{e}\left[\sqrt{1-\varepsilon^{2}}\sin\phi_{hLLT}G_{L,LLT}^{\sin\phi_{hLLT}}\right.\nonumber\\
    &\left.+\sqrt{2\varepsilon(1-\varepsilon)}\left(\sin\left(\phi_{h}-\phi_{hLLT}\right)G_{L,LLT}^{\sin\left(\phi_{h}-\phi_{hLLT}\right)}+\sin\left(\phi_{h}+\phi_{hLLT}\right)G_{L,LLT}^{\sin\left(\phi_{h}+\phi_{hLLT}\right)}\right)\right],\\
    \mathcal{F}_{L,LTT}=&\cos2\phi_{hLTT}\left(F_{L,LTT}^{T\cos2\phi_{hLTT}}+\varepsilon F_{L,LTT}^{L\cos2\phi_{hLTT}}\right)\nonumber\\
    &+\sqrt{2\varepsilon(1+\varepsilon)}\left(\cos\left(\phi_{h}-2\phi_{hLTT}\right)F_{L,LTT}^{\cos\left(\phi_{h}-2\phi_{hLTT}\right)}+\cos\left(\phi_{h}+2\phi_{hLTT}\right)F_{L,LTT}^{\cos\left(\phi_{h}+2\phi_{hLTT}\right)}\right)\nonumber\\
    &+\varepsilon\left(\cos\left(2\phi_{h}-2\phi_{hLTT}\right)F_{L,LTT}^{\cos\left(2\phi_{h}-2\phi_{hLTT}\right)}+\cos\left(2\phi_{h}+2\phi_{hLTT}\right)F_{L,LTT}^{\cos\left(2\phi_{h}+2\phi_{hLTT}\right)}\right)\nonumber\\
    &+\lambda_{e}\left[\sqrt{1-\varepsilon^{2}}\sin2\phi_{hLTT}G_{L,LTT}^{\sin2\phi_{hLTT}}\right.\nonumber\\
   &\left.+\sqrt{2\varepsilon(1-\varepsilon)}\left(\sin\left(\phi_{h}-2\phi_{hLTT}\right)G_{L,LTT}^{\sin\left(\phi_{h}-2\phi_{hLTT}\right)}+\sin\left(\phi_{h}+2\phi_{hLTT}\right)G_{L,LTT}^{\sin\left(\phi_{h}+2\phi_{hLTT}\right)}\right)\right],\\
    \mathcal{F}_{L,TTT}=&\cos3\phi_{hTTT}\left(F_{L,TTT}^{T\cos3\phi_{hTTT}}+\varepsilon F_{L,TTT}^{L\cos3\phi_{hTTT}}\right)\nonumber\\
    &+\sqrt{2\varepsilon(1+\varepsilon)}\left(\cos\left(\phi_{h}-3\phi_{hTTT}\right)F_{L,TTT}^{\cos\left(\phi_{h}-3\phi_{hTTT}\right)}+\cos\left(\phi_{h}+3\phi_{hTTT}\right)F_{L,TTT}^{\cos\left(\phi_{h}+3\phi_{hTTT}\right)}\right)\nonumber\\
    &+\varepsilon\left(\cos\left(2\phi_{h}-3\phi_{hTTT}\right)F_{L,TTT}^{\cos\left(2\phi_{h}-3\phi_{hTTT}\right)}+\cos\left(2\phi_{h}+3\phi_{hTTT}\right)F_{L,TTT}^{\cos\left(2\phi_{h}+3\phi_{hTTT}\right)}\right)\nonumber\\
    &+\lambda_{e}\left[\sqrt{1-\varepsilon^{2}}\sin3\phi_{hTTT}G_{L,TTT}^{\sin3\phi_{hTTT}}\right.\nonumber\\
    &\left.+\sqrt{2\varepsilon(1-\varepsilon)}\left(\sin\left(\phi_{h}-3\phi_{hTTT}\right)G_{L,TTT}^{\sin\left(\phi_{h}-3\phi_{hTTT}\right)}+\sin\left(\phi_{h}+3\phi_{hTTT}\right)G_{L,TTT}^{\sin\left(\phi_{h}+3\phi_{hTTT}\right)}\right)\right],\\
    \mathcal{F}_{T,LLL}=&\cos\left(\phi_{h}-\phi_{T}\right)\left(F_{T,LLL}^{T\cos\left(\phi_{h}-\phi_{T}\right)}+\varepsilon F_{T,LLL}^{L\cos\left(\phi_{h}-\phi_{T}\right)}\right)\nonumber\\
    &+\sqrt{2\varepsilon(1+\varepsilon)}\left(\cos\phi_{T}F_{T,LLL}^{\cos\phi_{T}}+\cos\left(2\phi_{h}-\phi_{T}\right)F_{T,LLL}^{\cos\left(2\phi_{h}-\phi_{T}\right)}\right)\nonumber\\
    &+\varepsilon\left(\cos\left(\phi_{h}+\phi_{T}\right)F_{T,LLL}^{\cos\left(\phi_{h}+\phi_{T}\right)}+\cos\left(3\phi_{h}-\phi_{T}\right)F_{T,LLL}^{\cos\left(3\phi_{h}-\phi_{T}\right)}\right)\nonumber\\
    &+\lambda_{e}\left[\sqrt{1-\varepsilon^{2}}\sin\left(\phi_{h}-\phi_{T}\right)G_{T,LLL}^{\sin\left(\phi_{h}-\phi_{T}\right)}\right.\nonumber\\
    &\left.+\sqrt{2\varepsilon(1-\varepsilon)}\left(\sin\phi_{T}G_{T,LLL}^{\sin\phi_{T}}+\sin\left(2\phi_{h}-\phi_{T}\right)G_{T,LLL}^{\sin\left(2\phi_{h}-\phi_{T}\right)}\right)\right],
\end{align}
\begin{align}
     &\mathcal{F}_{T,LLT}=
     \cos\left(\phi_{h}-\phi_{hLLT}-\phi_{T}\right)\left(F_{T,LLT}^{T\cos\left(\phi_{h}-\phi_{hLLT}-\phi_{T}\right)}+\varepsilon F_{T,LLT}^{L\cos\left(\phi_{h}-\phi_{hLLT}-\phi_{T}\right)}\right)\nonumber\\
     &+\cos\left(\phi_{h}+\phi_{hLLT}-\phi_{T}\right)\left(F_{T,LLT}^{T\cos\left(\phi_{h}+\phi_{hLLT}-\phi_{T}\right)}+\varepsilon F_{T,LLT}^{L\cos\left(\phi_{h}+\phi_{hLLT}-\phi_{T}\right)}\right)\nonumber\\
     &+\sqrt{2\varepsilon(1+\varepsilon)}\left(\cos\left(\phi_{hLLT}-\phi_{T}\right)F_{T,LLT}^{\cos\left(\phi_{hLLT}-\phi_{T}\right)}+\cos\left(\phi_{hLLT}+\phi_{T}\right)F_{T,LLT}^{\cos\left(\phi_{hLLT}+\phi_{T}\right)}\right.\nonumber\\
     &\left.+\cos\left(2\phi_{h}-\phi_{hLLT}-\phi_{T}\right)F_{T,LLT}^{\cos\left(2\phi_{h}-\phi_{hLLT}-\phi_{T}\right)}+\cos\left(2\phi_{h}+\phi_{hLLT}-\phi_{T}\right)F_{T,LLT}^{\cos\left(2\phi_{h}+\phi_{hLLT}-\phi_{T}\right)}\right)\nonumber\\
     &+\varepsilon\left(\cos\left(\phi_{h}-\phi_{hLLT}+\phi_{T}\right)F_{T,LLT}^{\cos\left(\phi_{h}-\phi_{hLLT}+\phi_{T}\right)}+\cos\left(\phi_{h}+\phi_{hLLT}+\phi_{T}\right)F_{T,LLT}^{\cos\left(\phi_{h}+\phi_{hLLT}+\phi_{T}\right)}\right.\nonumber\\
     &\left.+\cos\left(3\phi_{h}-\phi_{hLLT}-\phi_{T}\right)F_{T,LLT}^{\cos\left(3\phi_{h}-\phi_{hLLT}-\phi_{T}\right)}+\cos\left(3\phi_{h}+\phi_{hLLT}-\phi_{T}\right)F_{T,LLT}^{\cos\left(3\phi_{h}+\phi_{hLLT}-\phi_{T}\right)}\right)\nonumber\\
     &+\lambda_{e}\left[\sqrt{2\varepsilon(1-\varepsilon)}\left(\sin\left(\phi_{hLLT}-\phi_{T}\right)G_{T,LLT}^{\sin\left(\phi_{hLLT}-\phi_{T}\right)}+\sin\left(\phi_{hLLT}+\phi_{T}\right)G_{T,LLT}^{\sin\left(\phi_{hLLT}+\phi_{T}\right)}\right.\right.\nonumber\\
     &\left.+\sin\left(2\phi_{h}-\phi_{hLLT}-\phi_{T}\right)G_{T,LLT}^{\sin\left(2\phi_{h}-\phi_{hLLT}-\phi_{T}\right)}+\sin\left(2\phi_{h}+\phi_{hLLT}-\phi_{T}\right)G_{T,LLT}^{\sin\left(2\phi_{h}+\phi_{hLLT}-\phi_{T}\right)}\right)\nonumber\\
     &\left.+\sqrt{1-\varepsilon^{2}}\left(\sin\left(\phi_{h}-\phi_{hLLT}-\phi_{T}\right)G_{T,LLT}^{\sin\left(\phi_{h}-\phi_{hLLT}-\phi_{T}\right)}+\sin\left(\phi_{h}+\phi_{hLLT}-\phi_{T}\right)G_{T,LLT}^{\sin\left(\phi_{h}+\phi_{hLLT}-\phi_{T}\right)}\right)\right],\\  
     %
     &\mathcal{F}_{T,LTT}=
	 \cos\left(\phi_{h}-2\phi_{hLTT}-\phi_{T}\right)\left(F_{T,LTT}^{T\cos\left(\phi_{h}-2\phi_{hLTT}-\phi_{T}\right)}+\varepsilon F_{T,LTT}^{L\cos\left(\phi_{h}-2\phi_{hLTT}-\phi_{T}\right)}\right)\nonumber\\
     &+\cos\left(\phi_{h}+2\phi_{hLTT}-\phi_{T}\right)\left(F_{T,LTT}^{T\cos\left(\phi_{h}+2\phi_{hLTT}-\phi_{T}\right)}+\varepsilon F_{T,LTT}^{L\cos\left(\phi_{h}+2\phi_{hLTT}-\phi_{T}\right)}\right)\nonumber\\
     &+\sqrt{2\varepsilon(1+\varepsilon)}\left(\cos\left(2\phi_{hLTT}-\phi_{T}\right)F_{T,LTT}^{\cos\left(2\phi_{hLTT}-\phi_{T}\right)}+\cos\left(2\phi_{hLTT}+\phi_{T}\right)F_{T,LTT}^{\cos\left(2\phi_{hLTT}+\phi_{T}\right)}\right.\nonumber\\
     &\left.+\cos\left(2\phi_{h}-2\phi_{hLTT}-\phi_{T}\right)F_{T,LTT}^{\cos\left(2\phi_{h}-2\phi_{hLTT}-\phi_{T}\right)}+\cos\left(2\phi_{h}+2\phi_{hLTT}-\phi_{T}\right)F_{T,LTT}^{\cos\left(2\phi_{h}+2\phi_{hLTT}-\phi_{T}\right)}\right)\nonumber\\
     &+\varepsilon\left(\cos\left(\phi_{h}-2\phi_{hLTT}+\phi_{T}\right)F_{T,LTT}^{\cos\left(\phi_{h}-2\phi_{hLTT}+\phi_{T}\right)}+\cos\left(\phi_{h}+2\phi_{hLTT}+\phi_{T}\right)F_{T,LTT}^{\cos\left(\phi_{h}+2\phi_{hLTT}+\phi_{T}\right)}\right.\nonumber\\
     &\left.+\cos\left(3\phi_{h}-2\phi_{hLTT}-\phi_{T}\right)F_{T,LTT}^{\cos\left(3\phi_{h}-2\phi_{hLTT}-\phi_{T}\right)}+\cos\left(3\phi_{h}+2\phi_{hLTT}-\phi_{T}\right)F_{T,LTT}^{\cos\left(3\phi_{h}+2\phi_{hLTT}-\phi_{T}\right)}\right)\nonumber\\
     &+\lambda_{e}\left[\sqrt{2\varepsilon(1-\varepsilon)}\left(\sin\left(2\phi_{hLTT}-\phi_{T}\right)G_{T,LTT}^{\sin\left(2\phi_{hLTT}-\phi_{T}\right)}+\sin\left(2\phi_{hLTT}+\phi_{T}\right)G_{T,LTT}^{\sin\left(2\phi_{hLTT}+\phi_{T}\right)}\right.\right.\nonumber\\
     &\left.+\sin\left(2\phi_{h}-2\phi_{hLTT}-\phi_{T}\right)G_{T,LTT}^{\sin\left(2\phi_{h}-2\phi_{hLTT}-\phi_{T}\right)}+\sin\left(2\phi_{h}+2\phi_{hLTT}-\phi_{T}\right)G_{T,LTT}^{\sin\left(2\phi_{h}+2\phi_{hLTT}-\phi_{T}\right)}\right)\nonumber\\
     &\left.+\sqrt{1-\varepsilon^{2}}\left(\sin\left(\phi_{h}-2\phi_{hLTT}-\phi_{T}\right)G_{T,LTT}^{\sin\left(\phi_{h}-2\phi_{hLTT}-\phi_{T}\right)}+\sin\left(\phi_{h}+2\phi_{hLTT}-\phi_{T}\right)G_{T,LTT}^{\sin\left(\phi_{h}+2\phi_{hLTT}-\phi_{T}\right)}\right)\right],\\ 
     %
     &\mathcal{F}_{T,TTT}=
     \cos\left(\phi_{h}-3\phi_{hTTT}-\phi_{T}\right)\left(F_{T,TTT}^{T\cos\left(\phi_{h}-3\phi_{hTTT}-\phi_{T}\right)}+\varepsilon F_{T,TTT}^{L\cos\left(\phi_{h}-3\phi_{hTTT}-\phi_{T}\right)}\right)\nonumber\\
     &+\cos\left(\phi_{h}+3\phi_{hTTT}-\phi_{T}\right)\left(F_{T,TTT}^{T\cos\left(\phi_{h}+3\phi_{hTTT}-\phi_{T}\right)}+\varepsilon F_{T,TTT}^{L\cos\left(\phi_{h}+3\phi_{hTTT}-\phi_{T}\right)}\right)\nonumber\\
     &+\sqrt{2\varepsilon(1+\varepsilon)}\left(\cos\left(3\phi_{hTTT}-\phi_{T}\right)F_{T,TTT}^{\cos\left(3\phi_{hTTT}-\phi_{T}\right)}+\cos\left(3\phi_{hTTT}+\phi_{T}\right)F_{T,TTT}^{\cos\left(3\phi_{hTTT}+\phi_{T}\right)}\right.\nonumber\\
     &\left.+\cos\left(2\phi_{h}-3\phi_{hTTT}-\phi_{T}\right)F_{T,TTT}^{\cos\left(2\phi_{h}-3\phi_{hTTT}-\phi_{T}\right)}+\cos\left(2\phi_{h}+3\phi_{hTTT}-\phi_{T}\right)F_{T,TTT}^{\cos\left(2\phi_{h}+3\phi_{hTTT}-\phi_{T}\right)}\right)\nonumber\\
     &+\varepsilon\left(\cos\left(\phi_{h}-3\phi_{hTTT}+\phi_{T}\right)F_{T,TTT}^{\cos\left(\phi_{h}-3\phi_{hTTT}+\phi_{T}\right)}+\cos\left(\phi_{h}+3\phi_{hTTT}+\phi_{T}\right)F_{T,TTT}^{\cos\left(\phi_{h}+3\phi_{hTTT}+\phi_{T}\right)}\right.\nonumber\\
     &\left.+\cos\left(3\phi_{h}-3\phi_{hTTT}-\phi_{T}\right)F_{T,TTT}^{\cos\left(3\phi_{h}-3\phi_{hTTT}-\phi_{T}\right)}+\cos\left(3\phi_{h}+3\phi_{hTTT}-\phi_{T}\right)F_{T,TTT}^{\cos\left(3\phi_{h}+3\phi_{hTTT}-\phi_{T}\right)}\right)\nonumber\\
     &+\lambda_{e}\left[\sqrt{2\varepsilon(1-\varepsilon)}\left(\sin\left(3\phi_{hTTT}-\phi_{T}\right)G_{T,TTT}^{\sin\left(3\phi_{hTTT}-\phi_{T}\right)}+\sin\left(3\phi_{hTTT}+\phi_{T}\right)G_{T,TTT}^{\sin\left(3\phi_{hTTT}+\phi_{T}\right)}\right.\right.\nonumber\\
     &\left.+\sin\left(2\phi_{h}-3\phi_{hTTT}-\phi_{T}\right)G_{T,TTT}^{\sin\left(2\phi_{h}-3\phi_{hTTT}-\phi_{T}\right)}+\sin\left(2\phi_{h}+3\phi_{hTTT}-\phi_{T}\right)G_{T,TTT}^{\sin\left(2\phi_{h}+3\phi_{hTTT}-\phi_{T}\right)}\right)\nonumber\\
     &\left.+\sqrt{1-\varepsilon^{2}}\left(\sin\left(\phi_{h}-3\phi_{hTTT}-\phi_{T}\right)G_{T,TTT}^{\sin\left(\phi_{h}-3\phi_{hTTT}-\phi_{T}\right)}+\sin\left(\phi_{h}+3\phi_{hTTT}-\phi_{T}\right)G_{T,TTT}^{\sin\left(\phi_{h}+3\phi_{hTTT}-\phi_{T}\right)}\right)\right],\label{e.ATTTT}
\end{align}
where $F$ and $G$ correspond to the structure functions for unpolarized and polarized lepton beams, respectively. Apart from the trigonometric function superscripts, which indicate the angular modulations, the superscripts $L$ and $T$ in some terms are used to distinguish the longitudinal and the transverse polarization of the virtual photon. The structure functions $F$'s and $G$'s, as scalar functions of $x$, $Q^2$, $z$, and $P_{h\perp}^2$, are linear combinations of the $V$'s in Eq.~\eqref{e.totaltensor}. One may easily find that the number of the structure functions corresponding to different spin states is exactly the number of the basis tensors that depend on the corresponding spin tensors. 

Among the 288 structure functions, 126 rank-3 tensor polarized ones are newly defined and exist only for the produced hadron with $s\geq3/2$, while the 18 unpolarized ones, the 54 vector polarized ones, and the 90 rank-2 tensor polarized ones also exist when detecting a spin-1 hadron. It is also worthwhile to mention that there are 192 terms of the 288 structure functions corresponding to the unpolarized lepton, while the other 96 terms relate to the polarized lepton.

For the production of a spin-3/2 hadron in SIDIS off an unpolarized target, the differential cross section can be expressed in terms of 72 structure functions in Eqs.~\eqref{e.FUU}--\eqref{e.FUTTT}, where the 31 rank-3 tensor polarized ones are newly defined and survive only when the spin of the produced hadron is no less than 3/2. If the lepton is unpolarized, we find that the corresponding cross section is characterized by a total of 48 structure functions, which is consistent with the number of structure functions in the cross section for $e^+e^-\rightarrow \Omega h X$ in Ref.~\cite{Zhao:2022lbw}.

\section{The calculation in the parton model}
\label{s.partonmodel}

In this section, we derive the structure functions in the parton model at the leading twist. 
For convenience, we choose the nucleon and the hadron back-to-back frame, in which the momenta can be expressed as
\begin{align}
	P^\mu=P^+\bar{n}^\mu+\frac{M^2}{2P^+}n^\mu,
    \quad 
    P^\mu_h=\frac{M^2_h}{2P_h^-}\bar{n}^\mu+P^-_h n^\mu,\label{e.PPh}
\end{align}
where $\bar{n}^\mu = (1,0,{\bm 0}_\perp)$ and $n^\mu = (0,1,{\bm 0}_\perp)$ in light-cone coordinates. Correspondingly, the transverse metric tensor $g_T^{\mu\nu}$ and the transverse antisymmetric tensor $\epsilon_T^{\mu\nu}$ can be expressed by the momenta of the target and the produced hadron,
\begin{align}
	&g_T^{\mu\nu} =g^{\mu\nu}-\frac{(P_h\cdot P)(P_h^{\mu}P^{\nu}+P_h^{\nu}P^{\mu})}{(P_h\cdot P)^{2}-M_h^{2}M^{2}}+\frac{M_h^{2}P^{\mu}P^{\nu}+M^{2}P_h^{\mu}P_h}{(P_h\cdot P)^{2}-M_h^{2}M^{2}}, \label{e.gt} \\
	&\epsilon_T^{\mu\nu} =\epsilon^{\mu\nu\rho\sigma}\frac{P_{\rho}P_{h\sigma}}{\sqrt{(P_h\cdot P)^{2}-M_h^{2}M^{2}}}.\label{e.epsilont}
\end{align}

We consider the SIDIS process in the kinematic region $\bm{q}_T^2\ll Q^2$, where $\bm{q}_T$ is the transverse momentum of the virtual photon. At the leading order, we apply the TMD factorization in which the hadronic tensor can be expressed as the convolution of the parton distribution correlator $\Phi$ and the fragmentation correlator $\Delta$,
\begin{align}
	W^{\mu\nu}=2z\sum_{a}e_{a}^{2}\int d^{2}\boldsymbol{k}_{T}\int d^{2}\boldsymbol{p}_{T}\delta^{2}\left(\boldsymbol{k}_{T}+\boldsymbol{q}_{T}-\boldsymbol{p}_{T}\right){\rm Tr}\left[\Phi\left(x,k_{T}\right)\gamma^{\mu}\Delta\left(z,p_{T}\right)\gamma^{\nu}\right],
	\label{e.htpartonmodel}
\end{align}
where $\bm{k}_T$ and $\bm{p}_T$ are the momenta of the quark from the target and the quark that fragment to the final-state hadron, respectively. The sum runs over all active flavors of quark and antiquark $a$ with $e_a$ being the fractional charge of $a$.  
The light-cone longitudinal momentum fractions are defined by $x=k^+/P^+$ and $z=p^-/P_h^-$. The TMD distribution correlation function is defined by
\begin{align}
	\Phi(x,k_T)&=\int\frac{d\xi^- d^2\bm{\xi}_T}{(2\pi)^3} e^{ik\cdot \xi} \langle P| \bar{\psi}(0) \mathcal{L}(0,\xi)\psi(\xi)|P\rangle\bigg|_{\xi^+=0},\label{e.pdfcorrelator}
\end{align} 
and the fragmentation correlation function is defined by
\begin{align}
	\Delta(z,p_T)&=\frac{1}{2z}\sum_{X}\int\frac{d\eta^+ d^2\bm{\eta}_T}{(2\pi)^3}e^{ip\cdot \eta} \langle 0 |\mathcal{L}(\infty,\eta)\psi(\eta)|P_h,X \rangle \langle P_h,X|\bar{\psi}(0)\mathcal{L}^\dagger(\infty,0)|0 \rangle\bigg|_{\eta^-=0},\label{e.ffcorrelator}
\end{align}
where the symbol $\displaystyle{\sum_{X}}$ indicates the integration over the momenta of the undetected hadrons labeled by $X$. The gauge link, ensuring the gauge invariance of correlators, is defined by a path integral 
\begin{align}
	\mathcal{L}(y_2,y_1)=\mathcal{P}\exp{\left[-ig\int_{y_1}^{y_2}dy\cdot A(y)\right]},
\end{align}
where $g$ is the coupling constant in the strong interaction and $A(y)$ is the gluon field operator. 

The correlation functions are $4\times4$ matrices in the Dirac space, which can be parametrized by Dirac $\gamma$-matrices. Since $P^+$ is chosen as the large momentum component of the target, the leading-twist TMD PDFs can be projected out from the correlation function $\Phi(x,k_T)$ by the Dirac matrices $\gamma^+$, $\gamma^+\gamma_5$, and $i\sigma^{i+}\gamma_5$. Similarly, the large momentum component of the produced hadron $P_h^-$ is opposite to the target momentum, and thus one can pick out the leading-twist TMD FFs using the Dirac structures $\gamma^-$, $\gamma^-\gamma_5$, and $i\sigma^{i-}\gamma_5$. Therefore, the complete parametrization of the correlators $\Phi$ and $\Delta$ can be written in concise forms,
\begin{align}
	\Phi(x,k_{T})=&\frac{1}{2}\left\{ \Phi^{\left[\gamma^{+}\right]}(x,k_{T})\gamma^{-}-\Phi^{\left[\gamma^{+}\gamma_{5}\right]}(x,k_{T})\gamma^{-}\gamma_{5}+\Phi^{\left[i\sigma^{i+}\gamma_{5}\right]}(x,k_{T})i\sigma_{+i}\gamma_{5}\right\} ,\label{e.pdf}\\ 
	\Delta(z,p_{T})=&\frac{1}{2}\left\{ \Delta^{\left[\gamma^{-}\right]}(z,p_{T})\gamma^{+}-\Delta^{\left[\gamma^{-}\gamma_{5}\right]}(z,p_{T})\gamma^{+}\gamma_{5}+\Delta^{\left[i\sigma^{j-}\gamma_{5}\right]}(z,p_{T})i\sigma_{-j}\gamma_{5}\right\},\label{e.ff}
\end{align}
where the projections are given by
\begin{align}
	\Phi^{[\Gamma]}\left(x,k_{T}\right) & =\frac{1}{2}{\rm Tr}\left[\Phi\left(x,k_{T}\right)\Gamma\right],\\
	\Delta^{[\Gamma]}\left(z,p_{T}\right) & =\frac{1}{2}{\rm Tr}\left[\Delta\left(z,p_{T}\right)\Gamma\right].\label{e.DeltaGamma}
\end{align}
At the leading twist, the correlation function $\Phi(x,k_T)$ can be parametrized as eight quark TMD PDFs for a spin-1/2 target~\cite{Mulders:1995dh},
\begin{align}
	\Phi^{[\gamma^{+}]}& =f_1-\frac{\epsilon_T^{\rho\sigma}k_{T\rho}S_{T\sigma}}{M}f_{1T}^\perp,  \\
	\Phi^{[\gamma^+\gamma_5]}& =S_Lg_{1L}-\frac{k_T\cdot S_T}{M}g_{1T},  \\
	\Phi^{[i\sigma^{i+}\gamma_5]}& =S_T^i h_{1T} + S_L \frac{k_T^i}{M} h_{1L}^\perp -\frac{k_T^{ij}}{M^2}S_{Tj}h_{1T}^\perp-\frac{\epsilon_T^{ij} k_{Tj}}{M}h_1^\perp.
\end{align} 
A complete set of leading-twist quark TMD FFs for spin-3/2 hadrons has been defined according to the decomposition of the quark-quark correlation function in Ref.~\cite{Zhao:2022lbw}. The parametrization of $\Delta^{[\Gamma]}(z,p_T)$'s in terms of TMD FFs are given by 
\begin{align}
	\Delta^{\left[\gamma^{-}\right]}
	=&D_{1}+\Big(\epsilon_{T}^{\mu \nu} S_{h T \nu} \frac{p_{T \mu}}{M_h} D_{1 T}^{\perp}\Big)+S_{h L L} D_{1 L L}+\frac{\boldsymbol{S}_{h L T} \cdot \bm{p}_T}{M_h} D_{1 L T}^\perp+S_{hTT\mu\nu}\frac{p_T^{\mu\nu}}{M_h^{2}} D_{1 T T}^\perp\nonumber\\
	&+\Big(\epsilon_{T}^{\mu \nu} S_{h L L T \nu} \frac{p_{T \mu}}{M_h} D_{1 L L T}^\perp\Big)+\Big(\epsilon_{T\nu}^{\mu} S_{h L T T}^{ \nu \rho} \frac{p_{T\mu\rho}}{M_h^{2}} D_{1 L T T}^\perp\Big)+\Big(\epsilon_{T\nu}^\mu S_{hTTT}^{\nu\rho\sigma} \frac{p_{T\mu\rho\sigma}}{M_h^3} D_{1T T T}^\perp\Big),\\
    \Delta^{\left[\gamma^{-} \gamma_{5}\right]}
	=&S_{h L} G_{1 L}+\frac{\boldsymbol{S}_{h T} \cdot \boldsymbol{p}_{T}}{M_h} G_{1 T}^\perp
	-\Big(\epsilon_{T}^{\mu \nu} S_{h L T \nu} \frac{p_{T \mu}}{M_h} G_{1 L T}^\perp\Big)+\Big(\epsilon_{T\nu}^{\mu} S_{h T T}^{\nu \rho} \frac{p_{T\mu\rho} }{M_h^{2}} G_{1 T T}^\perp\Big)\nonumber\\
	&+S_{h L L L}G_{1 L L L}+\frac{\boldsymbol{S}_{h L L T} \cdot \boldsymbol{p}_{T}}{M_h} G_{1 L L T}^\perp+S_{hLTT}^{\mu\nu}\frac{p_{T\mu\nu}}{M_h^2}G_{1LTT}^\perp
	+S_{h T T T \mu\nu\rho} \frac{p_{T}^{\mu\nu\rho}}{M_h^3}G_{1T T T}^\perp,\\
	\Delta^{\left[i \sigma^{i-} \gamma_{5}\right]}
	=&\Big(\frac{\epsilon_{T}^{i j} p_{T j}}{M_h} H_{1}^{\perp}\Big)+S_{h L} \frac{p_{T}^{i}}{M_h} H_{1 L}^{\perp}+S_{h T}^{i} H_{1 T}- \frac{p_{T}^{ij}}{M_h^2} S_{hTj} H_{1 T}^{\perp}\nonumber\\
    &-\Big(S_{h L L} \frac{\epsilon_{T}^{i j} p_{T j}}{M_h} H_{1 L L}^{\perp}\Big)-\Big(\epsilon_{T}^{i j} S_{h L T j} H_{1 L T} \Big) + \Big( \epsilon_{Tj}^{i} \frac{p_{T}^{jl}}{M_h^2} S_{hLTl} H_{1 L T}^{\perp}\Big) \nonumber\\
	&-\left(\epsilon_{T}^{i j} S_{h L T j} H_{1 L T} \right) + \Big( \epsilon_{Tj}^{i} \frac{p_{T}^{jl}}{M_h^2} S_{hLTl} H_{1 L T}^{\perp}\Big)\nonumber\\
	&+S_{h L L L} \frac{ p_{T}^i}{M_h} H_{1 L L L}^{\perp}+S_{h L L T}^i H_{1L L T} -  \frac{ p_{T}^{ij}}{M_h^2} S_{hLLTj}H_{1 L L T}^{\perp} +S_{h L T T}^{i j} \frac{p_{T j}}{M_h} H_{1 L T T}^{  ^\perp}\nonumber\\
	&+ \frac{p_{T}^{ijl}}{M_h^3} S_{hLTTjl} H_{1 L T T}^{\perp\perp}+S_{hT T T}^{i j l} \frac{p_{T jl}}{M_h^2} H_{1 T T T}^{ ^\perp} + \frac{ p_{T}^{ijlm}}{M_h^4} S_{hTTT jlm}  H_{1 T T T}^{\perp\perp},
\end{align}
where $k_T^{\mu\nu}$, $k_T^{\mu\nu\rho }$, and $k_T^{\mu\nu\rho\sigma}$ are the completely symmetric and traceless tensors~\cite{Boer:2016xqr} as defined in Eqs.~\eqref{e.kij}--\eqref{e.kijkl}. At the leading twist, there are 32 TMD FFs for spin-3/2 hadrons, and 14 of them are for the rank-3 tensor polarized hadron states. Substituting the correlators $\Phi$ and $\Delta$ in Eq.~\eqref{e.htpartonmodel} with the expressions above, 
one can obtain the explicit expression of the hadronic tensor in terms of TMD PDFs and FFs. We note that the helicity conservation requires the chiral-odd term in TMD PDFs, which correspond to the $\Phi^{\left[i\sigma^{i+}\gamma_5\right]}$ term, only couples with the chiral-odd term in TMD FFs, which correspond to the $\Delta^{\left[i\sigma^{j-}\gamma_{5}\right]}$ term. Similarly the chiral-even terms that related to $\Phi^{\left[\gamma^{+}\right]}$ and $\Phi^{\left[\gamma^{+}\gamma_5\right]}$ in TMD PDFs only couple with the chiral-even terms that related to $\Delta^{\left[\gamma^{-}\right]}$ and $\Delta^{\left[\gamma^{-}\gamma_5\right]}$ in TMD FFs. 
As a result, there are three symmetric terms,
\begin{align}
	& {\rm Tr}\left[\Phi^{\left[\gamma^{+}\right]}(x,k_{T})\gamma^{-}\gamma^{\mu}\Delta^{\left[\gamma^{-}\right]}(z,p_{T})\gamma^{+}\gamma^{\nu}\right],\nonumber\\
	& {\rm Tr}\left[\Phi^{\left[\gamma^{+}\gamma_{5}\right]}(x,k_{T})\gamma^{-}\gamma_{5}\gamma^{\mu}\Delta^{\left[\gamma^{-}\gamma_{5}\right]}(z,p_{T})\gamma^{+}\gamma_{5}\gamma^{\nu}\right],\nonumber\\
	& {\rm Tr}\left[\Phi^{\left[i\sigma^{i+}\gamma_{5}\right]}(x,k_{T})i\sigma_{+i}\gamma_{5}\gamma^{\mu}\Delta^{\left[i\sigma^{j-}\gamma_{5}\right]}(z,p_{T})i\sigma_{-j}\gamma_{5}\gamma^{\nu}\right],
\end{align}
and two antisymmetric terms  
\begin{align}
	& {\rm -Tr}\left[\Phi^{\left[\gamma^{-}\right]}(x,k_{T})\gamma^{+}\gamma^{\mu}\Delta^{\left[\gamma^{+}\gamma_{5}\right]}(z,p_{T})\gamma^{-}\gamma_{5}\gamma^{\nu}\right],\nonumber\\
	& -{\rm Tr}\left[\Phi^{\left[\gamma^{-}\gamma_{5}\right]}(x,k_{T})\gamma^{+}\gamma_{5}\gamma^{\mu}\Delta^{\left[\gamma^{+}\right]}(z,p_{T})\gamma^{-}\gamma^{\nu}\right],
\end{align}
in the hadronic tensor. The complete expressions of the hadronic tensor in terms of the leading-twist TMD PDFs and FFs are given in Appendix~\ref{a.hadrontensor}.

Contracting the hadronic tensor obtained above with the leptonic tensor, one can express the cross section in terms of TMD PDFs and FFs in the parton model. We compare this expression with the cross section in terms of structure functions, and we obtain the one-to-one relations between the structure functions and the convolution of PDFs and FFs. We list the nonzero structure functions in the parton model at the leading twist below, but before that we introduce a transverse momentum convolution notation for the sake of legibility,
\begin{align}
	\mathcal{C}[w f D]=x \sum_a e_a^2 \int d^2 \boldsymbol{k}_T d^2 \boldsymbol{p}_T \delta^{(2)}\left(\boldsymbol{k}_T-\boldsymbol{p}_T-\boldsymbol{P}_{h \perp} / z\right) w\left(\boldsymbol{k}_T, \boldsymbol{p}_T\right) f^a\left(x, k_T^2\right) D^a\left(z, p_T^2\right),
\end{align}
where $f^a$ is a TMD PDF for the target and $D^a$ is a TMD FF for the produced hadron. For concise expressions, we define the following dimensionless scalar functions $w\left(k_T, p_T\right)$:
\begin{align}
	w_{1} & =\frac{\hat{h}\cdot k_{T}}{M},\quad
	\bar{w}_{1}  =\frac{\hat{h}\cdot p_{T}}{M_{h}},\quad
	w_{2}=\frac{2\big(\hat{h}\cdot k_{T}\big)\big(\hat{h}\cdot p_{T}\big)+\left(k_{T}\cdot p_{T}\right)}{MM_{h}},\quad
	w_{2}^{\prime} =\frac{k_{T}\cdot p_{T}}{MM_{h}},\nonumber\\
	w_{3} & =\frac{2k_{T}^{ij}\hat{h}_{i}\hat{h}_{j}\big(\hat{h}\cdot p_{T}\big)+k_{T}^{ij}\hat{h}_{i}p_{Tj}}{M^{2}M_{h}},\quad
    {w}_{3}^\prime =-\frac{k_{T}^{ij}\hat{h}_{i}p_{Tj}}{M^2M_{h}},\nonumber\\
	\bar{w}_{3}& =\frac{2p_{T}^{ij}\hat{h}_{i}\hat{h}_{j}\big(\hat{h}\cdot k_{T}\big)+p_{T}^{ij}\hat{h}_{i}k_{Tj}}{MM_{h}^{2}},\quad
    \bar{w}_{3}^\prime  =-\frac{p_{T}^{ij}\hat{h}_{i}k_{Tj}}{MM_{h}^{2}},\nonumber\\
	w_{4} & =\frac{k_{T}^{ij}\hat{h}_{i}\hat{h}_{j}}{M^{2}},\quad
	\bar{w}_{4} =\frac{p_{T}^{ij}\hat{h}_{i}\hat{h}_{j}}{M_{h}^{2}},\quad
	w_{5} =-\frac{2k_{T}^{ij}\hat{h}_{i}\hat{h}_{j}p_{T}^{lm}\hat{h}_{l}\hat{h}_{m}+k_{T}^{ij}\hat{h}_{i}p_{Tjl}\hat{h}^{l}}{M^{2}M_{h}^{2}},\quad
	w_{5}^{\prime} =\frac{2k_{T}^{ij}\hat{h}_{i}p_{Tjl}\hat{h}^{l}}{M^{2}M_{h}^{2}},\nonumber\\
	w_{6}&=-\frac{2\left(2p_{T}^{ijl}\hat{h}_{i}\hat{h}_{j}\hat{h}_{l}\big(k_{T}\cdot\hat{h}\big)+p_{T}^{ijl}\hat{h}_{i}\hat{h}_{j}k_{Tl}\right)}{MM_{h}^{3}},\quad
	w_{6}^{\prime} =\frac{2p_{T}^{ijl}k_{Ti}\hat{h}_{j}\hat{h}_{l}}{MM_{h}^{3}},\nonumber\\
	w_{7} &=\frac{2p_{T}^{ijl}\hat{h}_{i}\hat{h}_{j}\hat{h}_{l}}{M_{h}^{3}},\quad
    w_{8}=\frac{2\left(2k_{T}^{ij}\hat{h}_{i}\hat{h}_{j}p_{T}^{lmn}\hat{h}_{l}\hat{h}_{m}\hat{h}_{n}+k_{T}^{ij}\hat{h}_{i}p_{Tjlm}\hat{h}^{l}\hat{h}^{m}\right)}{M^{2}M_{h}^{3}},\nonumber\\
	w_{9} &=\frac{2\left(2k_{T}^{i}\hat{h}_{i}p_{T}^{jlmn}\hat{h}_{j}\hat{h}_{l}\hat{h}_{m}\hat{h}_{n}+k_{T}^{i}p_{Tijlm}\hat{h}^{j}\hat{h}^{l}\hat{h}^{m}\right)}{MM_{h}^{4}},\quad
	w_{10}=\frac{4p_{T}^{ijlm}\hat{h}_{i}\hat{h}_{j}\hat{h}_{l}\hat{h}_{m}}{M_{h}^{4}},\nonumber\\
	w_{11}&=\frac{4\left(2k_{T}^{ij}\hat{h}_{i}\hat{h}_{j}p_{T}^{klmn}\hat{h}_{k}\hat{h}_{l}\hat{h}_{m}\hat{h}_{n}+k_{Tij}\hat{h}^{i}p_{T}^{jlmn}\hat{h}_{l}\hat{h}_{m}\hat{h}_{n}\right)}{M^{2}M_{h}^{4}},
\end{align}
where $\hat{h}^\mu=P_{h\perp}^\mu/|\bm{P}_{h\perp}|$ is a unit vector giving the direction of the transverse momentum.

Through a leading order calculation in the parton model, we can obtain 128 nonzero structure functions at the leading twist. Among them, 96 terms contribute to the cross section for the unpolarized lepton beam, 
\begin{align} 
	F_{U,U}^{T} & =\mathcal{C}\left[f_{1}D_{1}\right],\label{e.FUUT} \\ 
	F_{U,U}^{\cos2\phi_{h}} & =\mathcal{C}\left[-w_{2}h_{1}^{\perp}H_{1}^{\perp}\right],
\\ 
	F_{U,L}^{\sin2\phi_{h}} & =\mathcal{C}\left[w_{2}h_{1}^{\perp}H_{1L}^{\perp}\right],
\\
	F_{U,T}^{T\sin\phi_{hT}} & =\mathcal{C}\left[\bar{w}_{1}f_{1}D_{1T}^{\perp}\right],\\
	F_{U,T}^{\sin\left(2\phi_{h}+\phi_{hT}\right)} & =\mathcal{C}\left[-w_{1}h_{1}^{\perp}H_{1T}\right],\\
	F_{U,T}^{\sin\left(2\phi_{h}-\phi_{hT}\right)} & =\mathcal{C}\left[-\bar{w}_{3}h_{1}^{\perp}H_{1T}^{\perp}\right],
\\
	F_{U,LL}^{T} & =\mathcal{C}\left[f_{1}D_{1LL}\right],\\
	F_{U,LL}^{\cos2\phi_h} & =\mathcal{C}\left[w_{2}h_{1}^{\perp}H_{1LL}^{\perp}\right],
\\
	F_{U,LT}^{T\cos\phi_{hLT}} & =\mathcal{C}\left[-\bar{w}_{1}f_{1}D_{1LT}^{\perp}\right],\\
	F_{U,LT}^{\cos(2\phi_h+\phi_{hLT})} & =\mathcal{C}\left[-w_{1}h_{1}^{\perp}H_{1LT}\right],\\
	F_{U,LT}^{\cos(2\phi_h-\phi_{hLT})} & =\mathcal{C}\left[-\bar{w}_{3}h_{1}^{\perp}H_{1LT}^{\perp}\right],
\\
	F_{U,TT}^{T\cos2\phi_{hTT}} & =\mathcal{C}\left[-2\bar{w}_{4}f_{1}D_{1TT}^{\perp}\right],\\
	F_{U,TT}^{\cos(2\phi_h+2\phi_{hTT})} & =\mathcal{C}\left[-w_{2}^{\prime}h_{1}^{\perp}H_{1TT}^{\perp}\right],\\
	F_{U,TT}^{\cos(2\phi_h-2\phi_{hTT})} & =\mathcal{C}\left[w_{6}h_{1}^{\perp}H_{1TT}^{\perp\perp}\right],
\\
	F_{U,LLL}^{\sin2\phi_h} & =\mathcal{C}\left[w_{2}h_{1}^{\perp}H_{1LLL}^{\perp}\right],
\\
	F_{U,LLT}^{T\sin\phi_{hLLT}} & =\mathcal{C}\left[\bar{w}_{1}f_{1}D_{1LLT}^{\perp}\right],\\
	F_{U,LLT}^{\sin\left(2\phi_h+\phi_{hLLT}\right)} & =\mathcal{C}\left[-w_{1}h_{1}^{\perp}H_{1LLT}\right],\\
	F_{U,LLT}^{\sin\left(2\phi_h-\phi_{hLLT}\right)} & =\mathcal{C}\left[-\bar{w}_{3}h_{1}^{\perp}H_{1LLT}^{\perp}\right],
\\
	F_{U,LTT}^{T \sin 2\phi_{hLTT}} & =\mathcal{C}\left[-2\bar{w}_{4}f_{1}D_{1LTT}^{\perp}\right],\\
	F_{U,LTT}^{\sin\left(2\phi_h+2\phi_{hLTT}\right)} & =\mathcal{C}\left[-w_{2}^{\prime}h_{1}^{\perp}H_{1LTT}^{\perp}\right],\\
	F_{U,LTT}^{\sin\left(2\phi_h-2\phi_{hLTT}\right)} & =\mathcal{C}\left[w_{6}h_{1}^{\perp}H_{1LTT}^{\perp\perp}\right],
\\
	F_{U,TTT}^{T \sin 3\phi_{hTTT}} & =\mathcal{C}\left[2w_{7}f_{1}D_{1TTT}^{\perp}\right],\\
	F_{U,TTT}^{\sin\left(2\phi_h+3\phi_{hTTT}\right)} & =\mathcal{C}\left[-2\bar{w}_{3}^\prime h_{1}^{\perp}H_{1TTT}^{\perp}\right],\\
	F_{U,TTT}^{\sin\left(2\phi_h-3\phi_{hTTT}\right)} & =\mathcal{C}\left[2w_{9}h_{1}^{\perp}H_{1TTT}^{\perp\perp}\right],\\
	F_{L,U}^{\sin2\phi_{h}} & =\mathcal{C}\left[-w_{2}h_{1L}^{\perp}H_{1}^{\perp}\right],\\
	F_{L,L}^{T} & =\mathcal{C}\left[g_{1L}G_{1L}\right],\\
	F_{L,L}^{\cos2\phi_{h}} & =\mathcal{C}\left[-w_{2}h_{1L}^{\perp}H_{1L}^{\perp}\right],\\
	F_{L,T}^{T\cos\phi_{hT}} & =\mathcal{C}\left[-\bar{w}_{1}g_{1L}G_{1T}^{\perp}\right],\\
	F_{L,T}^{\cos\left(2\phi_{h}+\phi_{hT}\right)} & =\mathcal{C}\left[w_{1}h_{1L}^{\perp}H_{1T}\right],\\
	F_{L,T}^{\cos\left(2\phi_{h}-\phi_{hT}\right)} & =\mathcal{C}\left[\bar{w}_{3}h_{1L}^{\perp}H_{1T}^{\perp}\right].\\
	F_{L,LL}^{\sin2\phi_h} & =\mathcal{C}\left[w_{2}h_{1L}^\perp H_{1LL}^{\perp}\right],\\
	F_{L,LT}^{T\sin\phi_{hLT}} & =\mathcal{\mathcal{C}}\left[-\bar{w}_{1}g_{1L}G_{1LT}^{\perp}\right],\\
	F_{L,LT}^{\sin(2\phi_h+\phi_{hLT})} & =\mathcal{C}\left[-w_{1}h_{1L}^{\perp}H_{1LT}\right],\\
	F_{L,LT}^{\sin(2\phi_h-\phi_{hLT})} & =\mathcal{C}\left[-\bar{w}_{3}h_{1L}^{\perp}H_{1LT}^{\perp}\right],\\
	F_{L,TT}^{T\sin2\phi_{hTT}} & =\mathcal{C}\left[-2\bar{w}_{4}g_{1L}G_{1TT}^{\perp}\right],\\
	F_{L,TT}^{\sin(2\phi_h+2\phi_{hTT})} & =\mathcal{C}\left[-w_{2}^{\prime}h_{1L}^{\perp}H_{1TT}^{\perp}\right],\\
	F_{L,TT}^{\sin(2\phi_h-2\phi_{hTT})} & =\mathcal{C}\left[w_{6}h_{1L}^{\perp}H_{1TT}^{\perp\perp}\right].\\
	F_{L,LLL}^{T} & =\mathcal{C}\left[g_{1L}G_{1LLL}\right],\\
	F_{L,LLL}^{\cos2\phi_h} & =\mathcal{C}\left[-w_{2}h_{1L}^{\perp}H_{1LLL}^{\perp}\right],\\
	F_{L,LLT}^{T\cos\phi_{hLLT}} & =\mathcal{C}\left[-\bar{w}_{1}g_{1L}G_{1LLT}^{\perp}\right],\\
	F_{L,LLT}^{\cos\left(2\phi_h+\phi_{hLLT}\right)} & =\mathcal{C}\left[w_{1}h_{1L}^{\perp}H_{1LLT}\right],\\
	F_{L,LLT}^{\cos\left(2\phi_h-\phi_{hLLT}\right)} & =\mathcal{C}\left[\bar{w}_{3}h_{1L}^{\perp}H_{1LLT}^{\perp}\right],\\
	F_{L,LTT}^{T\cos 2\phi_{hLTT}} & =\mathcal{C}\left[-2\bar{w}_{4}g_{1L}G_{1LTT}^{\perp}\right],\\
	F_{L,LTT}^{\cos\left(2\phi_h+2\phi_{hLTT}\right)} & =\mathcal{C}\left[w_{2}^{\prime}h_{1L}^{\perp}H_{1LTT}^{\perp}\right],\\
	F_{L,LTT}^{\cos\left(2\phi_h-2\phi_{hLTT}\right)} & =\mathcal{C}\left[-w_{6}h_{1L}^{\perp}H_{1LTT}^{\perp\perp}\right],\\
	F_{L,TTT}^{T\cos 3\phi_{hTTT}} & =\mathcal{C}\left[2w_{7}g_{1L}G_{1TTT}^{\perp}\right],\\
	F_{L,TTT}^{\cos\left(2\phi_h+3\phi_{hTTT}\right)} & =\mathcal{C}\left[2\bar{w}_{3}^\prime h_{1L}^{\perp}H_{1TTT}^{\perp}\right],\\
	F_{L,TTT}^{\cos\left(2\phi_h-3\phi_{hTTT}\right)} & =\mathcal{C}\left[-2w_{9}h_{1L}^{\perp}H_{1TTT}^{\perp\perp}\right],\\
%
%
	F_{T,U}^{T\sin(\phi_{h}-\phi_{T})} & =\mathcal{C}\left[w_{1}f_{1T}^{\perp}D_{1}\right],\\
	F_{T,U}^{\sin(\phi_{h}+\phi_{T})} & =\mathcal{C}\left[\bar{w}_{1}h_{1T}H_{1}^{\perp}\right],\\
	F_{T,U}^{\sin(3\phi_{h}-\phi_{T})} & =\mathcal{C}\left[w_{3}h_{1T}^{\perp}H_{1}^{\perp}\right], \\
	F_{T,L}^{T \cos\left(\phi_{h}-\phi_{T}\right)} & =\mathcal{C}\left[-w_{1}g_{1T}G_{1L}\right],\\
	F_{T,L}^{\cos\left(\phi_{h}+\phi_{T}\right)} & =\mathcal{C}\left[\bar{w}_{1}h_{1T}H_{1L}^{\perp}\right],\\
	F_{T,L}^{\cos\left(3\phi_{h}-\phi_{T}\right)} & =\mathcal{C}\left[w_{3}h_{1T}^{\perp}H_{1L}^{\perp}\right],\\
	F_{T,T}^{T\cos\left(\phi_{h}-\phi_{hT}-\phi_{T}\right)} & =\mathcal{C}\left[\frac{w_{2}}{2}\left(f_{1T}^{\perp}D_{1T}^{\perp}+g_{1T}^{\perp}G_{1T}^{\perp}\right)\right],\\
	F_{T,T}^{T\cos\left(\phi_{h}+\phi_{hT}-\phi_{T}\right)} & =\mathcal{C}\left[\frac{w_{2}^{\prime}}{2}\left(f_{1T}^{\perp}D_{1T}^{\perp}-g_{1T}^{\perp}G_{1T}^{\perp}\right)\right],\\
	F_{T,T}^{\cos\left(\phi_h+\phi_{hT}+\phi_{T}\right)} & =\mathcal{C}\left[-h_{1T}H_{1T}\right],\\
	F_{T,T}^{\cos\left(\phi_h-\phi_{hT}+\phi_{T}\right)} & =\mathcal{C}\left[-\bar{w}_{4}h_{1T}H_{1T}^{\perp}\right],\\
	F_{T,T}^{\cos\left(3\phi_h+\phi_{hT}-\phi_{T}\right)} & =\mathcal{C}\left[-w_{4}h_{1T}^{\perp}H_{1T}\right],\\
	F_{T,T}^{\cos\left(3\phi_h-\phi_{hT}-\phi_{T}\right)} & =\mathcal{C}\left[w_{5}h_{1T}^{\perp}H_{1T}^{\perp}\right],\\
	F_{T,LL}^{T\sin(\phi_h-\phi_{T})} & =\mathcal{C}\left[w_{1}f_{1T}^{\perp}D_{1LL}\right],\\
	F_{T,LL}^{\sin(\phi_h+\phi_{T})} & =\mathcal{C}\left[-\bar{w}_{1}h_{1T}H_{1LL}^{\perp}\right],\\
	F_{T,LL}^{\sin(3\phi_h-\phi_{T})} & =\mathcal{C}\left[-w_{3}h_{1T}^{\perp}H_{1LL}^{\perp}\right],\\
	F_{T,LT}^{T\sin(\phi_h-\phi_{hLT}-\phi_{T})} & =\mathcal{C}\left[-\frac{w_{2}}{2}\left(g_{1T}^{\perp}G_{1LT}^{\perp}+f_{1T}^{\perp}D_{1LT}^{\perp}\right)\right],\\
	F_{T,LT}^{T\sin(\phi_h+\phi_{hLT}-\phi_{T})} & =\mathcal{C}\left[-\frac{w_{2}^{\prime}}{2}\left(g_{1T}^{\perp}G_{1LT}^{\perp}-f_{1T}^{\perp}D_{1LT}^{\perp}\right)\right],\\
	F_{T,LT}^{\sin(\phi_h+\phi_{hLT}+\phi_{T})} & =\mathcal{C}\left[h_{1T}H_{1LT}\right],\\
	F_{T,LT}^{\sin(\phi_h-\phi_{hLT}+\phi_{T})} & =\mathcal{C}\left[\bar{w}_{4}h_{1T}H_{1LT}^{\perp}\right],\\
	F_{T,LT}^{\sin(3\phi_h+\phi_{hLT}-\phi_{T})} & =\mathcal{C}\left[w_{4}h_{1T}^{\perp}H_{1LT}\right],\\
	F_{T,LT}^{\sin(3\phi_h-\phi_{hLT}-\phi_{T})} & =\mathcal{C}\left[-w_{5}h_{1T}^{\perp}H_{1LT}^{\perp}\right],\\
	F_{T,TT}^{T\sin(\phi_h-2\phi_{hTT}-\phi_{T})} & =\mathcal{C}\left[-\bar{w}_{3}\left(g_{1T}^{\perp}G_{1TT}^{\perp}+f_{1T}^{\perp}D_{1TT}^{\perp}\right)\right],\\
	F_{T,TT}^{T\sin(\phi_h+2\phi_{hTT}-\phi_{T})} & =\mathcal{C}\left[\bar{w}_{3}^\prime\left(g_{1T}^{\perp}G_{1TT}^{\perp}-f_{1T}^{\perp}D_{1TT}^{\perp}\right)\right],\\
	F_{T,TT}^{\sin(\phi_h+2\phi_{hTT}+\phi_{T})} & =\mathcal{C}\left[-\bar{w}_{1}h_{1T}H_{1TT}^{\perp}\right],\\
	F_{T,TT}^{\sin(\phi_h-2\phi_{hTT}+\phi_{T})} & =\mathcal{C}\left[w_{7}h_{1T}H_{1TT}^{\perp\perp}\right],\\
	F_{T,TT}^{\sin(3\phi_h+2\phi_{hTT}-\phi_{T})} & =\mathcal{C}\left[-{w}_{3}^\prime h_{1T}^{\perp}H_{1TT}^{\perp}\right],\\
	F_{T,TT}^{\sin(3\phi_h-2\phi_{hTT}-\phi_{T})} & =\mathcal{C}\left[w_{8}h_{1T}^{\perp}H_{1TT}^{\perp\perp}\right],\\
	F_{T,LLL}^{T\cos\left(\phi_h-\phi_{T}\right)} & =\mathcal{C}\left[-w_{1}g_{1T}G_{1LLL}\right],\\
	F_{T,LLL}^{\cos\left(\phi_h+\phi_{T}\right)} & =\mathcal{C}\left[\bar{w}_{1}h_{1T}H_{1LLL}^{\perp}\right],\\
	F_{T,LLL}^{\cos\left(3\phi_h-\phi_{T}\right)} & =\mathcal{C}\left[w_{3}h_{1T}^{\perp}H_{1LLL}^{\perp}\right],\\
	F_{T,LLT}^{T\cos\left(\phi_h-\phi_{hLLT}-\phi_{T}\right)} & =\mathcal{C}\left[\frac{w_{2}}{2}\left(f_{1T}^{\perp}D_{1LLT}^{\perp}+g_{1T}G_{1LLT}^{\perp}\right)\right],\\
	F_{T,LLT}^{T\cos\left(\phi_h+\phi_{hLLT}-\phi_{T}\right)} & =\mathcal{C}\left[\frac{w_{2}^{\prime}}{2}\left(f_{1T}^{\perp}D_{1LLT}^{\perp}-g_{1T}G_{1LLT}^{\perp}\right)\right],\\
	F_{T,LLT}^{\cos\left(\phi_h+\phi_{hLLT}+\phi_{T}\right)} & =\mathcal{C}\left[-h_{1T}H_{1LLT}\right],\\
	F_{T,LLT}^{\cos\left(\phi_h-\phi_{hLLT}+\phi_{T}\right)} & =\mathcal{C}\left[-\bar{w}_{4}h_{1T}H_{1LLT}^{\perp}\right],\\
	F_{T,LLT}^{\cos\left(3\phi_h+\phi_{hLLT}-\phi_{T}\right)} & =\mathcal{C}\left[-w_{4}h_{1T}^{\perp}H_{1LLT}\right],\\
	F_{T,LLT}^{\cos\left(3\phi_h-\phi_{hLLT}-\phi_{T}\right)} & =\mathcal{C}\left[w_{5}h_{1T}^{\perp}H_{1LLT}^{\perp}\right],\\
	F_{T,LTT}^{T\cos\left(\phi_h-2\phi_{hLTT}-\phi_{T}\right)} & =\mathcal{C}\left[-\bar{w}_{3}\left(f_{1T}^{\perp}D_{1LTT}^{\perp}-g_{1T}G_{1LTT}^{\perp}\right)\right],\\
	F_{T,LTT}^{T\cos\left(\phi_h+2\phi_{hLTT}-\phi_{T}\right)} & =\mathcal{C}\left[\bar{w}_{3}^\prime\left(f_{1T}^{\perp}D_{1LTT}^{\perp}+g_{1T}G_{1LTT}^{\perp}\right)\right],\\
	F_{T,LTT}^{\cos\left(\phi_h+2\phi_{hLTT}+\phi_{T}\right)} & =\mathcal{C}\left[\bar{w}_{1}h_{1T}H_{1LTT}^{\perp}\right],\\
	F_{T,LTT}^{\cos\left(\phi_h-2\phi_{hLTT}+\phi_{T}\right)} & =\mathcal{C}\left[-w_{7}h_{1T}H_{1LTT}^{\perp\perp}\right],\\
	F_{T,LTT}^{\cos\left(3\phi_h+2\phi_{hLTT}-\phi_{T}\right)} & =\mathcal{C}\left[{w}_{3}^\prime h_{1T}^\perp H_{1LTT}^{\perp}\right],\\
	F_{T,LTT}^{\cos\left(3\phi_h-2\phi_{hLTT}-\phi_{T}\right)} & =\mathcal{C}\left[-w_{8}h_{1T}^{\perp}H_{1LTT}^{\perp\perp}\right],\\
	F_{T,TTT}^{T\cos\left(\phi_h-3\phi_{hTTT}-\phi_{T}\right)} & =\mathcal{C}\left[-w_{6}\left(f_{1T}^{\perp}D_{1TTT}^{\perp}-g_{1T}G_{1TTT}^{\perp}\right)\right],\\
	F_{T,TTT}^{T\cos\left(\phi_h+3\phi_{hTTT}-\phi_{T}\right)} & =\mathcal{C}\left[w_{6}^{\prime}\left(f_{1T}^{\perp}D_{1TTT}^{\perp}+g_{1T}G_{1TTT}^{\perp}\right)\right],\\
	F_{T,TTT}^{\cos\left(\phi_h+3\phi_{hTTT}+\phi_{T}\right)} & =\mathcal{C}\left[-2\bar{w}_{4}h_{1T}H_{1TTT}^{\perp}\right],\\
	F_{T,TTT}^{\cos\left(\phi_h-3\phi_{hTTT}+\phi_{T}\right)} & =\mathcal{C}\left[w_{10}h_{1T}H_{1TTT}^{\perp\perp}\right],\\
	F_{T,TTT}^{\cos\left(3\phi_h+3\phi_{hTTT}-\phi_{T}\right)} & =\mathcal{C}\left[w_{5}^{\prime}h_{1T}^{\perp}H_{1TTT}^{\perp}\right],\\
	F_{T,TTT}^{\cos\left(3\phi_h-3\phi_{hTTT}-\phi_{T}\right)} & =\mathcal{C}\left[w_{11}h_{1T}^{\perp}H_{1TTT}^{\perp\perp}\right],
\end{align}
and 32 terms contribute to the cross section for the polarized lepton beam,
\begin{align}
	G_{U,L} & =\mathcal{C}\left[f_{1}G_{1L}\right],\\
	G_{U,T}^{\cos\phi_{hT}} & =\mathcal{C}\left[-\bar{w}_{1}f_{1}G_{1T}^{\perp}\right],
\\
	G_{U,LT}^{\sin\phi_{hLT}} & =\mathcal{C}\left[-\bar{w}_{1}f_{1}G_{1LT}^{\perp}\right],\\
	G_{U,TT}^{\sin2\phi_{hTT}} & =\mathcal{C}\left[-2\bar{w}_{4}f_{1}G_{1TT}^{\perp}\right],
\\
	G_{U,LLL} & =\mathcal{C}\left[f_{1}G_{1LLL}\right],\label{e.GULLL}\\
	G_{U,LLT}^{\cos\phi_{hLLT}} & =\mathcal{C}\left[-\bar{w}_{1}f_{1}G_{1LLT}^{\perp}\right],\\
	G_{U,LTT}^{\cos 2\phi_{hLTT}} & =\mathcal{C}\left[-2\bar{w}_{4}f_{1}G_{1LTT}^{\perp}\right],\\ \label{e.GTTTc3}
	G_{U,TTT}^{\cos 3\phi_{hTTT}} & =\mathcal{C}\left[2w_{7}f_{1}G_{1TTT}^{\perp}\right],\\
      G_{L,U}&=\mathcal{C}\left[g_{1L}D_{1}\right],\\
	  G_{L,T}^{\sin\phi_{hT}} & =\mathcal{C}\left[\bar{w}_{1}g_{1L}D_{1T}^{\perp}\right],\\
	G_{L,LL} & =\mathcal{C}\left[g_{1L}D_{1LL}\right],\\
	G_{L,LT}^{\cos\phi_{hLT}} & =\mathcal{C}\left[-\bar{w}_{1}g_{1L}D_{1LT}^{\perp}\right],\\
	G_{L,TT}^{\cos 2\phi_{hTT}} & =\mathcal{C}\left[-2\bar{w}_{4}g_{1L}D_{1TT}^{\perp}\right],\\
	G_{L,LLT}^{\sin\phi_{hLLT}} & =\mathcal{C}\left[\bar{w}_{1}g_{1L}D_{1LLT}^{\perp}\right],\\
	G_{L,LTT}^{\sin 2\phi_{hLTT}} & =\mathcal{C}\left[-2\bar{w}_{4}g_{1L}D_{1LTT}^{\perp}\right],\\
	G_{L,TTT}^{\sin 3\phi_{hTTT}} & =\mathcal{C}\left[2w_{7}g_{1L}D_{1TTT}^{\perp}\right],\\
    G_{T,U}^{\cos\left(\phi_{h}-\phi_{T}\right)}&=\mathcal{C}\left[-w_{1}g_{1T}^{\perp}D_{1}\right],\\
    G_{T,L}^{\sin\left(\phi_{h}-\phi_{T}\right)} & =\mathcal{C}\left[w_{1}f_{1T}^{\perp}G_{1L}\right],\\
    G_{T,T}^{\sin\left(\phi_{h}-\phi_{hT}-\phi_{T}\right)} & =\mathcal{C}\left[-\frac{w_{2}}{2}\left(f_{1T}^{\perp}G_{1T}^{\perp}-g_{1T}^{\perp}D_{1T}^{\perp}\right)\right],\\
	G_{T,T}^{\sin\left(\phi_{h}+\phi_{hT}-\phi_{T}\right)} & =\mathcal{C}\left[\frac{w_{2}^{\prime}}{2}\left(f_{1T}^{\perp}G_{1T}^{\perp}+g_{1T}^{\perp}D_{1T}^{\perp}\right)\right],\\
	G_{T,LL}^{\cos(\phi_h-\phi_{T})} & =\mathcal{C}\left[-w_{1}g_{1T}D_{1LL}\right],\\
	G_{T,LT}^{\cos(\phi_h-\phi_{hLT}-\phi_{T})} & =\mathcal{C}\left[\frac{w_{2}}{2}\left(g_{1T}D_{1LT}^{\perp}-f_{1T}^{\perp}G_{1LT}^{\perp}\right)\right],\\
	G_{T,LT}^{\cos(\phi_h+\phi_{hLT}-\phi_{T})} & =\mathcal{C}\left[-\frac{w_{2}^{\prime}}{2}\left(g_{1T}D_{1LT}^{\perp}+f_{1T}^{\perp}G_{1LT}^{\perp}\right)\right],\\
	G_{T,TT}^{\cos(\phi_h-2\phi_{hTT}-\phi_{T})} & =\mathcal{C}\left[\bar{w}_{3}\left(g_{1T}D_{1TT}^{\perp}-f_{1T}^{\perp}G_{1TT}^{\perp}\right)\right],\\
	G_{T,TT}^{\cos(\phi_h+2\phi_{hTT}-\phi_{T})} & =\mathcal{C}\left[\bar{w}_{3}^\prime\left(g_{1T}D_{1TT}^{\perp}+f_{1T}^{\perp}G_{1TT}^{\perp}\right)\right],\\
	G_{T,LLL}^{\sin\left(\phi_h-\phi_{T}\right)} & =\mathcal{C}\left[w_{1}f_{1T}^{\perp}G_{1LLL}\right],\\
	G_{T,LLT}^{\sin\left(\phi_h-\phi_{hLLT}-\phi_{T}\right)} & =\mathcal{C}\left[\frac{w_{2}}{2}\left(g_{1T}D_{1LLT}^{\perp}-f_{1T}^{\perp}G_{1LLT}^{\perp}\right)\right],\\
	G_{T,LLT}^{\sin\left(\phi_h+\phi_{hLLT}-\phi_{T}\right)} & =\mathcal{C}\left[\frac{w_{2}^{\prime}}{2}\left(g_{1T}D_{1LLT}^{\perp}+f_{1T}^{\perp}G_{1LLT}^{\perp}\right)\right],\\
	G_{T,LTT}^{\sin\left(\phi_h-2\phi_{hLTT}-\phi_{T}\right)} & =\mathcal{C}\left[-\bar{w}_{3}\left(g_{1T}D_{1LTT}^{\perp}+f_{1T}^{\perp}G_{1LTT}^{\perp}\right)\right],\\
	G_{T,LTT}^{\sin\left(\phi_h+2\phi_{hLTT}-\phi_{T}\right)} & =\mathcal{C}\left[\bar{w}_{3}^\prime\left(g_{1T}D_{1LTT}^{\perp}-f_{1T}^{\perp}G_{1LTT}^{\perp}\right)\right],\\
	G_{T,TTT}^{\sin\left(\phi_h-3\phi_{hTTT}-\phi_{T}\right)} & =\mathcal{C}\left[-w_{6}\left(g_{1T}D_{1TTT}^{\perp}+f_{1T}^{\perp}G_{1TTT}^{\perp}\right)\right],\\
	G_{T,TTT}^{\sin\left(\phi_h+3\phi_{hTTT}-\phi_{T}\right)} & =\mathcal{C}\left[w_{6}^{\prime}\left(g_{1T}D_{1TTT}^{\perp}-f_{1T}^{\perp}G_{1TTT}^{\perp}\right)\right].
\end{align}
 We note that half of the 192 structure functions for the unpolarized lepton are nonvanishing at the leading twist, and 42 of them are for rank-3 tensor polarized hadron states. In the case of polarized leptons, one third of the 96 structure functions are nonvanishing, with 14 of them corresponding to rank-3 tensor polarized hadron states. We can in principle use these nonvanishing rank-3 tensor polarized structure functions to study the leading-twist spin-3/2 hadrons TMD FFs for rank-3 tensor polarized states. The other vanishing structure functions are the high twist or high order contributions. We also find that all $S_{hLLL}$-dependent structure functions exhibit the same expressions as those depending on $S_{hL}$ in the parton model, and this similarity can be extended to $S_{hT}$-dependent and $S_{hLLT}$-dependent structure functions. The reason is that both the polarization vector $S_h^\mu$ and the rank-3 polarization tensor $R_h^{\mu\nu\rho}$ are parity violated. 

\section{The numerical calculation in the spectator diquark model}
\label{s.numericalresult}

The semi-inclusive production of spin-3/2 hadrons in DIS provides additional experimental observables to study the nucleon spin structures. The observables related to the rank-3 tensor polarization of the hadron can also shed light to study the role of spin in the hadronization process. In this section, we take the spin transfer to a $S_{hLLL}$ polarized hadron as an example and estimate it in the spectator model.

\subsection{Spin-3/2 particle FFs in the spectator diquark model}

The fundamental concept of the spectator model is treating the intermediate states that can be incorporated into the definition of the correlation function $\Phi$ in Eq.~\eqref{e.pdfcorrelator} or $\Delta$ in Eq.~\eqref{e.ffcorrelator}, as a pointlike particle with the diquark quantum numbers. This model has been applied to calculate the quark TMD PDFs~\cite{Jakob:1997wg,Brodsky:2002cx,Gamberg:2003ey,Bacchetta:2003rz,Bacchetta:2008af,Bacchetta:2010si,Lu:2004hu,Meissner:2008ay,Ma:2019agv} and  FFs~\cite{Bacchetta:2007wc,Lu:2015wja,Yang:2017cwi}. In the naive picture of the quark structure of the nucleon, the diquark system can be a spin-0 particle (scalar diquark) or a spin-1 particle (axial vector diquark). Since the quark fragmentation can be modeled as $s\rightarrow \Omega(sss) + a(\bar{s}\bar{s})$, where $a$ denotes an axial vector diquark, we only consider the spin-1 diquark in the following.

The matrix element in Eq.~\eqref{e.ffcorrelator}, describing the intermediate unobserved states and the $\Omega$ baryon in the final state, can be written as
\begin{align}
    \left\langle P_{h}, X|\bar{\psi}(0)| 0\right\rangle=\bar{U}^\alpha\left(P_{h} \right) \Upsilon_{\alpha}^{a \mu } \frac{i}{\slashed{p}-m} \varepsilon_\mu(P_h-p,\lambda_a),
\end{align}
where $U^\alpha(P_h)$ is the spinor for a spin-3/2 hadron with momentum $P_h$, $\Upsilon_{\alpha}^{a\mu}$ is the hadron-quark-diquark vertex yet to be specific, and $i/(\slashed{p}-m)$ is a quark propagator for the untruncated quark line. The spin-3/2 hadron spinors have been introduced in Refs.~\cite{Christensen:2013aua,Huang:2003ym,Haberzettl:1998rw}, which can be expressed as the direct product of a basic spin-1/2 spinor and a spin-1 polarization vector. The explicit forms of the spin-3/2 spinor are given by
\begin{align}
   U_{-3 / 2}^\alpha(P_h)&=\varepsilon_{-}^\alpha(P_h) u_{-1/2 }(P_h),\label{e.U-3/2}\\
	U_{-1/ 2}^\alpha(P_h)&=\sqrt{\frac{2}{3}}\varepsilon_{0}^\alpha(P_h) u_{-1/2 }(P_h) + \sqrt{\frac{1}{3}}\varepsilon_{-}^\alpha(P_h) u_{1/2 }(P_h),\\
	U_{1/ 2}^\alpha(P_h)&=\sqrt{\frac{1}{3}}\varepsilon_{+}^\alpha(P_h) u_{-1/2 }(P_h) + \sqrt{\frac{2}{3}}\varepsilon_{0}^\alpha(P_h) u_{1/2 }(P_h),\\
	U_{3/ 2}^\alpha(P_h)&=\varepsilon_{+}^\alpha(P_h) u_{1/2 }(P_h),\label{e.U3/2} 
\end{align}
which satisfy the spin sum rules
\begin{align}
	\sum_{m=-3/2}^{m=3/2} U_{m}^\mu \bar{U}_{m}^\nu
		= & -(\slashed{P}_h+M_h) \Bigl[g^{\mu \nu}-\frac{2P_h^\mu P_h^\nu}{3 M_h^2}-\frac{1}{3} \gamma^\mu \gamma^\nu
		-\frac{1}{3 M_h}\left(P_h^\nu \gamma^\mu-P_h^\mu \gamma^\nu\right)\Bigl],\label{e.spinorsum}
\end{align}
where the subscript $m$ represents the magnetic quantum number. $\varepsilon_\mu(P_h-p,\lambda_a)$ is the polarization vector of a spin-1 axial vector diquark with momentum $P_h-p$ and helicity state $\lambda_a$. Various choices for the summation over all polarized states have been employed  in Refs.~\cite{Gamberg:2007wm,Jakob:1997wg,Goldstein:2002vv,Brodsky:2000ii}, and in this calculation, we choose
\begin{align}
		\sum_{\lambda_a} \varepsilon_\mu^\ast(P_h-p , \lambda_a) \varepsilon_\nu(P_h-p , \lambda_a) =-g_{\mu \nu}+\frac{(P_h-p)_\mu(P_h-p)_\nu}{M_a^2},\label{e.vectorsum}
\end{align} 
where the diquark is treated as a particle with the mass $M_a$. With the constraints $P_h^\alpha U_\alpha =0$ and $\gamma^\alpha U_\alpha =0$~\cite{Christensen:2013aua,Huang:2003ym,Haberzettl:1998rw}, the simplest form of the hadron-quark-diquark vertex can be written as
\begin{align}
	\Upsilon_\alpha^{a \mu}=\frac{g_a(p^2)}{\sqrt{3}}\gamma_{5} \gamma^\mu \frac{p_\alpha}{m},\label{e.vertex}
\end{align}
where the function $g_a(p^2)$ is a form factor that describes the composite structure of the produced hadron and the diquark spectator, and the index $\alpha$ is used to balance the spin-3/2 spinor index. 

For convenience, the large momentum of the produced hadron is taken along the ``+" direction when we calculate the TMD FFs in the diquark model. Applying the spectator diquark model, one can express the correlator for the production of the unpolarized hadron as
\begin{align}
	\Delta_U(p,P_h)=&\frac{|g_a (p^2)|^2}{(2\pi)^3}\frac{\slashed{p}+m}{18(1-z) P_h^{+} m^2 M_h^2 M_a^2(p^2-m^2)^2}
	\Big[(p\cdot P_h)^2-M_h^2 p^2\Big]  \nonumber \\
	&\times\Big[(p^2-M_h^2-2M_a^2)\slashed{P}_h + M_h(M_h^2-4M_a^2-\slashed{p}\slashed{P}_h-\slashed{P}_h\slashed{p}) + 2\slashed{p}(M_h^2-p\cdot P_h)\Big] (\slashed{p}+m),
\end{align}
with 
\begin{align}
   p^2=\frac{z}{1-z}\bm{p}_T^2+\frac{M_a^2}{1-z}+\frac{M_h^2}{z}, \quad
    p\cdot P_h=\frac{M_h^2}{2z}+\frac{z(p^2+\bm{p}_T^2)}{2}.
\end{align}
Since TMD FFs are defined through projections by Dirac matrices as defined in Eq.~\eqref{e.DeltaGamma},
the unpolarized fragmentation function $D_1$ in the spectator model can be expressed as
\begin{align}
		D_1(z,\bm{p}_T^2)=\Delta^{[\gamma^-]}_U\left(z,\bm{p}_T^2\right)=&\frac{|g_a (p^2)|^2}{(2\pi)^3}\frac{1}{36m^2 M_h^2 M_a^2 (1-z)^3 z^6 (L^2+\bm{p}_T^2)^2}\nonumber\\
		&\times\Bigg\{
		\Big[(1-z)^2\big[M_h^4(1-z)^2-2M_h^2 z^2(M_a^2-\bm{p}_T^2)\big]+z^4(M_a^2+\bm{p}_T^2)^2\Big]\nonumber\\
		&\times \Big[(1-z)^2 \big[m^2M_h^2(1-z)^2+2M_a^2(m^2z^2+M_h^2+3mM_hz) \nonumber\\
		&+z^2\bm{p}_T^2(m^2+M_h^2)\big]+z^2\big[2M_a^2\bm{p}_T^2(z^2+1)+(M_a^2-\bm{p}_T^2z)^2\big]
		\Big]
		\Bigg\},
		\label{e.D1}
\end{align}
where 
\begin{align}
	L^2=\frac{1-z}{z^2}M_h^2+m^2+\frac{M_a^2-m^2}{z}.
\end{align}

To compute the rank-3 tensor polarized TMD FFs, we employ the light-cone formalism spinors and polarization vectors. This approach is similar to using the light-cone wave functions to calculate the TMD FFs in the spectator model~\cite{Bacchetta:2008af}. Following the conventions in Refs.~\cite{Meissner:2007rx,Lepage:1980fj}, the Dirac spinors in standard representation are given by
\begin{equation}
	u(P_h,+)=\frac{1}{\sqrt{2^{3 / 2} P_h^{+}}}
	\begin{pmatrix}
		\sqrt{2} P_h^{+}+M_h \\
		P_{hx}+i P_{hy} \\
		\sqrt{2} P_h^{+}-M_h \\
		P_{hx}+i P_{hy}	
	\end{pmatrix},
	\quad
	u(P_h,-)=\frac{1}{\sqrt{2^{3 / 2} P_h^{+}}}
	\begin{pmatrix}
		-P_{hx}+i P_{hy} \\
			\sqrt{2} P_h^{+}+M_h \\
			P_{hx}-i P_{hy} \\
			-\sqrt{2} P_h^{+}+M_h
	\end{pmatrix},
\end{equation}
where $u(P_h,+)$ and $u(P_h,-)$ refer to the spinors with helicity ``$+$'' and helicity ``$-$,'' respectively. The light-cone polarization vectors are given by
\begin{align}
		\varepsilon  (P_h-p,+) 		
		 =&\left(0, -\frac{(P_h-p)_x+i(P_h-p)_y}{\sqrt{2}(P_h-p)^{+}}, -\frac{1}{\sqrt{2}},-\frac{i}{\sqrt{2}}\right) 
		 =\left(0,\frac{z(p_x+i p_y)}{\sqrt{2}(z-1) P_h^{+}}, -\frac{1}{\sqrt{2}},-\frac{i}{\sqrt{2}}\right), \\
		\varepsilon(P_h-p,-)=&\left( 0, -\frac{z(p_x-i p_y)}{\sqrt{2}(z-1) P_h^{+}}, \frac{1}{\sqrt{2}},-\frac{i}{\sqrt{2}}\right) ,\\
		\varepsilon(P_h-p,0)=&\dfrac{1}{M_a}\bigg(\frac{(z-1)}{z}P_h^+, \frac{z(\bm{p}_T^2-M_a^2)}{2(z-1)P_h^+}, -p_x-p_y\bigg).
\end{align}
Similarly, the polarization vectors appearing in spin-3/2 spinors are given by 
\begin{align}
	\varepsilon (P_h,+) =&\left(0, 0,-\frac{1}{\sqrt{2}},-\frac{i}{\sqrt{2}}\right), \\
	\varepsilon(P_h,-)=&\left(0, 0, \frac{1}{\sqrt{2}},-\frac{i}{\sqrt{2}}\right) ,\\
    \varepsilon(P_h,0)=&\frac{1}{M_a}\left(P_h^+, -\frac{M_a^2}{2P_h^+},0,0\right).
\end{align}
The spin component $S_{hLLL}$ can be expressed in terms of the combination of the probabilities $P(m_{(\theta,\phi)})$~\cite{Zhao:2022lbw}, 
\begin{align}
	S_{hLLL}=\dfrac{3}{10}\left[P\left(\dfrac{3}{2}_{(0,0)}\right)-P\left(-\dfrac{3}{2}_{(0,0)}\right)\right]-\dfrac{9}{10}\left[P\left(\dfrac{1}{2}_{(0,0)}\right)-P\left(-\dfrac{1}{2}_{(0,0)}\right)\right].
\end{align}
As a result, the $S_{hLLL}$-dependent correlator in the spectator model can be expressed as
\begin{align}
	\Delta_{LLL}(p,P_h) =&\frac{|g_a (p^2)|^2}{12(2\pi)^3(1-z) P_h^{+}}\nonumber\\
 \times\Bigg[&
	\frac{3}{10}\bigg( \varepsilon_\nu^\ast \frac{-i}{\slashed{p}-m} \Upsilon_\beta^\nu U_{3/2}^\beta
	\bar{U}_{3/2}^\alpha \Upsilon_\alpha^\mu \frac{-i}{\slashed{p}-m}\varepsilon_\mu   
    - \varepsilon_\nu^\ast \frac{-i}{\slashed{p}-m} \Upsilon_\beta^\nu U_{-3/2}^\beta\bar{U}_{-3/2}^\alpha \Upsilon_\alpha^\mu \frac{-i}{\slashed{p}-m}\varepsilon_\mu  \bigg) \nonumber\\ 
	&-\frac{9}{10}\bigg( \varepsilon_\nu^\ast \frac{-i}{\slashed{p}-m} \Upsilon_\beta^\nu U_{1/2}^\beta \bar{U}_{1/2}^\alpha \Upsilon_\alpha^\mu \frac{-i}{\slashed{p}-m}\varepsilon_\mu   
    - \varepsilon_\nu^\ast \frac{-i}{\slashed{p}-m} \Upsilon_\beta^\nu U_{-1/2}^\beta \bar{U}_{-1/2}^\alpha \Upsilon_\alpha^\mu \frac{-i}{\slashed{p}-m}\varepsilon_\mu   
	\bigg)\Bigg],\label{e.DeltaLLL}
\end{align}
where, for conciseness, the momenta of the hadron and diquark in the spinors and polarization vectors are not written. 
The fragmentation function $G_{1LLL}$ describes the longitudinal polarized quark fragmenting to the $S_{hLLL}$ polarized produced hadron. After performing the calculation of the definition of the FFs, one can obtain the analytic expression of $G_{1LLL}$ depending on $z$ and $\bm{p}_T^2$,
\begin{align}
	G_{1LLL}(z,\bm{p}_T^2)=&\Delta^{[\gamma^- \gamma_5 ]}_{LLL}\left(z, \bm{p}_T^2\right)\nonumber\\
 =&\frac{|g_a(p^2)|^2}{(2\pi)^3}\frac{1}{80 m^2 M_h^2 M_a^2 (1-z)^3 z^6 (L^2+\bm{p}_T^2)^2} \nonumber\\ 
	&\times\Bigg\{
	8M_h  \bm{p}_T^2 (1-z)^2 z^2 \Big[(1-z)(M_h^2-m^2z^2)-z^2(zL^2+\bm{p}_T^2)\Big]\nonumber\\
	&\times \Big[(m+M_h)\Big(mM_h+z^2(mM_h-\bm{p}_T^2)-2zmM_h\Big) - M_a^2\Big(z(m-M_h)+2M_h\Big)\Big]\nonumber\\
	&+\Big[4M_h^2 \bm{p}_T^2 (1-z)^2z^2-2\Big(z^2(L^2z+\bm{p}_T^2)-(1-z)(M_h^2-z^2m^2)\Big)^2\Big]\nonumber\\
	&\times\Big[(1-z)^2\Big(m^2M_h^2(1-z)^2-2M_a^2(z^2m^2+zmM_h+M_h^2)-z^2\bm{p}_T^2(m^2+4mM_h+M_h^2)\Big)\nonumber\\
	&+z^2\Big(2M_a^2\bm{p}_T^2(z(z-1)+1)+M_a^4+\bm{p}_T^4z^2\Big)	\Big]
	\Bigg\}.\label{e.G1LLL}
\end{align}
\subsection{Numerical estimates}
The spin transfer to a $S_{hLLL}$ polarized hadron can be defined by the ratio of the $S_{hLLL}$-dependent structure function~\eqref{e.GULLL} and the unpolarized structure function~\eqref{e.FUUT}. In the parton model, the spin transfer is written as the convolutions of TMD PDFs and FFs,
\begin{align}
	A_{LLL}=\frac{G_{U,LLL}}{F_{U,U}^T}=\frac{\mathcal{C}\Big[f_1(x,\bm{k}_T^2) G_{1LLL}(z,\bm{p}_T^2)\Big]}{\mathcal{C}\Big[f_1(x,\bm{k}_T^2) D_1(z,\bm{p}_T^2)\Big]},\label{e.asymmetry}
\end{align}
where the unpolarized TMD distribution function can usually be taken using the Gaussian ansatz~\cite{DAlesio:2020wjq,Callos:2020qtu},
\begin{align}
	f_{1q}(x,\bm{k}_T^2)=f_{1q}(x)\dfrac{1}{\pi\Delta_k^2}e^{-\bm{k}_T^2/\Delta_k^2},\label{e.Gaussian}
\end{align}
with the Gaussian width $\Delta_k$ chosen as $\Delta_k^2=0.61\,\rm{GeV}^2$. The adoption of this ansatz here will induce the cancellation of $f_{1q}(x)$ in the numerator and denominator after the integration on $\bm{k}_T$, 
\begin{align}
	A_{LLL}=\frac{\int |\bm{p}_T| d\left|\bm{p}_T\right| d\theta \exp{\Big[-\frac{1}{\Delta_p^2}\big(\bm{p}_T^2+\frac{\bm{P}_{h\perp}^2}{z^2}+\frac{2}{z} \left|\bm{p}_T\right| \left|\bm{P}_{h\perp}\right| \cos{\theta}\big)\Big]} G_{1LLL}(z,\bm{p}_T^2)} {\int \left|\bm{p}_T\right| d\left|\bm{p}_T\right| d\theta \exp{\Big[-\frac{1}{\Delta_p^2}\big(\bm{p}_T^2+\frac{\bm{P}_{h\perp}^2}{z^2}+\frac{2}{z} \left|\bm{p}_T\right| \left|\bm{P}_{h\perp}\right| \cos{\theta}\big)\Big]} D_{1}(z,\bm{p}_T^2)}. \label{e.ALLL}
\end{align}
To obtain the final numerical result for $A_{LLL}$, we first need the inputs of the diquark model, some of which are still unknown, to determine $D_1(z,\bm{p}_T^2)$ and $G_{1LLL}(z,\bm{p}_T^2)$ whose analytic expressions have been given in Eqs.~\eqref{e.D1} and~\eqref{e.G1LLL}, and then implement the integration on $\bm{p}_T$ during which divergence may arise at large $|\bm{p}_T|$. To cut off the large $|\bm{p}_T|$ contribution, one may directly impose a cutoff on $|\bm{p}_T|$, as in Ref.~\cite{Amrath:2005gv}, or introduce a Gaussian form factor at the quark-hadron vertex, which can also effectively cut off the divergence at the large $|\bm{p}_T|$ region, as in Refs.~\cite{Bacchetta:2007wc,Gamberg:2003eg}. We follow Ref.~\cite{Bacchetta:2007wc} and choose the Gaussian form factor as
\begin{align}
	g_a(p^2) \mapsto \dfrac{g_a}{z} e^{-\frac{p^2}{\Lambda^2}}\label{e.formfactor}
\end{align}
with $\Lambda^2=\lambda^2 z^{a}(1-z)^{b}$. To sum up, the parameters required for the numerical calculation are $g_a$, $\lambda$, $a$, $b$ together with the masses of the axial vector diquark $M_a$ and the initial quark $m$.

\begin{figure}
    \centering
    \includegraphics[width=0.45\linewidth]{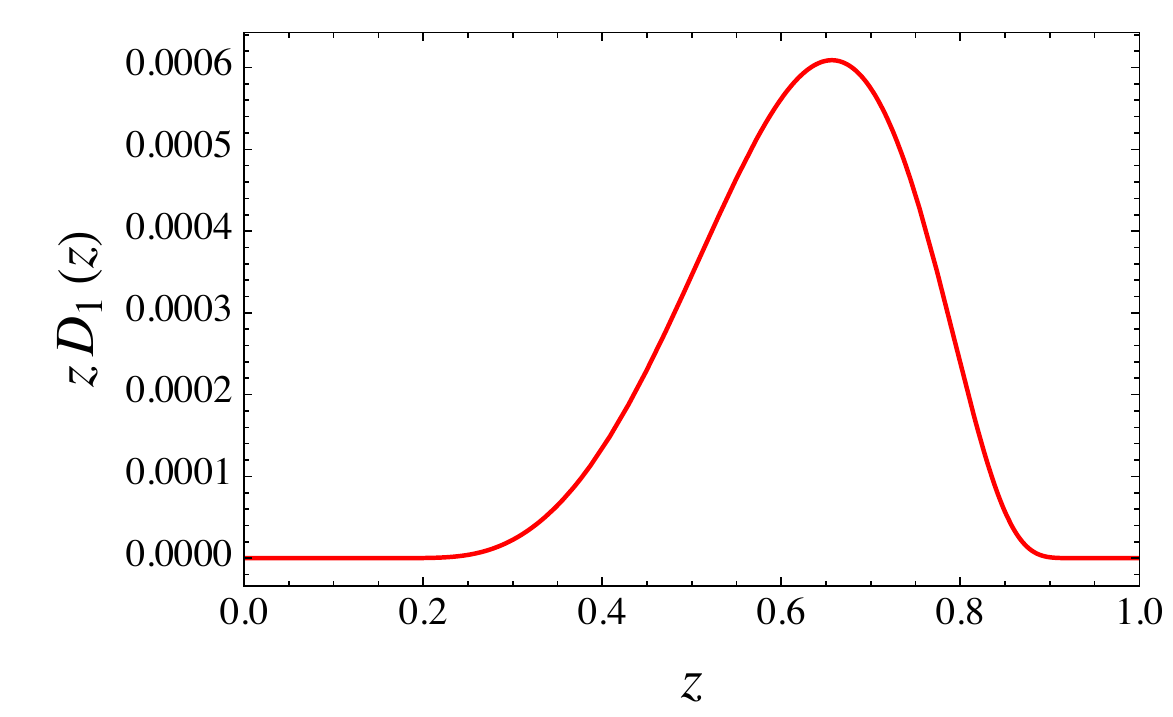}
    \includegraphics[width=0.45\linewidth]{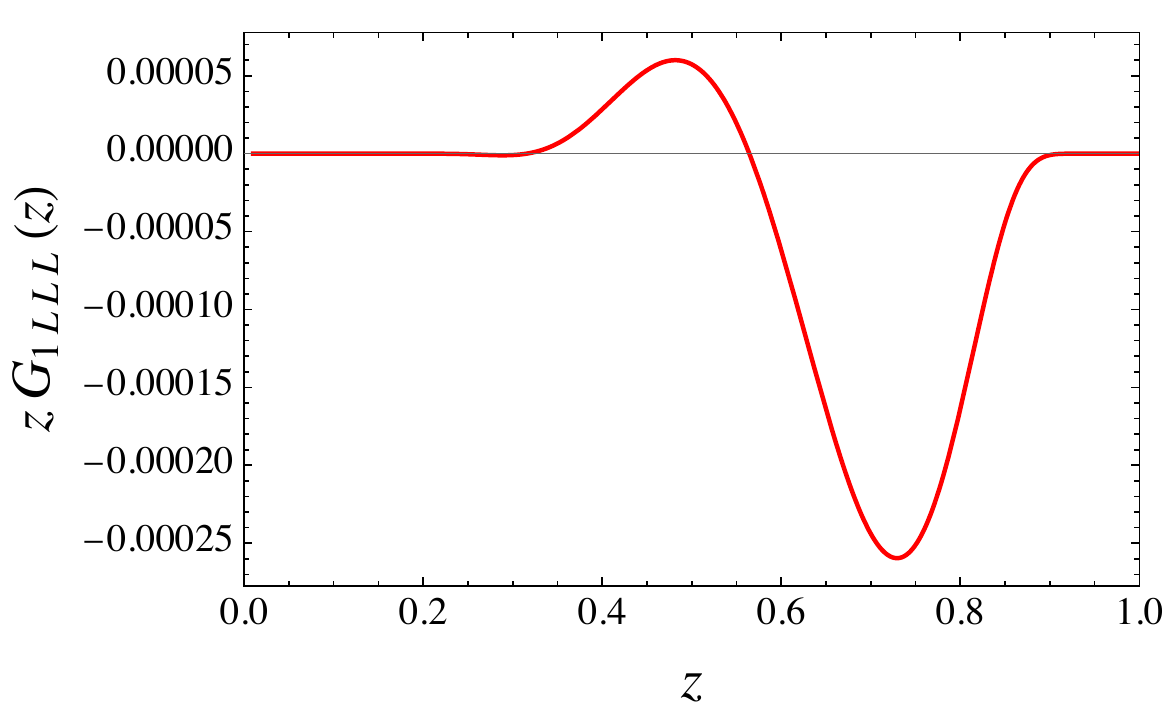}
    \caption{The unpolarized fragmentation function $zD_1(z)$ (left) and the $S_{hLLL}$-dependent fragmentation function $zG_{1LLL}(z)$ (right) calculated from the spectator model.}
    \label{f.zD1}
\end{figure}

We take the constituent quark mass as $m=0.5\,$GeV for the $s$ quark, and the $\Omega$ mass as $M_h=1.672\,$GeV. In Ref.~\cite{Bacchetta:2007wc}, the mass of the spectator composed of one $s$ quark is taken as $m_s=1.12\,$GeV, so it is quite reasonable to assume the mass of the spectator diquark composed of two $\bar{s}$ quarks to be $M_a=2\,$GeV. To determine the parameters related to the Gaussian form factor, we fit the unpolarized FF $D_1(z)$ with events generated by colliding $e^+ e^-$ pairs using $\scalebox{0.8}{PYTHIA}$ at $\sqrt{s}=10\,$GeV, where the unpolarized  collinear fragmentation function $D_1(z)$ can be obtained by integrating over the transverse momentum $\bm{P}_{hT}=-z\bm{p}_T$ of the produced hadron with respect to the direction of the quark, 
\begin{align}
	D_1(z)=\int d^2 \bm{P}_{hT} D_1(z,\boldsymbol P_{hT}^2)=\pi z^2\int_0^\infty d\bm{p}_T^2D_1(z, z^2\boldsymbol p_T^2).
\end{align}
By fitting the events generated by $\scalebox{0.8}{PYTHIA}$, we fix the model parameters as
\begin{align}
	g_a=0.32,\quad \lambda=5.967\,{\rm GeV},\quad a=1.35,\quad b=0.6.\label{e.parameters}
\end{align}
The corresponding unpolarized collinear FF $D_1(z)$ and the integrated $S_{hLLL}$-dependent fragmentation function $G_{1LLL}(z)$ are shown in Fig.~\ref{f.zD1}, where the $G_{1LLL}(z)$ is given by 
\begin{align}
	G_{1LLL}(z)=\int d^2 \bm{P}_{hT} G_{1LLL}(z,\boldsymbol P_{hT}^2)=\pi z^2\int_0^\infty d\bm{p}_T^2G_{1LLL}(z, z^2\boldsymbol p_T^2).
\end{align}

\begin{figure}[ht]
	\centering
	\includegraphics[width=0.5\textwidth]{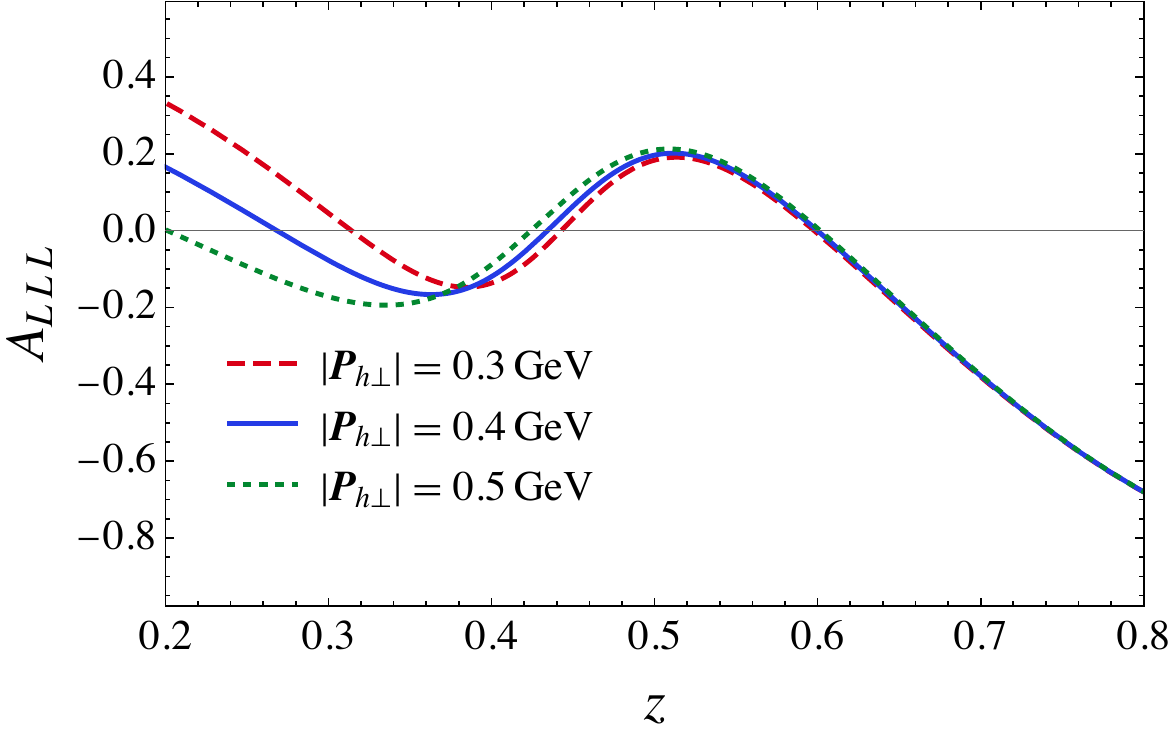}
	\caption{The $S_{hLLL}$-dependent spin transfer in the spectator model.}
	\label{f.asymmetry}
\end{figure}

Substituting the parameters above to Eq.~\eqref{e.ALLL}, we obtain the numerical results of $A_{LLL}$. As shown in Fig.~\ref{f.asymmetry}, the spin transfer can reach a few percent level, and in the large-$z$ region, $\gtrsim 0.4$, the $A_{LLL}$ has mild $|\bm{P}_{h\perp}|$ dependence. By varying the values of the model parameters in Eq.~\eqref{e.parameters}, we find the spin transfer exhibits insensitivity to these parameters. Based on the model estimation, the $A_{LLL}$ is expected to be sizable and would be measurable in future experiments.

\section{Summary}
\label{s.summary}

In this paper, we have studied the semi-inclusive production of a spin-3/2 hadron in the electron-nucleon deep inelastic scattering process. 

Taking into account the polarization of both the nucleon and the produced hadron, as well as the lepton beam, we derive the complete expression of the differential cross section in terms of structure functions. In this kinematic analysis, we decompose the hadronic tensor by constructing basis Lorentz tensors. We further demonstrated that all polarized basis Lorentz tensors can be systematically given by unpolarized basic Lorentz tensors multiplied by a spin-dependent scalar or pseudoscalar. Restricted to the parity conserving case, the SIDIS differential cross sections for the production of a spin-3/2 hadron are expressed in terms of 288 structure functions according to various angular modulations and spin states of the nucleon and the hadron. Among them, 126 rank-3 tensor polarized ones are newly defined and exist only when the spin of the produced hadron is no less than 3/2.

Considering the kinematic region $\bm{q}_T^2 \ll Q^2 $, we perform a leading order calculation in the parton model. At the leading twist, half of the 192 structure functions for an unpolarized lepton beam are nonzero, among which 42 of the rank-3 tensor polarized TMD FFs contribute. For a polarized lepton beam, one third of the 96 structure functions are nonzero and 14 of them are for rank-3 tensor polarized hadron states. Hence the SIDIS process with the production of a spin-3/2 hadron provide many new observables for the study of nucleon structures. 

Taking a rank-3 tensor spin transfer, $A_{LLL}$, as an example, we perform a numerical estimation of the new observables related to the rank-3 tensor polarization of the produced hadron. 
Based on the spectator model calculation, we find the spin transfer to a $S_{hLLL}$ polarized $\Omega$, with the spin state analyzed via its weak decay, can be sizable, up to a few percentages. In addition, the measurement of the spin state of the produced hadron will also allow us to learn the spin-dependent FFs. Therefore, the experimental measurements of the production of spin-3/2 hadron in SIDIS process at future facilities, such as the EIC and the EicC, will be useful for us to understand the nucleon spin structures and the spin effects in hadronization process.

\acknowledgments{
We thank Kai-bao Chen, Xiaoyan Zhao, and Yongjie Deng for useful discussions. This work was supported by the National Natural Science Foundation of China (Grants No.~12175117 and No.~12321005) and Shandong Province Natural Science Foundation (Grants No.~ZR2020MA098 and No.~ZFJH202303).
}

\newpage

\appendix
\section{DECOMPOSITION OF SPIN TENSORS}
\label{a.Sdecompositon}

In this appendix we show that the spin tensors can be decomposed in the coordinate constructed by $P^\mu$, $P_h^\mu$, $q^\mu$, and $\epsilon^{\mu \nu\rho\sigma}$; for example, the spin vector for a spin-1/2 target can be written as
\begin{align}
    S^{\mu}=&\bigg[\frac{1}{A_{1}}(P\cdot q)(q\cdot S)+\frac{A_{3}}{A_{1}A_{4}}\Big(A_{2}(q\cdot S)+A_{1}(P_{h}\cdot S)\Big)\bigg]P^{\mu}\notag\\
    &+\bigg[\frac{1}{A_{1}Q^{2}}(P\cdot q)^{2}(q\cdot S)-\frac{A_{2}}{A_{1}A_{4}}\Big(A_{2}(q\cdot S)+A_{1}(P_{h}\cdot S)\Big)-\frac{q\cdot S}{Q^{2}}\bigg]q^{\mu}\notag\\
    &-\frac{1}{A_{4}}\Big(A_{2}(q\cdot S)+A_{1}(P_{h}\cdot S)\Big)P_{h}^{\mu}+\frac{1}{A_{4}}\epsilon^{PP_{h}qS}\epsilon^{\mu PP_{h}q},
\end{align} 
where the factors $A_1$, $A_2$, $A_3$, and $A_4$, as scalar functions of $Q^2$, $P\cdot q$, $P_h\cdot q$, and $P\cdot q$, are defined by
\begin{align}
    A_{1}&=M^{2}Q^{2}+(P\cdot q)^{2},\\
    A_{2}&=M^{2}(P_{h}\cdot q)-(P\cdot P_{h})(P\cdot q),\\
    A_{3}&=(P\cdot q)(P_{h}\cdot q)+Q^{2}(P\cdot P_{h}),\\
    A_{4}&=Q^{2}(P\cdot P_{h})^{2}+2(P\cdot P_{h})(P\cdot q)(P_{h}\cdot q)-M^{2}(P_{h}\cdot q)^{2}-M_{h}^{2}\big((P\cdot q)^{2}+M^{2}Q^{2}\big)^{2}.
\end{align}
One can easily verify that the relation $P_\mu S^\mu=0$ holds with this decomposition. Similarly, the rank-2 spin tensor for the produced hadron can be decomposed as
\begin{align}
    T_{h}^{\mu\nu}=&\frac{q^{\mu}q^{\nu}}{Q^{2}}\bigg[\frac{T_{h}^{qq}}{Q^{2}}-\frac{2A_{2}}{A_{1}A_{4}}\big(A_{3}T_{h}^{qP}-A_{2}T_{h}^{qq}\big)+\frac{A_{2}^{2}Q^{2}}{A_{1}^{2}A_{4}^{2}}\Big(A_{2}^{2}T_{h}^{qq}-2A_{2}A_{3}T_{h}^{qP}+A_{3}^{2}T_{h}^{PP}\Big)\nonumber\\%
&\quad-\frac{2P\cdot q}{A_{1}Q^{2}}\Big(Q^{2}T_{h}^{qP}+(P\cdot q)T_{h}^{qq}\Big)+\frac{(P\cdot q)^{2}}{A_{1}^{2}Q^{2}}\Big(Q^{4}T_{h}^{PP}+2Q^{2}(P\cdot q)T_{h}^{qP}+(P\cdot q)^{2}T_{h}^{qq}\Big)\nonumber\\%
&\quad+\frac{2A_{2}P\cdot q}{A_{1}^{2}A_{4}}\Big((P\cdot q)\big(A_{3}T_{h}^{qP}-A_{2}T_{h}^{qq}\big)+Q^{2}\big(A_{3}T_{h}^{PP}-A_{2}T_{h}^{qP}\big)\Big)\bigg]\nonumber\\%
&+\frac{P^{\mu}P^{\nu}}{Q^{2}}\bigg[\frac{A_{3}^{2}Q^{2}}{A_{1}^{2}A_{4}^{2}}\Big(A_{2}^{2}T_{h}^{qq}-2A_{2}A_{3}T_{h}^{qP}+A_{3}^{2}T_{h}^{PP}\Big)\nonumber\\%
&\quad+\frac{2P\cdot q}{A_{1}A_{4}}\Big((P\cdot q)\big(A_{3}T_{h}^{qP}-A_{2}T_{h}^{qq}\big)+Q^{2}\big(A_{3}T_{h}^{PP}-A_{2}T_{h}^{qP}\big)\Big)\nonumber\\%
&\quad+\frac{Q^{2}}{A_{1}^{2}}\Big(Q^{4}T_{h}^{PP}+2Q^{2}(P\cdot q)T_{h}^{qP}+(P\cdot q)^{2}T_{h}^{qq}\Big)\bigg]\nonumber\\%
&+\frac{P_{h}^{\mu}P_{h}^{\nu}}{Q^{2}}\bigg[\frac{Q^{2}}{A_{4}^{2}}\Big(A_{2}^{2}T_{h}^{qq}-2A_{2}A_{3}T_{h}^{qP}+A_{3}^{2}T_{h}^{PP}\Big)\bigg]\nonumber\\%
&+\frac{P^{\{\mu}q^{\nu\}}}{Q^{2}}\bigg[\frac{A_{3}}{A_{1}A_{4}}\big(A_{3}T_{h}^{qP}-A_{2}T_{h}^{qq}\big)-\frac{A_{2}A_{3}Q^{2}}{A_{1}^{2}A_{4}^{2}}\Big(A_{2}^{2}T_{h}^{qq}-2A_{2}A_{3}T_{h}^{qP}+A_{3}^{2}T_{h}^{PP}\Big)\nonumber\\%
&\quad-\frac{1}{A_{1}}\Big(Q^{2}T_{h}^{qP}+(P\cdot q)T_{h}^{qq}\Big)-\frac{A_{2}P\cdot q}{A_{1}^{2}A_{4}}\Big((P\cdot q)\big(A_{3}T_{h}^{qP}-A_{2}T_{h}^{qq}\big)+Q^{2}\big(A_{3}T_{h}^{PP}-A_{2}T_{h}^{qP}\big)\Big)\nonumber\\%
&\quad+\frac{P\cdot q}{A_{1}^{2}}\Big(Q^{4}T_{h}^{PP}+2Q^{2}(P\cdot q)T_{h}^{qP}+(P\cdot q)^{2}T_{h}^{qq}\Big)\bigg]\nonumber\\%
&+\frac{P^{\{\mu}P_{h}^{\nu\}}}{Q^{2}}\bigg[-\frac{A_{3}Q^{2}}{A_{1}A_{4}^{2}}\Big(A_{2}^{2}T_{h}^{qq}-2A_{2}A_{3}T_{h}^{qP}+A_{3}^{2}T_{h}^{PP}\Big)\nonumber\\%
&\quad-\frac{A_{1}Q^{2}}{A_{1}^{2}A_{4}}\Big((P\cdot q)\big(A_{3}T_{h}^{qP}-A_{2}T_{h}^{qq}\big)+Q^{2}\big(A_{3}T_{h}^{PP}-A_{2}T_{h}^{qP}\big)\Big)\bigg]\nonumber\\%
&+\frac{P_{h}^{\{\mu}q^{\nu\}}}{Q^{2}}\bigg[-\frac{1}{A_{4}}\big(A_{3}T_{h}^{qP}-A_{2}T_{h}^{qq}\big)-\frac{A_{2}Q^{2}}{A_{1}A_{4}^{2}}\Big(A_{2}^{2}T_{h}^{qq}-2A_{2}A_{3}T_{h}^{qP}+A_{3}^{2}T_{h}^{PP}\Big)\nonumber\\%
&\quad+\frac{A_{1}P\cdot q}{A_{1}^{2}A_{4}}\Big((P\cdot q)\big(A_{3}T_{h}^{qP}-A_{2}T_{h}^{qq}\big)+Q^{2}\big(A_{3}T_{h}^{PP}-A_{2}T_{h}^{qP}\big)\Big)\bigg]\nonumber\\%
&+\frac{Q^{2}}{A_{4}^{2}}\epsilon^{\mu PP_{h}q}\epsilon^{\nu PP_{h}q}\bigg[\big(M_{h}^{2}M^{2}-(P\cdot P_{h})^{2}\big)T_{h}^{qq}-\big(M_{h}^{2}Q^{2}+(P_{h}\cdot q)^{2}\big)T_{h}^{PP}\nonumber\\%
&\quad+2\big((P\cdot P_{h})(P_{h}\cdot q)-P\cdot q\big)T_{h}^{Pq}\bigg]\nonumber\\%
&+\frac{1}{A_{4}}q^{\{\mu}\epsilon^{\nu\}PP_{h}q}\epsilon^{T_{h}^{q}PP_{h}q}\nonumber\\%
&+\frac{Q^{2}}{A_{1}A_{4}^{2}}\epsilon^{\mu PP_{h}q}\big(A_{3}P^{\nu}-A_{1}P_{h}^{\nu}-A_{2}q^{\nu}\big)\big(A_{3}\epsilon^{T_{h}^{P}PP_{h}q}-A_{2}\epsilon^{T_{h}^{q}PP_{h}q}\big)\nonumber\\%
&+\frac{Q^{2}}{A_{1}A_{4}^{2}}\epsilon^{\nu PP_{h}q}\big(A_{3}P^{\mu}-A_{1}P_{h}^{\mu}-A_{2}q^{\mu}\big)\big(A_{3}\epsilon^{T_{h}^{P}PP_{h}q}-A_{2}\epsilon^{T_{h}^{q}PP_{h}q}\big)\nonumber\\%
&-\frac{1}{A_{1}A_{4}}\epsilon^{\mu PP_{h}q}\big(Q^{2}P^{\nu}+(P\cdot q)q^{\nu}\big)\big(Q^{2}\epsilon^{T_{h}^{P}PP_{h}q}+(P\cdot q)\epsilon^{T_{h}^{q}PP_{h}q}\big)\nonumber\\%
&-\frac{1}{A_{1}A_{4}}\epsilon^{\nu PP_{h}q}\big(Q^{2}P^{\mu}+(P\cdot q)q^{\mu}\big)\big(Q^{2}\epsilon^{T_{h}^{P}PP_{h}q}+(P\cdot q)\epsilon^{T_{h}^{q}PP_{h}q}\big).
\end{align}
Again, the orthogonal relation between the spin vector and momentum $P_{h\mu}T_h^{\mu\nu}=0$ can be verified using the above expression. The similar decomposition applied to $S_h^\mu$ and $R_h^{\mu\nu\rho}$ and the lengthy expressions will not be listed again. In conclusion, the spin tensors can always be decomposed in the basis building from $P^\mu$, $P_h^\mu$, $q^\mu$, and $\epsilon^{\mu \nu\rho\sigma}$ with the spin information being in the spin-dependent scalar coefficients. Thus, one can construct the polarized basis tensor by multiplying the basic Lorentz tensors by a spin-dependent scalar or pseudoscalar.

\section{THE HADRONIC TENSOR IN THE PARTON MODEL}
\label{a.hadrontensor}
The TMD FFs are defined as functions of $z$ and $k_T^2=-\bm{k}_T^2$, with inhomogeneous $k_T$ dependence factored out. The completely symmetric and traceless tensors $k_T^{i_1\cdots i_n}$ are defined as~\cite{Boer:2016xqr}
\begin{align}
k_{T}^{i j}&= k_{T}^{i} k_{T}^{j}-\frac{1}{2} k_{T}^{2} g_{T}^{i j}, \label{e.kij}\\
k_{T}^{i j k}&= k_{T}^{i} k_{T}^{j} k_{T}^{k}-\frac{1}{4} k_{T}^{2}\left(g_{T}^{i j} k_{T}^{k}+g_{T}^{i k} k_{T}^{j}+g_{T}^{j k} k_{T}^{i}\right), \label{e.kijk}\\
k_{T}^{i j k l}&= k_{T}^{i} k_{T}^{j} k_{T}^{k} k_{T}^{l} \nonumber\\
&\quad-\frac{1}{6} k_{T}^{2}\left(g_{T}^{i j} k_{T}^{k l}+g_{T}^{i k} k_{T}^{j l}+g_{T}^{i l} k_{T}^{j k}+g_{T}^{j k} k_{T}^{i l}+g_{T}^{j l} k_{T}^{i k}+g_{T}^{k l} k_{T}^{i j}\right) \nonumber\\
&\quad-\frac{1}{8}\left(k_{T}^{2}\right)^{2}\left(g_{T}^{i j} g_{T}^{k l}+g_{T}^{i k} g_{T}^{j l}+g_{T}^{i l} g_{T}^{j k}\right) , \label{e.kijkl}
\end{align}
which satisfy
\begin{align}
g_{T i j} k_{T}^{i j}=g_{T i j} k_{T}^{i j k}=g_{T i j} k_{T}^{i j k l}=0.
\end{align}

At the leading twist, the hadronic tensor for the production of the spin-3/2 hadrons in SIDIS can be expressed in terms of 128 combinations of the spin-1/2 TMD PDFs and spin-3/2 TMD FFs. We decompose the hadronic tensor into two parts, denoted as $W_S$ and $W_A$, where $W_S$ is the symmetric part,
\begin{align}
	W^{\mu\nu}_S=&2z \sum_a e_a^2 \int d^2 \boldsymbol{k}_T d^2 \boldsymbol{p}_T \delta^{(2)}\left(\boldsymbol{k}_T-\boldsymbol{p}_T-\boldsymbol{P}_{h \perp} / z\right)\nonumber\\
	&\bigg\{ -g_T^{\mu\nu}f_{1}D_{1}-\frac{k_T^{\{\mu}p_T^{\nu\}}-(k_T\cdot p_T)g_T^{\mu\nu}}{MM_{h}}h_{1}^{\perp}H_{1}^{\perp}+\frac{k_T^{\{\mu}\epsilon_T^{\nu\}\rho}p_{T\rho}+p_T^{\{\mu}\epsilon_T^{\nu\}\rho}k_{T\rho}}{2MM_{h}}S_{hL}h_{1}^{\perp}H_{1L}^{\perp}\nonumber\\
	&-g_T^{\mu\nu}\epsilon_{T}^{\alpha\beta}\frac{p_{T\alpha}S_{hT\beta}}{M_{h}}f_{1}D_{1T}^{\perp}+\frac{k_T^{\{\mu}\epsilon_T^{\nu\}\rho}S_{hT\rho}+S_{hT}^{\{\mu}\epsilon_T^{\nu\}\rho}k_{T\rho}}{2M}h_{1}^{\perp}H_{1T}  \nonumber\\
	&-\frac{k_T^{\{\mu}\epsilon_T^{\nu\}\rho}p_{T\rho\alpha}S_{hT}^\alpha + p_T^{\alpha\{\mu}\epsilon_T^{\nu\}\rho}S_{hT\alpha}k_{T\rho}}{2MM_{h}^{2}}h_{1}^{\perp}H_{1T}^{\perp} \nonumber\\
    &-g_{T}^{\mu\nu}S_{hLL}f_{1}D_{1LL}+\frac{k_T^{\{\mu}p_T^{\nu\}}-(k_T\cdot p_T)g_T^{\mu\nu}}{MM_{h}}S_{hLL}h_{1}^{\perp}H_{1LL}^{\perp}-g_{T}^{\mu\nu}\frac{\bm{S}_{hT}\cdot\bm{p}_{T}}{M_{h}}f_{1}D_{1LT}^{\perp}\nonumber\\
	&+\frac{k_T^{\{\mu}S_{hLT}^{\nu\}}-(k_T\cdot S_{hLT})g_T^{\mu\nu}}{M}h_{1}^{\perp}H_{1LT}+\frac{k_T^{\{\mu} p_T^{\nu\}\alpha}S_{hLT\alpha}-p_T^{\alpha\beta}k_{T\alpha}S_{hLT\beta}g_T^{\mu\nu}}{MM_{h}^{2}}\left(-h_{1}^{\perp}H_{1LT}^{\perp}\right)\nonumber\\
	&-g_{T}^{\mu\nu}S_{hTT\alpha\beta}\frac{p_{T}^{\alpha\beta}}{M_{h}^{2}}f_{1}D_{1TT}^{\perp}+\frac{k_T^{\{\mu}S_{hTT}^{\nu\}\rho}p_{T\rho}-S_{hTT}^{\alpha\beta}k_{T\alpha}p_{T\beta}g_T^{\mu\nu}}{MM_{h}}h_{1}^{\perp}H_{1TT}^{\perp}\nonumber\\
	&+\frac{k_T^{\{\mu}p_T^{\nu\}\alpha\beta}S_{hTT\alpha\beta}-p_T^{\alpha\beta\rho}S_{hTT\alpha\beta}k_{T\rho} g_T^{\mu\nu}}{MM_{h}^{3}}h_{1}^{\perp}H_{1TT}^{\perp\perp}\nonumber\\
 %
&+\frac{k_T^{\{\mu}\epsilon_T^{\nu\}\rho}p_{T\rho} + p_T^{\{\mu}\epsilon_T^{\nu\}\rho}k_{T\rho}}{2MM_{h}}S_{hLLL}h_{1}^{\perp}H_{1LLL}^{\perp}-g_{T}^{\mu\nu}\frac{\epsilon_{T}^{\alpha\beta}p_{T\alpha}S_{hLLT\beta}}{M_{h}}f_{1}D_{1LLT}^{\perp}\nonumber\\
 &+\frac{k_T^{\{\mu}\epsilon_T^{\nu\}\rho}S_{hLLT\rho} + S_{hLLT}^{\{\mu}\epsilon_T^{\nu\}\rho}k_{T\rho}
 }{2M}h_{1}^{\perp}H_{1LLT}-\frac{k_T^{\{\mu}\epsilon_T^{\nu\}\rho}p_{T\rho\alpha}S_{hLLT}^\alpha + p_T^{\alpha\{\mu}\epsilon_T^{\nu\}\rho}S_{hLLT\alpha}k_{T\rho}}{2MM_{h}^{2}}h_{1}^{\perp}H_{1LLT}^{\perp}\nonumber\\
	&-g_{T}^{\mu\nu}\frac{\epsilon_{T\beta}^{\alpha}p_{T\alpha\rho}S_{hLTT}^{\beta\rho}}{M_{h}^{2}}f_{1}D_{1LTT}^{\perp}+\frac{k_T^{\{\mu}\epsilon_T^{\nu\}\rho}S_{hLTT\rho\alpha}p_T^\alpha + S_{hLTT}^{\alpha\{\mu}\epsilon_T^{\nu\}\rho}p_{T\alpha} k_{T\rho}}{2MM_{h}}h_{1}^{\perp}H_{1LTT}^{\perp}\nonumber\\
	&+\frac{k_T^{\{\mu}\epsilon_T^{\nu\}\rho}p_{T\rho\alpha\beta}S_{hLTT}^{\alpha\beta}+ p_T^{\alpha\beta\{\mu}\epsilon_T^{\nu\}\rho}k_{T\rho}S_{hLTT\alpha\beta}}{2MM_{h}^{3}}h_{1}^{\perp}H_{1LTT}^{\perp\perp}-g_{T}^{\mu\nu}\frac{\epsilon_{T\beta}^{\alpha}p_{T\alpha\rho\sigma}S_{hTTT}^{\beta\rho\sigma}}{M_{h}^{3}}f_{1}D_{1TTT}^{\perp}\nonumber\\
  &+\frac{k_T^{\{\mu}\epsilon_T^{\nu\}\rho}S_{hTTT\rho\alpha\beta}p_{T}^{\alpha\beta}+ S_{hTTT}^{\alpha\beta\{\mu}\epsilon_T^{\nu\}\rho}k_{T\rho}p_{T\alpha\beta}}{2MM_{h}^{2}}h_{1}^{\perp}H_{1TTT}^{\perp}\nonumber\\
&+\frac{
k_T^{\{\mu}\epsilon_T^{\nu\}\rho}p_{T\rho\alpha\beta\tau}S_{hTTT}^{\alpha\beta\tau}+ p_{T}^{\alpha\beta\tau\{\mu}\epsilon_T^{\nu\}\rho}S_{hTTT\alpha\beta\tau}k_{T\rho}}{2MM_{h}^{4}}h_{1}^{\perp}H_{1TTT}^{\perp\perp}\nonumber\\
%
%
&+g_T^{\mu\nu}\epsilon_{T}^{\alpha\beta}\frac{k_{T\alpha}S_{T \beta}}{M}f_{1T}^{\perp}D_{1}-\frac{k_T^{\{\mu}\epsilon_T^{\nu\}\rho}p_{T\rho}+p_T^{\{\mu}\epsilon_T^{\nu\}\rho}k_{T\rho}}{2MM_{h}}S_{L}h_{1L}^{\perp}H_{1}^{\perp}\nonumber\\
	&-\frac{S_T^{\{\mu}\epsilon_T^{\nu\}\rho}p_{T\rho}+S_T^{\{\mu}\epsilon_T^{\nu\}\rho}p_{T\rho}}{2M_{h}}h_{1T}H_{1}^{\perp}+\frac{
 k_T^{\alpha\{\mu}\epsilon_T^{\nu\}\rho}S_{T\alpha}p_{T\rho} +p_T^{\{\mu}\epsilon_T^{\nu\}\rho}k_{T\rho\alpha}S_T^\alpha}{2M^{2}M_{h}}h_{1T}^{\perp}H_{1}^{\perp}\nonumber\\
	&-g_T^{\mu\nu}\bigg(S_{L}S_{hL}g_{1L}G_{1L}+S_{L}\frac{\bm{S}_{hT}\cdot\bm{p}_{T}}{M_{h}}g_{1L}G_{1T}^{\perp}-\epsilon_{T}^{\alpha\beta}\frac{k_{T\alpha}S_{T\beta}}{M}\epsilon_{T}^{\rho\sigma}\frac{p_{T\rho}S_{hT\sigma}}{M_{h}}f_{1T}^{\perp}D_{1T}^{\perp}\nonumber\\
	&+\frac{\bm{k}_{T}\cdot\bm{S}_{T}}{M}S_{hL}g_{1T}G_{1L}+\frac{\bm{k}_{T}\cdot\bm{S}_{T}}{M}\frac{\bm{p}_{T}\cdot\bm{S}_{hT}}{M_{h}}g_{1T}^{\perp}G_{1T}^{\perp}\bigg)\nonumber\\
	& -\frac{k_T^{\{\mu}p_T^{\nu\}}-(k_T\cdot p_T)g_T^{\mu\nu}}{MM_{h}}S_{L}S_{hL}h_{1L}^{\perp}H_{1L}^{\perp}-\frac{k_T^{\{\mu}S_{hT}^{\nu\}}-(k_T\cdot S_{hT})g_T^{\mu\nu}}{M}S_{L}h_{1L}^{\perp}H_{1T}\nonumber\\
 &+\frac{k_T^{\{\mu}p_T^{\nu\}\rho}S_{hT\rho}-p_T^{\alpha\beta}S_{hT\alpha}k_{T\beta}g_T^{\mu\nu}}{MM_{h}^{2}}S_{L}h_{1L}^{\perp}H_{1T}^{\perp}
	-\frac{p_T^{\{\mu}S_T^{\nu\}}-(p_T\cdot S_T)g_T^{\mu\nu}}{M_{h}}S_{hL}h_{1T}H_{1L}^{\perp}\nonumber\\
	&-\left(S_T^{\{\mu}S_{hT}^{\nu\}}-(S_T\cdot S_{hT})g_T^{\mu\nu}\right)h_{1T}H_{1T}+\frac{
 S_T^{\{\mu}p_T^{\nu\}\rho}S_{hT\rho}-p_T^{\alpha\beta}S_{T\alpha}S_{hT\beta} g_T^{\mu\nu}}{M_{h}^{2}}h_{1T}H_{1T}^{\perp}\nonumber\\
	&+\frac{k_T^{\alpha\{\mu}p_T^{\nu\}}S_{T\alpha}-k_T^{\alpha\beta}p_{T\alpha}S_{T\beta} g_T^{\mu\nu}}{M^{2}M_{h}}S_{hL}h_{1T}^{\perp}H_{1L}^{\perp}+\frac{k_T^{\alpha\{\mu}S_{hT}^{\nu\}}S_{T\alpha}-k_T^{\alpha\beta}S_{hT\alpha}S_{T\beta} g_T^{\mu\nu}}{M^{2}}h_{1T}^{\perp}H_{1T}\nonumber\\
	&-\frac{k_T^{\alpha\{\mu}p_T^{\nu\}\beta}S_{T\alpha}S_{hT\beta}-k_T^{\alpha\beta}S_{T\alpha}p_{T\beta\rho}S_{hT}^\rho g_T^{\mu\nu}}{M^{2}M_{h}^{2}}h_{1T}^{\perp}H_{1T}^{\perp}\nonumber\\
 %
&+\frac{k_T^{\{\mu}\epsilon_T^{\nu\}\rho}p_{T\rho}+p_T^{\{\mu}\epsilon_T^{\nu\}\rho}k_{T\rho}}
{2MM_{h}}S_{L}S_{hLL}h_{1L}^{\perp}H_{1LL}^{\perp}+g_{T}^{\mu\nu}S_{L}\frac{\epsilon_{T}^{\alpha\beta}p_{T\alpha}S_{hLT\beta}}{M_{h}}g_{1L}G_{1LT}^{\perp}\nonumber\\
&+\frac{k_T^{\{\mu}\epsilon_T^{\nu\}\rho}S_{hLT\rho}+S_{hLT}^{\{\mu}\epsilon_T^{\nu\}\rho}k_{T\rho}}{2M}S_{L}h_{1L}^{\perp}H_{1LT}-\frac{k_T^{\{\mu}\epsilon_T^{\nu\}\rho}p_{T\rho\alpha}S_{hLT}^\alpha+p_{T}^{\alpha\{\mu}\epsilon_T^{\nu\}\rho}S_{hLT\alpha}k_{T\rho}}{2MM_{h}^{2}}S_{L}h_{1L}^{\perp}H_{1LT}^{\perp}\nonumber\\
&-g_{T}^{\mu\nu}S_{L}\frac{\epsilon_{T\beta}^{\alpha}p_{T\alpha\rho}S_{hTT}^{\beta\rho}}{M_{h}^{2}}g_{1L}G_{1TT}^{\perp}+\frac{
 k_T^{\{\mu}\epsilon_T^{\nu\}\rho}S_{hTT\rho\alpha}p_T^\alpha+S_{hTT}^{\alpha\{\mu}\epsilon_T^{\nu\}\rho}p_{T\alpha}k_{T\rho}}{2MM_{h}}S_{L}h_{1L}^{\perp}H_{1TT}^{\perp}\nonumber\\
&+\frac{k_T^{\{\mu}\epsilon_T^{\nu\}\rho}p_{T\rho\alpha\beta}S_{hTT}^{\alpha\beta}+p_T^{\alpha\beta\{\mu}\epsilon_T^{\nu\}\rho}S_{hTT\alpha\beta}k_{T\rho}}{2MM_{h}^{3}}S_{L}h_{1L}^{\perp}H_{1TT}^{\perp\perp}\nonumber\\
	&+\frac{S_T^{\{\mu}\epsilon_T^{\nu\}\rho}p_{T\rho}+p_T^{\{\mu}\epsilon_T^{\nu\}\rho}S_{T\rho}}{2M_{h}}S_{hLL}h_{1T}H_{1LL}^{\perp}-\frac{k_T^{\alpha\{\mu}\epsilon_T^{\nu\}\rho}S_{T\alpha}p_{T\rho} +p_T^{\{\mu}\epsilon_T^{\nu\}\rho}k_{T\rho\alpha}S_T^\alpha}{2M^{2}M_{h}}S_{hLL}h_{1T}^{\perp}H_{1LL}^{\perp}\nonumber\\
	&+g_{T}^{\mu\nu}\frac{\epsilon_{T}^{\alpha\beta}k_{T\alpha}S_{T\beta}}{M}S_{hLL}f_{1T}^{\perp}D_{1LL}+g_{T}^{\mu\nu}\frac{\bm{S}_{T}\cdot\bm{k}_{T}}{M}\frac{\epsilon_{T}^{\alpha\beta}p_{T\alpha}S_{hLT\beta}}{M_{h}}g_{1T}^{\perp}G_{1LT}^{\perp}\nonumber\\
	&+
	g_{T}^{\mu\nu}\frac{\epsilon_{T}^{\alpha\beta}k_{T\alpha}S_{T\beta}}{M}\frac{\bm{S}_{hLT}\cdot\bm{p}_{T}}{M_{h}}f_{1T}^{\perp}D_{1LT}^{\perp}+\left(S_T^{\{\mu}\epsilon_T^{\nu\}\rho}S_{hLT\rho}+S_{hLT}^{\{\mu}\epsilon_T^{\nu\}\rho}S_{T\rho}\right)
    h_{1T}H_{1LT}\nonumber\\
    &-\frac{S_T^{\{\mu}\epsilon_T^{\nu\}\rho}p_{T\rho\alpha}S_{hLT}^\alpha+p_T^{\alpha\{\mu}\epsilon_T^{\nu\}\rho}S_{hLT\alpha}S_{T\rho}}{2M_{h}^{2}}h_{1T}H_{1LT}^{\perp}\nonumber\\
	&-\frac{k_T^{\alpha\{\mu}\epsilon_T^{\nu\}\rho}S_{T\alpha}S_{hLT\rho}+S_{hLT}^{\{\mu}\epsilon_T^{\nu\}\rho}k_{T\rho\alpha} S_{T}^\alpha}{2M^{2}}h_{1T}^{\perp}H_{1LT}\nonumber\\
    &+\frac{
    k_T^{\alpha\{\mu}\epsilon_T^{\nu\}\rho}S_{T\alpha} p_{T\rho\beta} S_{hLT}^\beta+p_T^{\alpha\{\mu}\epsilon_T^{\nu\}\rho}S_{hLT\alpha} k_{T\rho\beta} S_{T}^\beta}{2M^{2}M_{h}^{2}}h_{1T}^{\perp}H_{1LT}^{\perp}\nonumber\\
	&-g_{T}^{\mu\nu}\frac{\bm{S}_{T}\cdot\bm{k}_{T}}{M}\frac{\epsilon_{T\beta}^{\alpha}p_{T\alpha\rho}S_{hTT}^{\beta\rho}}{M_{h}^{2}}g_{1T}^{\perp}G_{1TT}^{\perp}+g_{T}^{\mu\nu}\frac{\epsilon_{T}^{\alpha\beta}k_{T\alpha}S_{T\beta}}{M}\frac{p_{T}^{\rho\tau}S_{hTT\rho\tau}}{M_{h}^{2}}f_{1T}^{\perp}D_{1TT}^{\perp}\nonumber\\
	&+\frac{S_T^{\{\mu}\epsilon_T^{\nu\}\rho}S_{hTT\rho\alpha}p_T^\alpha+S_{hTT}^{\alpha\{\mu}\epsilon_T^{\nu\}\rho}p_{T\alpha}S_{T\rho}}{2M_{h}}h_{1T}H_{1TT}^{\perp}\nonumber\\
   &+\frac{S_T^{\{\mu}\epsilon_T^{\nu\}\rho}p_{T\rho\alpha\beta}S_{hTT}^{\alpha\beta}+p_T^{\alpha\beta\{\mu}\epsilon_T^{\nu\}\rho}S_{hTT\alpha\beta}S_{T\rho}}{2M_{h}^{3}}h_{1T}H_{1TT}^{\perp\perp}\nonumber\\
	&-\frac{k_T^{\alpha\{\mu}\epsilon_T^{\nu\}\rho}S_{T\alpha}S_{hTT\rho\beta}p_{T}^\beta+S_{hTT}^{\alpha\{\mu}\epsilon_T^{\nu\}\rho}p_{T\alpha}k_{T\rho\beta}S_{T}^\beta}{2M^{2}M_{h}}h_{1T}^{\perp}H_{1TT}^{\perp}\nonumber\\
   &-\frac{k_T^{\alpha\{\mu}\epsilon_T^{\nu\}\rho}S_{T\alpha}p_{T\rho\beta\tau}S_{hTT}^{\beta\tau}+p_T^{\alpha\beta\{\mu}\epsilon_T^{\nu\}\rho}S_{hTT\alpha\beta}k_{T\rho\tau}S_T^\tau}{2M^{2}M_{h}^{3}}h_{1T}^{\perp}H_{1TT}^{\perp\perp}\nonumber\\
	&-g_{T}^{\mu\nu}S_{L}S_{hLLL}g_{1L}G_{1LLL}-\frac{k_T^{\{\mu}p_T^{\nu\}}-(k_T\cdot p_T)g_T^{\mu\nu}}{MM_{h}}S_{L}S_{hLLL}h_{1L}^{\perp}H_{1LLL}^{\perp}\nonumber\\
	& -g_{T}^{\mu\nu}\frac{\bm{S}_{hLLT}\cdot\bm{p}_{T}}{M_{h}}S_{L}g_{1L}G_{1LLT}^{\perp}-\frac{
   k_T^{\{\mu}S_{hLLT}^{\nu\}}-(k_T\cdot S_{hLLT})g_T^{\mu\nu}}{M}S_{L}h_{1L}^{\perp}H_{1LLT}\nonumber\\
	& +\frac{k_T^{\{\mu}p_T^{\nu\}\alpha}S_{hLLT\alpha}-p_T^{\alpha\beta}S_{hLLT\alpha}k_{T\beta}g_T^{\mu\nu}}{MM_{h}^{2}}S_{L}h_{1L}^{\perp}H_{1LLT}^{\perp}-g_{T}^{\mu\nu}\frac{S_{hLTT}^{\alpha\beta}p_{T\alpha\beta}}{M_{h}^{2}}S_{L}g_{1L}G_{1LTT}^{\perp}\nonumber\\
	&-\frac{k_T^{\{\mu}S_{hLTT}^{\nu\}\alpha}p_{T\alpha}-S_{hLTT}^{\alpha\beta}p_{T\alpha}k_{T\beta}g_{T}^{\mu\nu}}{MM_{h}}S_{L}h_{1L}^{\perp}H_{1LTT}^{\perp}\nonumber\\
 &-\frac{k_T^{\{\mu}p_T^{\nu\}\alpha\beta}S_{hLTT\alpha\beta}-p_T^{\alpha\beta\rho}S_{hLTT\alpha\beta}k_{T\rho}g_T^{\mu\nu}}{MM_{h}^{3}}S_{L}h_{1L}^{\perp}H_{1LTT}^{\perp\perp}\nonumber\\
	& -g_{T}^{\mu\nu}\frac{S_{hTTT}^{\alpha\beta\rho}p_{T\alpha\beta\rho}}{M_{h}^{3}}S_{L}g_{1L}G_{1TTT}^{\perp}-\frac{k_T^{\{\mu}S_{hTTT}^{\nu\}\alpha\beta}p_{T\alpha\beta}-S_{hTTT}^{\alpha\beta\rho}p_{T\alpha\beta}k_{T\rho}g_{T}^{\mu\nu}}{MM_{h}^{2}}S_{L}h_{1L}^{\perp}H_{1TTT}^{\perp}\nonumber\\
	& -\frac{k_T^{\{\mu}p_{T}^{\nu\}\alpha\beta\rho}S_{hTTT\alpha\beta\rho}-p_{T}^{\alpha\beta\rho\tau}S_{hTTT\alpha\beta\rho}k_{T\tau}g_{T}^{\mu\nu}}{MM_{h}^{4}}S_{L}h_{1L}^{\perp}H_{1TTT}^{\perp\perp}
  -g_{T}^{\mu\nu}\frac{\bm{k}_{T}\cdot\bm{S}_{T}}{M}S_{hLLL}g_{1T}G_{1LLL}\nonumber\\
	& -\frac{S_T^{\{\mu}p_{T}^{\nu\}}-(S_T\cdot p_T)g_{T}^{\mu\nu}}{M_{h}}S_{hLLL}h_{1T}H_{1LLL}^{\perp}+\frac{k_T^{\alpha\{\mu}p_T^{\nu\}}S_{T\alpha}-k_T^{\alpha\beta}S_{T\alpha}p_{T\beta}g_{T}^{\mu\nu}}{M_{h}}h_{1T}^{\perp}H_{1LLL}^{\perp}\nonumber\\
	%
	&+g_{T}^{\mu\nu}\frac{\epsilon_{T}^{\rho\sigma}k_{T\rho}S_{T\sigma}}{M}\frac{\epsilon_{T}^{\alpha\beta}p_{T\alpha}S_{hLLT\beta}}{M_{h}}f_{1T}^{\perp}D_{1LLT}^{\perp}-g_{T}^{\mu\nu}\frac{\bm{k}_{T}\cdot\bm{S}_{T}}{M}\frac{\bm{p}_{T}\cdot\bm{S}_{hLLT}}{M_{h}}g_{1T}G_{1LLT}^{\perp}\nonumber\\
   &-\left(S_T^{\{\mu}S_{hLLT}^{\nu\}}-(S_T\cdot S_{hLLT})g_{T}^{\mu\nu}\right)h_{1T}H_{1LLT}
   +\frac{S_T^{\{\mu}p_T^{\nu\}\alpha}S_{hLLT\alpha}-p_T^{\alpha\beta}S_{hLLT\alpha}S_{T\beta}}{M_{h}^{2}}h_{1T}H_{1LLT}^{\perp}\nonumber\\
    &+\frac{k_T^{\alpha\{\mu}S_{hLLT}^{\nu\}}S_{T\alpha}-k_T^{\alpha\beta}S_{T\alpha}S_{hLLT\beta}g_{T}^{\mu\nu}}{M^{2}}h_{1T}^{\perp}H_{1LLT}\nonumber\\
	& -\frac{k_T^{\alpha\{\mu}p_{T}^{\nu\}\beta}S_{T\alpha}S_{hLLT\beta}-k_T^{\alpha\beta}S_{T\alpha}p_{T\beta\rho}S_{hLLT}^\rho g_{T}^{\mu\nu}}{M^{2}M_{h}^{2}}h_{1T}^{\perp}H_{1LLT}^{\perp}\nonumber\\
	%
	& +g_{T}^{\mu\nu}\frac{\epsilon_{T}^{\rho\sigma}k_{T\rho}S_{T\sigma}}{M}\frac{\epsilon_{T\beta}^{\alpha}p_{T\alpha\tau}S_{hLTT}^{\beta\tau}}{M_{h}^{2}}f_{1T}^{\perp}D_{1LTT}^{\perp}-g_{T}^{\mu\nu}\frac{\bm{k}_{T}\cdot\bm{S}_{T}}{M}\frac{S_{hLTT\alpha\beta}p_{T}^{\alpha\beta}}{M_{h}^{2}}g_{1T}G_{1LTT}^{\perp}\nonumber\\
	& -\frac{S_T^{\{\mu}S_{hLTT}^{\nu\}\alpha}p_{T\alpha}-S_{hLTT}^{\alpha\beta}p_{T\alpha}S_{T\beta}
    g_{T}^{\mu\nu}}{M_{h}}h_{1T}H_{1LTT}^{\perp}\nonumber\\
    &-\frac{S_T^{\{\mu}p_{T}^{\nu\}\alpha\beta}S_{hLTT\alpha\beta}-p_{T}^{\alpha\beta\rho}S_{hLTT\alpha\beta}S_{T\rho}g_{T}^{\mu\nu}}{M_{h}^{3}}h_{1T}H_{1LTT}^{\perp\perp}\nonumber\\
	&+\frac{k_T^{\alpha\{\mu}S_{hLTT}^{\nu\}\beta}S_{T\alpha}p_{T\beta}-k_T^{\alpha\beta}S_{T\alpha}S_{hLTT\beta\rho}p_T^{\rho}g_{T}^{\mu\nu}}{M^{2}M_{h}}h_{1T}^{\perp}H_{1LTT}^{\perp}\nonumber\\
    &+\frac{k_T^{\alpha\{\mu}p_T^{\nu\}\beta\rho}S_{T\alpha}S_{hLTT\beta\rho}-p_T^{\alpha\beta\rho}S_{hLTT\alpha\beta}k_{T\rho\tau}S_T^\tau g_{T}^{\mu\nu}}{M^{2}M_{h}^{3}}h_{1T}^{\perp}H_{1LTT}^{\perp\perp}\nonumber\\
	%
	& +g_{T}^{\mu\nu}\frac{\epsilon_{T}^{\rho\sigma}k_{T\rho}S_{T\sigma}}{M}\frac{\epsilon_{T\beta}^{\alpha}p_{T\alpha\tau\gamma}S_{hTTT}^{\beta\tau\gamma}}{M_{h}^{3}}f_{1T}^{\perp}D_{1TTT}^{\perp}-g_{T}^{\mu\nu}\frac{\bm{k}_{T}\cdot\bm{S}_{T}}{M}\frac{S_{hTTT\alpha\beta\rho}p_{T}^{\alpha\beta\rho}}{M_{h}^{3}}g_{1T}G_{1TTT}^{\perp}\nonumber\\
	& -\frac{S_T^{\{\mu}S_{hTTT}^{\nu\}\alpha\beta}p_{T\alpha\beta}-S_{hTTT}^{\alpha\beta\rho}p_{T\alpha\beta}S_{T\rho}g_{T}^{\mu\nu}}{M_{h}^{2}}h_{1T}H_{1TTT}^{\perp}\nonumber\\
    &-\frac{S_T^{\{\mu}p_T^{\nu\}\alpha\beta\rho}S_{hTTT\alpha\beta\rho}-p_T^{\alpha\beta\rho\tau}S_{hTTT\alpha\beta\rho}S_{T\tau}g_{T}^{\mu\nu}}{M_{h}^{4}}h_{1T}H_{1TTT}^{\perp\perp}\nonumber\\
	& +\frac{k_T^{\alpha\{\mu}S_{hTTT}^{\nu\}\beta\rho}S_{T\alpha}p_{T\beta\rho}-S_{hTTT}^{\alpha\beta\rho}p_{T\alpha\beta}k_{T\rho\tau}S_T^\tau g_{T}^{\mu\nu}}{M^{2}M_{h}^{2}}h_{1T}^{\perp}H_{1TTT}^{\perp}\nonumber\\
 &+\frac{k_T^{\alpha\{\mu}p_T^{\nu\}\beta\rho\tau}S_{T\alpha}S_{hTTT\beta\rho\tau}-p_T^{\alpha\beta\rho\sigma}S_{hTTT\alpha\beta\rho}k_{T\sigma\tau}S_T^\tau g_{T}^{\mu\nu}}{M^{2}M_{h}^{4}}h_{1T}^{\perp}H_{1TTT}^{\perp\perp}
	\bigg\},
\end{align}
and $W_A$ is the antisymmetric part, 
\begin{align}
	W_A^{\mu\nu}=&2z \sum_a e_a^2 \int d^2 \boldsymbol{k}_T d^2 \boldsymbol{p}_T \delta^{(2)}\left(\boldsymbol{k}_T-\boldsymbol{p}_T-\boldsymbol{P}_{h \perp} / z\right)\nonumber\\
	&\bigg\{i\epsilon_{T}^{\mu\nu}\Big(S_{hL}f_{1}G_{1L}+\frac{\bm{S}_{hT}\cdot\bm{p}_{T}}{M_{h}}f_{1}G_{1T}^{\perp} \Big)-i\frac{p_T^{[\mu}S_{hLT}^{\nu]}}{M_{h}}f_{1}G_{1LT}^{\perp}+i\epsilon_{T}^{\mu\nu}\frac{\epsilon_{T\beta}^{\alpha}p_{T\alpha\rho}S_{hTT}^{\beta\rho}}{M_{h}^{2}}f_{1}G_{1TT}^{\perp}\nonumber\\
	&+i\epsilon_{T}^{\mu\nu}\Big(S_{hLLL}f_{1}G_{1LLL}+\frac{\bm{S}_{hLLT}\cdot\bm{p}_{T}}{M_{h}}f_{1}G_{1LLT}^{\perp}+\frac{S_{hLTT\alpha\beta}p_{T}^{\alpha\beta}}{M_{h}^{2}}f_{1}G_{1LTT}^{\perp}+\frac{S_{hTTT\alpha\beta\rho}p_{T}^{\alpha\beta\rho}}{M_{h}^{3}}f_{1}G_{1TTT}^{\perp}\Big)\nonumber\\
    &+iS_{L}\frac{p_T^{[\mu}S_{hT}^{\nu]}}
 {M_{h}}g_{1L}D_{1T}^{\perp}-i\frac{k_T^{[\mu}S_{T}^{\nu]}}{M}S_{hL}f_{1T}^{\perp}G_{1L}-i\frac{k_T^{[\mu}S_{T}^{\nu]}}{M}\frac{\bm{S}_{hT}\cdot\bm{p}_{T}}{M_{h}}f_{1T}^{\perp}G_{1T}^{\perp}+i\frac{\bm{S}_{T}\cdot\bm{k}_{T}}{M}\frac{p_T^{[\mu}S_{hT}^{\nu]}}{M_{h}}g_{1T}^{\perp}D_{1T}^{\perp}\nonumber\\
	&+i\epsilon_{T}^{\mu\nu}\bigg(S_{L}g_{1L}D_{1}+\frac{\bm{S}_{T}\cdot\bm{k}_{T}}{M}g_{1T}^{\perp}D_{1}+S_{L}S_{hLL}g_{1L}D_{1LL}+\frac{\bm{k}_{T}\cdot\bm{S}_{T}}{M}S_{hLL}g_{1T}D_{1LL}\nonumber\\
	&+S_{L}\frac{\bm{S}_{hLT}\cdot\bm{p}_{T}}{M_{h}}g_{1L}D_{1LT}^{\perp}+\frac{\bm{k}_{T}\cdot\bm{S}_{T}}{M}\frac{\bm{S}_{hLT}\cdot\bm{p}_{T}}{M_{h}}g_{1T}D_{1LT}^{\perp}+\frac{\epsilon_{T}^{\rho\sigma}k_{T\rho}S_{T\sigma}}{M}\frac{\epsilon_{T}^{\alpha\beta}p_{T\alpha}S_{hLT\beta}}{M_{h}}f_{1T}^{\perp}G_{1LT}^{\perp}\nonumber\\
	&+S_{L}\frac{p_{T}^{\alpha\beta}S_{hTT\alpha\beta}}{M_{h}^{2}}g_{1L}D_{1TT}^{\perp}+\frac{\bm{k}_{T}\cdot\bm{S}_{T}}{M}\frac{p_{T}^{\alpha\beta}S_{hTT\alpha\beta}}{M_{h}^{2}}g_{1T}D_{1TT}^{\perp}-\frac{\epsilon_{T}^{\rho\sigma}k_{T\rho}S_{T\sigma}}{M}\frac{\epsilon_{T\beta}^{\alpha}p_{T\alpha\rho}S_{hTT}^{\beta\rho}}{M_{h}^{2}}f_{1T}^{\perp}G_{1TT}^{\perp}\bigg)\nonumber\\
	&+i\frac{p_T^{[\mu}S_{hLLT}^{\nu]}}{M_{h}}S_{L}g_{1L}D_{1LLT}^{\perp}+i\epsilon_{T}^{\mu\nu}\bigg(\frac{\epsilon_{T\beta}^{\alpha}p_{T\alpha\rho}S_{hLTT}^{\beta\rho}}{M_{h}^{2}}S_{L}g_{1L}D_{1LTT}^{\perp}+\frac{\epsilon_{T\beta}^{\alpha}p_{T\alpha\rho\sigma}S_{hTTT}^{\beta\rho\sigma}}{M_{h}^{3}}S_{L}g_{1L}D_{1TTT}^{\perp}\bigg)\nonumber\\
	&-i\frac{k_T^{[\mu}S_{T}^{\nu]}}{M}\bigg(
		S_{hLLL}f_{1T}^{\perp}G_{1LLL}+\frac{\bm{S}_{hLLT}\cdot\bm{p}_{T}}{M_{h}}f_{1T}^{\perp}G_{1LLT}^{\perp}
		+\frac{S_{hLTT\alpha\beta}p_{T}^{\alpha\beta}}{M_{h}^{2}}f_{1T}^{\perp}G_{1LTT}^{\perp}\nonumber\\
	&+\frac{S_{hTTT\alpha\beta\rho}p_{T}^{\alpha\beta\rho}}{M_{h}^{3}}f_{1T}^{\perp}G_{1TTT}^{\perp}\bigg)+i\frac{p_T^{[\mu}S_{hLLT}^{\nu]}}{M_{h}}\frac{\bm{k}_{T}\cdot\bm{S}_{T}}{M}g_{1T}D_{1LLT}^{\perp}\nonumber\\
	&+i\epsilon_{T}^{\mu\nu}\bigg(
		\frac{\text{\ensuremath{\epsilon_{T\beta}^{\alpha}p_{T\alpha\rho}S_{hLTT}^{\beta\rho}}}}{M_{h}^{2}}\frac{\bm{k}_{T}\cdot\bm{S}_{T}}{M}g_{1T}D_{1LTT}^{\perp}+\frac{\text{\ensuremath{\epsilon_{T\beta}^{\alpha}p_{T\alpha\rho\sigma}S_{hTTT}^{\beta\rho\sigma}}}}{M_{h}^{3}}\frac{\bm{k}_{T}\cdot\bm{S}_{T}}{M}g_{1T}D_{1TTT}^{\perp}
	\bigg)
	\bigg\}.
\end{align}

\end{document}